\documentclass[apj]{emulateapj}

\usepackage{graphicx}
\usepackage{amsmath,amssymb}
\usepackage{mathptmx}
\usepackage[english]{babel}
\usepackage{booktabs}
\usepackage{multirow}
\usepackage{natbib}
\usepackage[breaklinks,colorlinks,
   urlcolor=blue,citecolor=blue,linkcolor=blue]{hyperref}

\def\FSMII{FSM17}

\def\FSMIII{Faerman~et~al., in prep.}

\newcommand{\beq}	{\begin{equation}}
\newcommand{\eeq}	{\end{equation}}
\newcommand{\beqs}	{\begin{displaymath}}
\newcommand{\eeqs}	{\end{displaymath}}
\newcommand{\beqa}{\begin{eqnarray}}
\newcommand{\eeqa}{\end{eqnarray}}
\newcommand{\beqas}	{\begin{eqnarray*}}
\newcommand{\eeqas}	{\end{eqnarray*}}

\def\vmax{\ifmmode {v_{\rm max}} \else $v_{\rm max}$\fi}
\def\rmax{\ifmmode {r_{\rm max}} \else $r_{\rm max}$\fi}

\def\vs{\ifmmode {v_s} \else $v_s$\fi}
\def\rs{\ifmmode {r_s} \else $r_s$\fi}
\def\mds{\ifmmode {M_{ds}} \else $M_{ds}$\fi}
\def\nds{\ifmmode {n_{ds}} \else $n_{ds}$\fi}
\def\rods{\ifmmode {\rho_{ds}} \else $\rho_{ds}$\fi}
\def\m300{\ifmmode {M_{300}} \else $M_{300}$\fi}
\def\mm150{\ifmmode {M_{150}} \else $M4_{150}$\fi}
\def\x300{\ifmmode {x_{300}} \else $x_{300}$\fi}

\def\alp{\ifmmode {\alpha} \else $\alpha$\fi}
\def\sig{\ifmmode {\sigma} \else $\sigma$\fi}
\def\tvir{\ifmmode {{\it T}_{\rm vir}} \else ${\it T}_{\rm vir}$\fi}
\def\tshock{\ifmmode {{\it T}_{\rm vir}} \else ${\it T}_{\rm vir}$\fi}

\def\nvirp{\ifmmode {n_{\rm H}(\rvir)} \else $n_{\rm H}(\rvir)$\fi}
\def\tcgm{\ifmmode {T_{\rm th}(\rcgm)} \else $T_{\rm th}(\rcgm)$\fi}
\def\ncgm{\ifmmode {n_{\rm H}(\rcgm)} \else $n_{\rm H}(\rcgm)$\fi}
\def\alpcgm{\ifmmode {\alpha(\rcgm)} \else $\alpha(\rcgm)$\fi}
\def\mvir{\ifmmode {M_{\rm vir}} \else $M_{\rm vir}$\fi}
\def\rvir{\ifmmode {r_{\rm vir}} \else $r_{\rm vir}$\fi}
\def\xvir{\ifmmode {x_{\rm vir}} \else $x_{\rm vir}$\fi}
\def\xcgm{\ifmmode {x_{\rm cgm}} \else $x_{\rm cgm}$\fi}
\def\rcgm{\ifmmode {r_{\rm CGM}} \else $r_{\rm CGM}$\fi}
\def\mdm{\ifmmode {M_{\rm DM}} \else $M_{\rm DM}$\fi}
\def\mdisk{\ifmmode {M_{\rm disk}} \else $M_{\rm disk}$\fi}
\def\mb{\ifmmode {M_{\rm b}} \else $M_{\rm b}$\fi}
\def\mbm{\ifmmode {M_{\rm b,miss}} \else $M_{\rm b,miss}$\fi}
\def\mbo{\ifmmode {M_{\rm b,obs}} \else $M_{\rm b,obs}$\fi}
\def\fb{\ifmmode {f_{\rm b}} \else $f_{\rm b}$\fi}
\def\mhw{\ifmmode {M_{\rm hot/warm}} \else $M_{\rm hot/warm}$\fi}
\def\mhot{\ifmmode {M_{\rm hot}} \else $M_{\rm hot}$\fi}
\def\mwarm{\ifmmode {M_{\rm warm}} \else $M_{\rm warm}$\fi}
\def\mcool{\ifmmode {M_{\rm cool}} \else $M_{\rm cool}$\fi}
\def\mgas{\ifmmode {M_{\rm CGM}} \else $M_{\rm CGM}$\fi}
\def\mcor{\ifmmode {M_{\rm CGM}} \else $M_{\rm CGM}$\fi}
\def\mhalo{\ifmmode {M_{\rm halo}} \else $M_{\rm halo}$\fi}
\def\rsun{\ifmmode {R_{\rm 0}} \else $R_{\rm 0}$\fi}

\def\Tpeak{\ifmmode {T_{\rm peak}} \else $T_{\rm peak}$\fi}
\def\Twarm{\ifmmode {T_{\rm warm}} \else $T_{\rm warm}$\fi}
\def\tgas{\ifmmode {T_{\rm gas}} \else $T_{\rm gas}$\fi}
\def\rgas{\ifmmode {r_{\rm gas}} \else $r_{\rm gas}$\fi}
\def\rogas{\ifmmode {\rho_{\rm gas}} \else $\rho_{\rm gas}$\fi}
\def\mhi{\ifmmode {M_{HI}} \else $M_{HI}$\fi}
\def\nhi{\ifmmode {N_{HI}} \else $N_{HI}$\fi}
\def\rhi{\ifmmode {R_{\rm HI}} \else $R_{\rm HI}$\fi}
\def\rhalf{\ifmmode {r_{1/2}} \else $r_{1/2}$\fi}
\def\thalf{\ifmmode {\theta_{1/2}} \else $\theta_{1/2}$\fi}
\def\tgc{\ifmmode {\theta_{\rm GC}} \else $\theta_{\rm GC}$\fi}
\def\npeak{\ifmmode {N_{\rm peak}} \else $N_{\rm peak}$\fi}
\def\tdyn{\ifmmode {t_{\rm dyn}} \else $t_{\rm dyn}$\fi}
\def\tcool{\ifmmode {t_{\rm cool}} \else $t_{\rm cool}$\fi}
\def\fcool{\ifmmode {f_{\rm cool}} \else $f_{\rm cool}$\fi}
\def\trec{\ifmmode {t_{\rm rec}} \else $t_{\rm rec}$\fi}
\def\Tmin{\ifmmode {T_{\rm min}} \else $T_{\rm min}$\fi}
\def\mbar{\ifmmode {\bar{m}} \else $\bar{m}$\fi}
\def\vturb{\ifmmode {\sigma_{\rm turb}} \else $\sigma_{\rm turb}$\fi}
\def\Psun{\ifmmode {P_{\rm th}(\rsun)} \else $P_{\rm th}(\rsun)$\fi}
\def\s04{\ifmmode {S_{0.4-2.0}} \else $S_{0.4-2.0}$\fi}
\def\Lcool{\ifmmode {L_{\rm cool}} \else $L_{\rm cool}$\fi}
\def\Lrad{\ifmmode {L_{\rm rad}} \else $L_{\rm rad}$\fi}
\def\cgm{\ifmmode {\rm CGM} \else ${\rm CGM}$\fi}
\def\zmean{\ifmmode {\left< Z' \right>} \else $\left< Z' \right>$\fi}
\def\lscale{\ifmmode {{L_s} } \else ${L_s}$\fi}
\def\aL{\ifmmode {a_{\Lambda} } \else $a_{\Lambda}$\fi}
\def\UH{\ifmmode {\left<U\right>_{H}} \else $\left<U\right>_{H}$\fi}
\def\zmin{\ifmmode {\zeta_{min}} \else $\zeta_{min}$\fi}
\def\ypar{\ifmmode {\Tilde{Y}_{500}} \else $\Tilde{Y}_{500}$\fi}
\def\fion{\ifmmode {f_{\rm ion}} \else $f_{\rm ion}$\fi}

\def\kal{\ifmmode {K_{\alpha}} \else $K_{\alpha}$\fi}
\def\kbe{\ifmmode {K_{\beta}} \else $K_{\beta}$\fi}

\def\nh{\ifmmode {n_{\rm H}} \else $n_{\rm H}$\fi}
\def\ca{\ifmmode {\chi_{\rm EM}} \else $\chi_{\rm EM}$\fi}
\def\cb{\ifmmode {\chi_{\rm N}} \else $\chi_{\rm N}$\fi}
\def\lia{\ifmmode {22~{\rm \AA}} \else $22~{\rm \AA}$\fi}
\def\lib{\ifmmode {19~{\rm \AA}} \else $19~{\rm \AA}$\fi}

\def\cg{\ifmmode {c_g} \else $c_g$\fi}
\def\c6{\ifmmode {c_{g,6}} \else $c_{g,6}$\fi}
\def\phim{\ifmmode {P_{\rm HIM}} \else $P_{\rm HIM}$\fi}

\def\chisq{\ifmmode {\chi_{mod}^2} \else $\chi_{mod}^2$\fi}

\def\kb{\ifmmode {k_{\rm B}} \else $k_{\rm B}$\fi}
\def\mp{\ifmmode {m_{\rm p}} \else $m_{\rm p}$\fi}

\def\msun{\ifmmode {\rm M_{\odot}} \else $\rm M_{\odot}$\fi}
\def\lsun{\ifmmode {\rm L_{\odot}} \else $\rm L_{\odot}$\fi}
\def\msuny{\ifmmode {\rm M_{\odot}\:yr^{-1}} \else $\rm M_{\odot}\:yr^{-1}$\fi}
\def\ergs{\ifmmode {\rm erg\:s^{-1}} \else $\rm erg\:s^{-1}$\fi}

\def\kms{\ifmmode {\rm km\:s^{-1}} \else $\rm km\:s^{-1}$\fi}
\def\cmc{\ifmmode {\rm cm^{-2}} \else $\rm cm^{-2}$\fi}
\def\cmv{\ifmmode {\rm cm^{-3} \:} \else $\rm cm^{-3}$\fi}
\def\kel{\ifmmode {\rm \:\: K} \else $\rm \:\: K$\fi}
\def\pc{\ifmmode {\rm \:\: pc} \else $\rm \:\: pc$\fi}
\def\kpc{\ifmmode {\rm \:\: kpc} \else $\rm \:\: kpc$\fi}
\def\amu{\ifmmode {\rm \:\: amu} \else $\rm \:\: amu$\fi}
\def\fluxun{\ifmmode {\rm \: erg~s^{-1}~cm^{-2}~deg^{-2}} \else $\rm \: erg~s^{-1}~cm^{-2}~deg^{-2}$\fi}
\def\liun{\ifmmode {\rm \:\: photons~s^{-1}~cm^{-2}~sr^{-1}} \else $\rm \:\: photons~s^{-1}~cm^{-2}~sr^{-1}$\fi}

\def\aox{\ifmmode {a_{\rm O}} \else $a_{\rm O}$\fi}

\def\gv{\ifmmode {g_{\mbox{\tiny {\it V}}}} \else $g_{\mbox{\tiny {\it V}}}$\fi}
\def\gM{\ifmmode {g_{\mbox{\tiny {\it M}}}} \else $g_{\mbox{\tiny {\it M}}}$\fi}

\def\ppv{\ifmmode {p_{\mbox{\tiny {\it V}}}} \else $p_{\mbox{\tiny {\it V}}}$\fi}
\def\ppm{\ifmmode {p_{\mbox{\tiny {\it M}}}} \else $p_{\mbox{\tiny {\it M}}}$\fi}
\def\pfv{\ifmmode {f_{\mbox{\tiny {\it V}}}} \else $f_{\mbox{\tiny {\it V}}}$\fi}
\def\pfm{\ifmmode {f_{\mbox{\tiny {\it M}}}} \else $f_{\mbox{\tiny {\it M}}}$\fi}

\def\pvw{\ifmmode {p_{\mbox{\tiny {\it V}},w}} \else $p_{\mbox{\tiny {\it V}},w}$\fi}
\def\pmw{\ifmmode {p_{\mbox{\tiny {\it M}},w}} \else $p_{\mbox{\tiny {\it M}},w}$\fi}
\def\fvw{\ifmmode {f_{\mbox{\tiny {\it V}},w}} \else $f_{\mbox{\tiny {\it V}},w}$\fi}
\def\fmw{\ifmmode {f_{\mbox{\tiny {\it M}},w}} \else $f_{\mbox{\tiny {\it M}},w}$\fi}

\shorttitle{Massive Coronae - Isentropic Model}
\shortauthors{Faerman et al.}
\slugcomment{Accepted to ApJ}

\begin{document}

\title{Massive Warm/Hot Galaxy Coronae: II. Isentropic Model}

\author{
Yakov Faerman \altaffilmark{1 *},
Amiel Sternberg \altaffilmark{2,3,4},
and Christopher F. McKee \altaffilmark{5}}

\altaffiltext{*}{e-mail: \href{mailto:yakov.faerman@mail.huji.ac.il}{yakov.faerman@mail.huji.ac.il}}
\altaffiltext{1}
{Racah Institute of Physics,
The Hebrew University, Jerusalem 91904, Israel}
\altaffiltext{2}
{School of Physics and Astronomy,
Tel Aviv University, Ramat Aviv 69978, Israel}
\altaffiltext{3}
{Center for Computational Astrophysics,
Flatiron Institute, 162 5th Avenue, 10010, New York, NY, USA}
\altaffiltext{4}
{Max-Planck-Institut fur Extraterrestrische Physik (MPE),
Giessenbachstr., 85748 Garching, FRG}
\altaffiltext{5}
{Department of Physics and Department of Astronomy,
University of California at Berkeley, Berkeley CA 94720}

\begin{abstract}
We construct a new analytic phenomenological model for the extended circumgalactic material (CGM) of $L^*$ galaxies. Our model reproduces the OVII/OVIII absorption observations of the Milky Way (MW) and the OVI measurements reported by the COS-Halos and eCGM surveys. The warm/hot gas is in hydrostatic equilibrium in a MW gravitational potential, and we adopt a barotropic equation of state, resulting in a temperature variation as a function of radius. A pressure component with an adiabatic index of $\gamma=4/3$ is included to approximate the effects of a magnetic field and cosmic rays. We introduce a metallicity gradient motivated by the enrichment of the inner CGM by the Galaxy. We then present our fiducial model for the corona, tuned to reproduce the observed OVI-OVIII column densities, and with a total mass of $\mgas \approx 5.5 \times 10^{10}$~\msun~inside $\rcgm \approx 280$~kpc. The gas densities in the CGM are low ($\nh = 10^{-5} - 3 \times 10^{-4}$~\cmv) and its collisional ionization state is modified by the metagalactic radiation field (MGRF). We show that for OVI-bearing warm/hot gas with typical observed column densities $N_{\rm OVI} \sim 3 \times 10^{14}$~\cmc~at large ($\gtrsim 100$~kpc) impact parameters from the central galaxies, the ratio of the cooling to dynamical times, $\tcool/\tdyn$, has a model-independent upper limit of $\lesssim 4$. In our model, $\tcool/\tdyn$ at large radii is $\sim 2-3$. We present predictions for a wide range of future observations of the warm/hot CGM, from UV/X-ray absorption and emission spectroscopy, to dispersion measure (DM) and Sunyaev-Zeldovich CMB measurements. We provide the model  outputs in machine-readable data files, for easy comparison and analysis.
\end{abstract}


\section{Introduction}
\label{sec_intro}

Observations of diffuse matter around galaxies, the circumgalactic medium (CGM), provide evidence for substantial reservoirs of ``warm/hot" ($10^5-10^6$~K) gas extending to large radii from the central galaxies \citep{P11,Tumlinson11,Gupta12,Johnson15,Burchett19}. The warm/hot CGM is traced by absorption and emission lines of highly ionized species in the UV and X-ray \citep{Bregman07,Henley10,Henley10lines}. Observations also find a cool ($\sim 10^{4}$~K) phase in the CGM, detected through absorption features from hydrogen and lower metal ions \citep{Werk13,Prochaska17}. Many questions remain open, such as what are the density and temperature distributions of the CGM, its metallicity, and ionization state, and total mass \citep{Bregman07rev,Putman12rev,TPW17}. Numerical simulations addressing these questions are challenging, due to the high resolution required and the computational cost \citep{Hummels19,Peeples19}. The properties of the simulated CGM are also sensitive to the assumed physical models, such as the feedback prescriptions and physical processes on small scales \citep{McCourt12,Fielding17b,Ji19,LiB20}. Analytic models provide a different avenue to address the open questions regarding the structure of the CGM \citep{MB04,Anderson10,Miller13,MP17,Stern18,MW18,QB18a, V19}.

In \citet[hereafter FSM17]{FSM17} we presented a two-phase model, with separate warm and hot components, for the circumgalactic corona, with the mean gas temperature constant (isothermal model) as a function of radius in each phase. We assumed that the metallicity is constant throughout the corona and found that a value of $Z'=0.5$ solar is needed to reproduce the oxygen column densities that are measured in absorption. Large CGM gas masses, comparable to those required for ``baryonic closure" of the parent galaxy halos, are also needed. Our isothermal model in FSM17 is successful in reproducing the highly ionized oxygen columns, but with some challenges, such as high gas temperature and pressure in the hot phase, and a short cooling time of the warm phase\footnote {~In this paper we adopt the terminology used for the CGM in the literature - ``warm/hot" for gas temperatures between $10^{5}$ and $10^{7}$~K, and ``cool" for $\sim 10^4$~K gas (see also \citealp{Werk16,Prochaska17})}.

In this paper, we construct an alternate model for the CGM in which we assume constant entropy (isentropic model) leading to a single phased structure with a large scale temperature gradient, from hot to warm. First, in \S\ref{sec_model}, we present the framework of our model. We solve the equation of hydrostatic equilibrium assuming a constant entropy adiabatic relation between the gas density and temperature, resulting in a temperature variation as a function of radius. We introduce a metallicity gradient and discuss the values for boundary conditions of the gas distributions. 
In \S\ref{sec_fiducial} we present our fiducial isentropic model, defined by a specific set of parameters chosen to reproduce absorption measurements of highly ionized oxygen ions (OVI-OVIII). As in FSM17, we focus on the MW and external galaxies for which OVI has been detected in the CGM. We show the gas density and temperature distributions in the model, discuss the gas ionization mechanisms and calculate the spatial distributions of ions and gas emission properties. We then address the different timescales in the model in \S\ref{sec_timescales}. We also derive a model-independent upper limit for the cooling to dynamical time ratio for OVI-bearing gas. 
In \S\ref{sec_comparison} we compare the model properties to observational data measured in the MW and other, low-redshift $L^*$ galaxies, and provide predictions for future observations in \S\ref{sec_predict}. We compare our current model to \FSMII~in \S\ref{sec_fsm17}, discuss the differences between our work and other models of the CGM in \S\ref{sec_disc}, and summarize in \S\ref{sec_summary}.

\section{Isentropic Model}
\label{sec_model}

In this section we introduce our model framework for setting the spatial distributions of the gas density, temperature and metallicity. As in \FSMII, we assume that the coronal gas is in hydrostatic equilibrium (HSE) within the gravitational potential of the central Galaxy and dark matter halo, with negligible self-gravity for the gas. We assume that the gas is supported by thermal pressure, magnetic fields and cosmic rays, and turbulence. Given the evidence for turbulence in the CGM \citep{Tumlinson2011b,Genel14,Werk16}, we do not imagine a perfect HSE. However, in the absence of large scale coherent motions (inflows or outflows) there can exist a close-to-equilibrium steady state \citep{Nelson16,Fielding17a,Loch20}. As in \FSMII~we assume a spherical version of the Milky Way potential presented by \citet{Klypin02}. In \S\ref{sec_timescales} we discuss the dynamical and cooling timescales in the corona.

In \FSMII~we assumed a constant (isothermal) mean temperature throughout the corona, and we invoked isobaric density and temperature fluctuations to enable simultaneous production of OVII and OVIII, and a cooling component for the OVI. \FSMII~is thus a multiphased model, hot and at constant mean temperature for OVII and OVIII, and warm for OVI cooling out of the hot. In our new isentropic model, the altered (adiabatic) equation of state (EoS) leads to a temperature gradient, enabling production of OVI, OVII and OVIII at differing radii, but in a single phase. In our new model we no longer require local temperature fluctuations. However, we still include turbulent motions as one of the sources of hydrostatic support. Furthermore, in our current model we adopt a varying metallicity profile, motivated by enrichment of the CGM by the galaxy. In \FSMII~we assumed constant metallicity. Finally, the gas temperature and density at the virial radius in our new model are lower than in \FSMII, leading to a lower CGM pressure at the boundary with the intergalactic medium (IGM). This is more consistent with our assumption of a large scale equilibrium, and low accretion rates onto the MW halo in the recent past.

We present the HSE equation with our new EoS in \S\ref{subsec_frame}, add a metallicity gradient in \S\ref{subsec_metallicity}, and in \S\ref{subsec_bounds} discuss the boundary conditions, needed to compute the actual gas distributions.

\subsection{Equation of State and Hydrostatic Equilibrium}
\label{subsec_frame}

Since the Galactic corona may be heated by AGN feedback and star formation, we imagine that it evolves toward a convective equilibrium. We therefore adopt an adiabatic EoS, relating the gas pressure and mass density,
\begin{equation}\label{eq:adiab1}
P(r) = K \rho(r)^{\gamma} ~~~,
\end{equation}
where $r$ is the radius and $K$ is the entropy parameter, which we assume is constant with radius. Using the ideal gas law allows us to relate the temperature to the density
\begin{equation}\label{eq:adiab2}
T(r) = K \frac{\mbar}{\kb} \rho(r)^{\gamma-1} ~~~,
\end{equation}
where $\mbar$ is the mean mass per particle.

For a mixture of $n$ fluids, we can write the HSE equation as the sum of the pressures for the different components
\begin{equation}\label{eq:hse1}
dP = \sum_{i=1}^n{dP_i} = -\rho d\varphi ~~~,
\end{equation}
where $\varphi$ is the gravitational potential. We include three pressure components, similar to those in \FSMII~- (i) thermal, (ii) non-thermal, from cosmic rays and magnetic fields, and (iii) turbulent support. We assume that the density of each component is proportional to the total gravitating gas mass density $\rho$. For the first two components we use the adiabatic EoS, with $\gamma_1=5/3$ and $\gamma_2=4/3$, respectively, and assume that the entropy parameter is constant with radius. For each component $dP_i = \gamma_i K_i\rho^{\gamma_i-1} d\rho$. For the turbulent component we assume a constant velocity scale, \vturb, as we did in \FSMII, and write $dP_3 = \vturb^2 d\rho$. Equation~\eqref{eq:hse1} is then
\begin{equation}\label{eq:hse2}
\left(\vturb^2 + \sum_{i=1,2}{\gamma_i K_i\rho^{\gamma_i-1}} \right) \rho^{-1}d\rho = - d\varphi ~~~.
\end{equation}
Integration then gives
\begin{equation}\label{eq:hse3}
\vturb^2\ln{\rho(r)} + \sum_{i=1,2}{\frac{\gamma_i}{\gamma_i-1} K_i\rho(r)^{\gamma_i-1}} = D_{b} - \int_{r_{b}}^r{\frac{GM(r)dr}{r^2}} ~~~,
\end{equation}
where $r_{b}$ is a reference point, which we normally take at the outer boundary, and $D_{b}$ is an integration constant.

To solve this equation for $\rho(r)$ for a given mass profile $M(r)$, we must specify~\vturb~and $K_i$.  The former is taken from observations of oxygen line velocities and widths (see \citealp{Tumlinson11}, the discussion in \FSMII~and \S\ref{subsec_bounds} here). For the latter - since in our model $K_i$ are constant with radius, they can be expressed as functions of the gas properties at the boundary $r_{b}$ - the temperature, $T_{{\rm th},b}$ and density, $\rho_{b}$. For the thermal component this is simply
\begin{equation}\label{eq:k1}
K_1 = \frac{\kb}{\mbar^{\gamma_1} } \frac{T_{{\rm th},b}}{n_{b}^{\gamma_1-1} } ~~~,
\end{equation}
where $n_b \equiv \rho_b/\mbar$ is the particle volume density. To obtain $K_2$, we use the \alp~parameter from \FSMII, defined as $\alp \equiv (P_{\rm th}+P_{\rm nth})/P_{\rm th}  = (T_{\rm th}+T_{\rm nth})/T_{\rm th}$. For isothermal conditions, $\alp$ is constant with radius. In our new model, the relative fractions of pressure support from each component vary with radius, and $\alp$ is not constant. We define $\alp_{b} \equiv \alp(r_{b}) = (T_{{\rm th},b}+T_{{\rm nth},b})/T_{{\rm th},b}$, allowing us to write 
\begin{equation}\label{eq:k2}
K_2 = \frac{\kb}{\mbar^{\gamma_2}} \frac{(\alp_{b} - 1)T_{{\rm th},b}}{n_{b}^{\gamma_2-1}} ~~~.
\end{equation}
Thus, given \vturb, and for the gas density, temperature and \alp~at the reference point, we can solve Equations \eqref{eq:hse2} or \eqref{eq:hse3} for the density profile, $\rho(r)$. We can then use the EoS (Eq. \ref{eq:adiab1}-\ref{eq:adiab2}) to find the pressure and temperature profiles for each of the corona components and the total pressure profile.

\subsection{Metallicity Distribution}
\label{subsec_metallicity}

The metal content of the CGM and its distribution are interesting for two reasons. First, the total metal content provides information on the cumulative metal production in the galaxy by star formation \citep{Peeples14}.  Second, observations of the CGM probe the gas properties, such as density and temperature, mainly through absorption and emission of radiation by metal ions \citep{Spitzer56,Bregman07rev,Prochaska09,TPW17}. Thus, metals are important as tracers of the gas distribution.

In \FSMII~we assumed a uniform metallicity distribution. In a more realistic scenario, the central region of the Galactic halo is expected to be enriched by metals, created in supernovae explosions and ejected from the disk by Galactic winds. The outer regions, close to the virial radius, may be dominated by metal-poor gas accreted from the cosmic web, resulting in a decreasing metallicity profile across the corona. Some of the accreted gas may also be pre-enriched. The level and extent of metal enrichment by outflows from the disk and the enrichment of the accreted gas depends on feedback energetics, the star formation history and distribution in the galaxy and the physics of gas mixing and diffusion in the corona (see \citealp{Fielding18} and \citealp{LiB20}).

The main observational constraints of our model in this work are oxygen absorption measurements, probing the gas phase metallicity. The mass of metals in the CGM locked in solid-state dust grains is an additional component (\citealp{Peek15}), and we do not address it here. As we discuss in \S\ref{sec_fiducial}, the mass if metals locked in dust is small compared to gas in our fiducial model, and we do not model the dust.

We adopt a metallicity profile given by
\begin{equation}\label{eq:zprofile}
Z'(r)  = Z'_0 \left[ 1+\left( \frac{r}{r_Z} \right)^2 \right]^{-1/2} ~~~,
\end{equation}
where $Z'_0$ is the Galactic metallicity and $r_Z$ is an adjustable metallicity length-scale within which the metallicity is equal to the inner metallicity $Z'_0$, and beyond which the metallicity decreases smoothly to the outer boundary of the CGM, which we denote by \rcgm. The length scale $r_Z$ can be set by estimating the maximal extent of outflows from the disk. Alternatively, we can set the metallicity at \rcgm, and then the length-scale is given by
\begin{equation}\label{eq:zlength}
r_Z  = \rcgm \left[ \left( \frac{Z'(\rcgm)}{Z'_0} \right)^2 - 1 \right]^{-1/2} ~~~.
\end{equation}

The mean metallicity is given by
\begin{equation}\label{eq:zmean1}
\zmean_{V} = \frac{\mbar}{\mcor} \int_{R_0}^{r_{\rm CGM}}{Z'(r)n(r)dV} ~~~,
\end{equation}
where $\mcor$ is the CGM gas mass. The mean metallicity is calculated over the corona volume, from the inner radius, $R_0$, to $r_{\rm CGM}$. The total mass of metals in the corona is then 
\begin{equation}
M_{\rm metals} = f_{\rm Z} \zmean_{V} \mcor ~~~, 
\end{equation}
where $f_{\rm Z} = 0.012$ is the mass fraction of metals at a solar metallicity, adopting the individual abundances from \citet{Asplund09}. The average line-of-sight metallicity is
\begin{equation}\label{eq:zmean3}
\bar{Z'} = \frac{1}{N}\int{Z'(r)n(r)ds} ~~~,
\end{equation}
where $ds$ is the path element and $N$ is the total gas column density along this sightline. The sightline can be calculated for an observer inside the galaxy (for MW observations) or an external observer at a given impact parameter (for other galaxies).

\subsection{Boundary Conditions}
\label{subsec_bounds}

In solving Equations (1)-(7) we set $r_{b}$, the reference point for the boundary conditions of the gas distribution, at the outer radius of the corona \rcgm. We now discuss the value ranges we consider for \rcgm, and the gas properties there, such as the density and temperature, by estimating them for the Milky Way.

Structure formation calculations and simulations predict that matter that falls onto the galaxy is shocked and heated (\citealp{White78,Birnboim03}). We define the boundary between the IGM and CGM as the location of this accretion shock. Simulations indicate that this occurs roughly, but not exactly, at the virial radius \citep{Schaal15}, which is estimated from the halo total mass. The mass of the MW has been measured over the last decade using a variety of methods, resulting in $\mvir = 1.3 \pm 0.3 \times 10^{12}$~\msun~\citep{BHG16}. In \FSMII~we used the gravitational potential profile from \cite{Klypin02} (model~B, see their Table~2), which has $\rvir = 258$~kpc, and $\mvir = 10^{12}$~\msun. These values are consistent with the range estimated by \cite{BHG16}, and we use the same gravitational potential and virial radius in this work. 

We combine the uncertainties regarding (i) the size (and mass) of the MW halo (i.e.~\rvir) and (ii) the location of the accretion shock, into the range for \rcgm, and examine values between the virial radius and $1.3$~\rvir, or $\sim 260-330$~kpc. Smaller CGM radii are not implausible in theory, but they may be inconsistent with measurements of OVI in other $L^*$ galaxies, as we discuss in \S \ref{subsec_obsext} (see \citealp{Johnson15}).

We now turn to the gas properties at this radius. First we set the temperature, $\tcgm$, to the virial temperature, defined by $2E_{\rm k} = E_{\rm pot}$, where $E_{\rm pot}$ is the potential energy of the mean particle evaluated at the outer boundary. The gas temperature is then given by
\begin{equation}\label{eq:tvir1}
\tshock = \frac{G \mbar}{3\kb} \frac{M(r)}{r} ~~~,
\end{equation}
Scaling this to the Galaxy mass and \rcgm, we get
\begin{equation}\label{eq:tvir2}
\tshock \approx 3.4 \times 10^{5}~{\rm K} \left( \frac{\mbar}{0.59~\mp} \right) \left( \frac{\mvir}{10^{12}~\msun}
\right) \left( \frac{\rcgm}{300~\kpc} \right)^{-1} ~~,
\end{equation}
where $0.59 \mp$ is the mean particle mass for fully ionized gas with the primordial abundance of helium.
\cite{Birnboim03} perform a detailed calculation of the gas temperature behind the virial shock, and find a similar value. In this work we consider temperatures in the range $\tcgm \approx 2-4 \times 10^5$~K, accounting for the uncertainty in the MW mass and the location of the shock. At these temperatures the OVI ion fraction, $f_{\rm OVI}$, is close to its peak in collisional ionization equilibrium (CIE), with $f_{\rm OVI} \approx 0.25$ at $T_{\rm peak} \approx 3 \times 10^5$~K (\citealp{GS07}, and \S\ref{subsec_ionization} here).

In our new model, each of the components providing pressure support behaves differently with radius, due to a different EoS or adiabatic index, and the \alp~parameter is a function of radius. For \alp~at \rcgm~we consider a range between 1 and 3, as we did in \FSMII. For $\alp=1$ there is only thermal and turbulent support, while pressure equipartition between thermal, magnetic and cosmic rays gives $\alp=3$ (see also \citealp{KQ20}).

For the turbulent velocity scale, we adopt $\vturb\sim 60$~\kms, similar to \FSMII~(see Section~2.1 and Table~3 there). This velocity was estimated from the velocity dispersion of the OVI absorption features in the COS-Halos star-forming galaxies, reported by \cite{Tumlinson11}. In our model, the OVI traces the extended warm/hot CGM.

To estimate the gas density at \rcgm~we consider the conditions inside and outside the MW halo.  \cite{McCon07} infer a lower limit of $10^{-5}-10^{-6}$~\cmv~for the LG intragroup medium density, from ram pressure stripping of the Pegasus dwarf galaxy, at $d\approx 920$~kpc from the MW. \cite{FSM13} used the HI distribution in Leo~T to estimate an upper limit for the gas pressure in the Local Group. They find that at $d=420$~kpc from the MW, $P_{\rm IGM}/\kb \lesssim 150$~K~\cmv. Assuming that the pressure of the intragroup medium in the Local Group (LG) does not vary significantly with position on a $100$~kpc scale gives an estimate for the CGM density
\begin{equation}
n(\rcgm) \sim  \frac{P_{\rm IGM}}{\alp(\rcgm) \tcgm + \sigma_{\rm turb}^2 \mbar/\kb} ~~~.
\end{equation}
For the chosen \tcgm, the adopted range of \alpcgm~and $\vturb$, this gives an upper limit of $\nh(\rcgm) < 0.5-2\times 10^{-4}$~\cmv, where $\nh$ is the hydrogen volume density. Another estimate is obtained at smaller distances from the Galaxy. As discussed in \FSMII~(see Section 5.1 there), studies of ram-pressure stripping in the LMC and MW dwarf satellite galaxies find CGM densities of $\sim 10^{-4}$~\cmv~at $50-100$~kpc (\citealp{Grcevich09,Gatto13,Salem15}), and \citet[hereafter BR00]{Blitz00} find an average density of $\sim 2.4 \times 10^{-5}$~\cmv~inside $250$~kpc. These values serve as upper limits for the density at the outer boundary and we consider densities of $\nh(\rcgm) \sim 1-5\times 10^{-5}$~\cmv.

For the metallicity, we examine values in the range of $Z'_0=0.5-1$ at the solar radius, and $0.1-0.5$ at \rcgm. We set the upper limit at \rcgm~as $Z'=0.5$ to allow for a constant metallicity profile, for comparison with \FSMII. The length-scale increases from $r_Z\sim 30$~kpc for a large metallicity gradient, ranging between $Z=0.1$ and $1$, to $r_Z>250$~kpc for flat metallicity profiles, changing by $\lesssim 25\%$. For a metallicity profile that varies by a factor of $3-5$ between small radii~and~\rcgm, the length-scale is $r_Z \approx 50-100$~kpc. These scales are similar to the extent of galactic winds in numerical simulations \citep{Salem15,Fielding17a}, and we prefer them in our model.

To summarize, the combination of \ncgm, \tcgm, \alpcgm~and~\vturb~allows us to compute the entropy parameters (Eq. \ref{eq:k1}-\ref{eq:k2}), numerically solve Equation \eqref{eq:hse2} and obtain the gas density profile, $\rho(r)$. Then, using the EoS we get the individual and total pressure and temperature profiles, from the outer boundary to the inner radius at the solar circle, at $\rsun= 8.5$~kpc. This radius is the inner boundary in our model. In \FSMII~we estimated that the thermal pressure above the Galactic disk, \Psun, is between $\sim 1000$ and $3000$~K~\cmv~\citep{Wolfire03,Dedes10}. \cite{Putman12rev} find pressures of $P/\kb \sim 500-1300$~K~\cmv~using observations of High Velocity Clouds (HVCs) at distances of $10-15$~kpc from the GC and $3-9$~kpc above the disk.
With the above observational constraints in mind we set the boundary conditions by fixing the temperature at the outer radius (\rcgm) and varying the density and non-thermal support there to set the inner pressure, at \rsun. Then, the metallicities at \rsun~and \rcgm~determine the metallicity length-scale and the distribution of metals is given by Equation~\eqref{eq:zprofile}.

\section{Fiducial Model}
\label{sec_fiducial}

In this section we present our fiducial, constant entropy model, for a specific set of boundary conditions, chosen to reproduce observations of the warm/hot CGM, as traced by highly ionized oxygen absorption measured in the MW and other low-redshift galaxies (see \S\ref{sec_comparison}).  Table~\ref{tab:mod_prop} summarizes the input parameters and the main properties of our fiducial model.

First, we discuss the basic gas properties, density and temperature (\S\ref{subsec_dist}), and the gas ionization state (\S\ref{subsec_ionization}). We show that for the gas densities in our fiducial model, photoionization by the metagalactic radiation field (MGRF) affects the metal ion fractions in addition to collisional ionization. This is in contrast with \FSMII, in which the gas densities and temperatures are higher, and photoionization is negligible. We calculate the ion fractions in the CGM and the gas radiative properties using Cloudy 17.00 \citep{Ferland17} and the \cite{HM12} MGRF. We then present the spatial distribution of selected metal ions (\S\ref{subsec_metalions}) and the gas emission properties (\S\ref{subsec_spectrum}).


\subsection{Gas Distributions}
\label{subsec_dist}

\bgroup
\def\arraystretch{1.3}

\begin{table}
\centering
	\caption{Fiducial model - Summary of properties}
	\label{tab:mod_prop}
		\begin{tabular}{| l | c |}
			\midrule
			\multicolumn{2}{| c |}{Input Parameters}					\\
			\midrule
 			$\mvir$ 				    & 	$1.0 \times 10^{12}$~\msun		    	\\
			$\rvir$ 					& 	$258$~kpc				    \\
			\midrule
			$\rsun$ 					& 	$8.5$~kpc				    \\
			$\rcgm$ 					& 	$283$~kpc ($1.1 \rvir$)	    \\
			$T(\rcgm)$			        & 	$2.4 \times 10^5$~K         \\
			$n_{\rm H}(\rcgm)$  	    &	$1.1 \times 10^{-5}$~\cmv	\\
			$\sigma_{\rm turb}$         & 	$60$~\kms	                \\
		    $\alp$~($\alpha_{\rm OML}$)      \footnote{~$\alpha-1$ gives the ratio of cosmic ray and magnetic field pressure to the thermal pressure. $\alpha_{\rm OML}$ also includes the turbulent pressure (see \S\ref{subsec_dist}).}
		                                    & 	$2.1$~($3.2$)	    	\\
			$Z'$					        &  	$0.3-1.0$		        \\
			$r_Z$						& 	$90$~kpc				    \\
			\midrule
			\multicolumn{2}{|c|}{Results}							    \\
			\midrule
			$M_{\rm gas}(\rvir)$	& 	$4.6 \times 10^{10}$~\msun      \\
			$f_{\rm b}$ (w/o disk)\footnote{~With and without the Galactic disk mass included, assuming $\mvir=10^{12}$~\msun~and $\mdisk=6.0 \times 10^{10}$~\msun.} 	& 	$0.68 / 0.29$	\\
			$M_{\rm gas}(\rcgm)$	& 	$5.5 \times 10^{10}$~\msun	    \\
			$M_{\rm metals}(\rcgm)$	& 	$3.1 \times 10^{8}$~\msun	    \\
			$\Psun$ 		        &   $1350$~K~\cmv	                \\
			$\Lrad$ 			        & 	$9.4 \times 10^{40}$~\ergs      \\
			$\Lcool$ 			        & 	$7.6 \times 10^{40}$~\ergs      \\
			$\tcool(\rcgm)$	        & 	$7.4 \times 10^9$~Gyr           \\
			$\tdyn(\rcgm)$	        & 	$3.1 \times 10^9$~Gyr           \\
			$\zeta(\rcgm)$          & 	$2.4 $          	            \\
			$\left<\tcool\right>$   & 	$3.6 \times 10^9$~Gyr  	        \\
			\midrule
			\multicolumn{2}{|c|}{Approximations - $\tilde{p} \times \left( r/\rcgm \right)^{-a}$}		\\
			\midrule
			$\tilde{T}_{\rm th}$~,~$a_{\rm T}$	  	& 	$2.7 \times 10^{5}$~K~~,~~$0.62$		    \\
			$\tilde{n}_{\rm H}$~,~$a_{\rm n}$	  	& 	$1.3 \times 10^{-5}$~\cmv~~,~~$0.93$	\\
			$\tilde{P}_{\rm tot}$~,~$a_{\rm P}$	  	& 	$22.1$~K~\cmv~~,~~$1.35$	           \\
			\bottomrule
	\end{tabular}
\end{table}

\begin{figure*}
\begin{tabular}{c c}
 \includegraphics[width=0.45\textwidth]{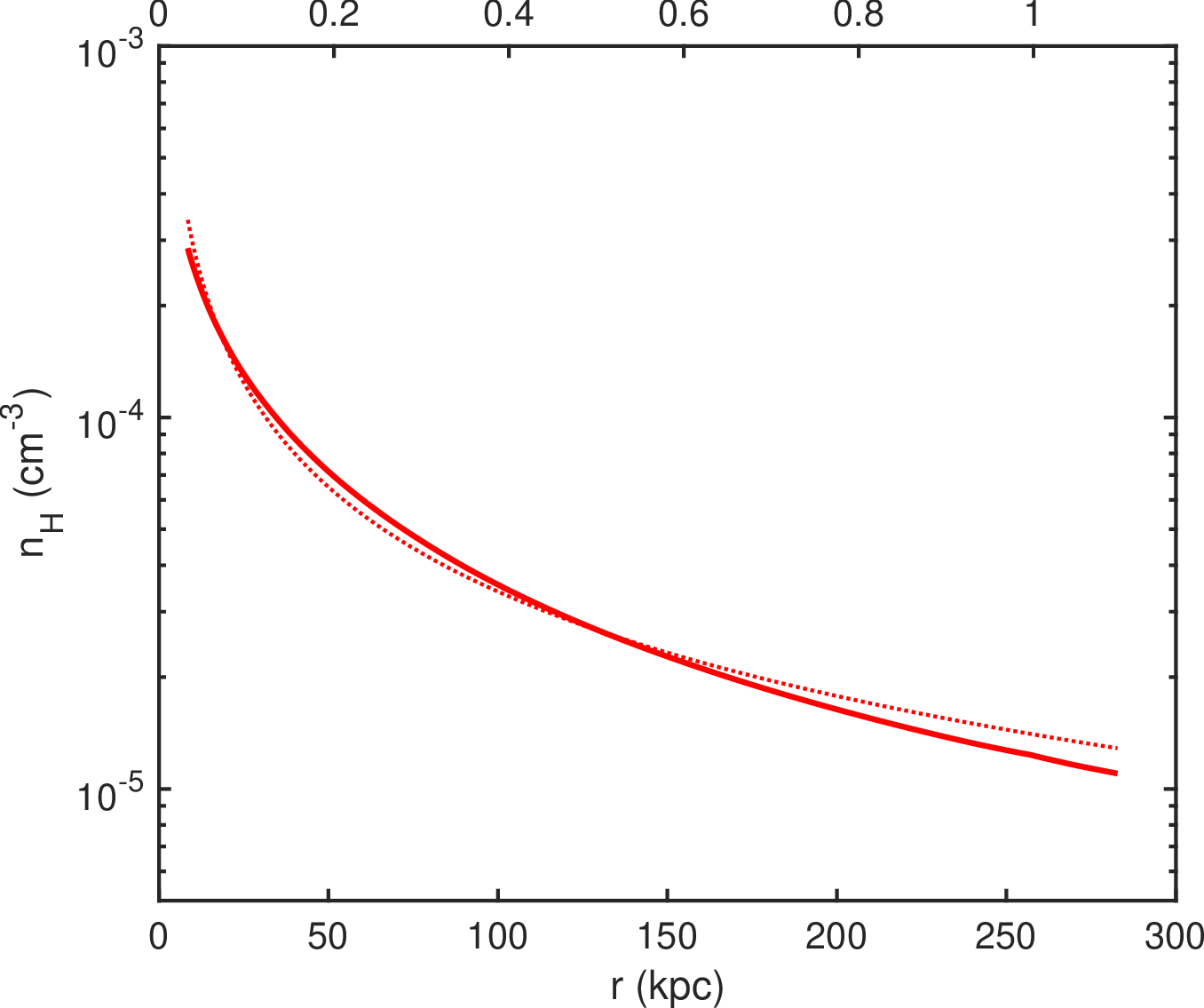} &
 \includegraphics[width=0.44\textwidth]{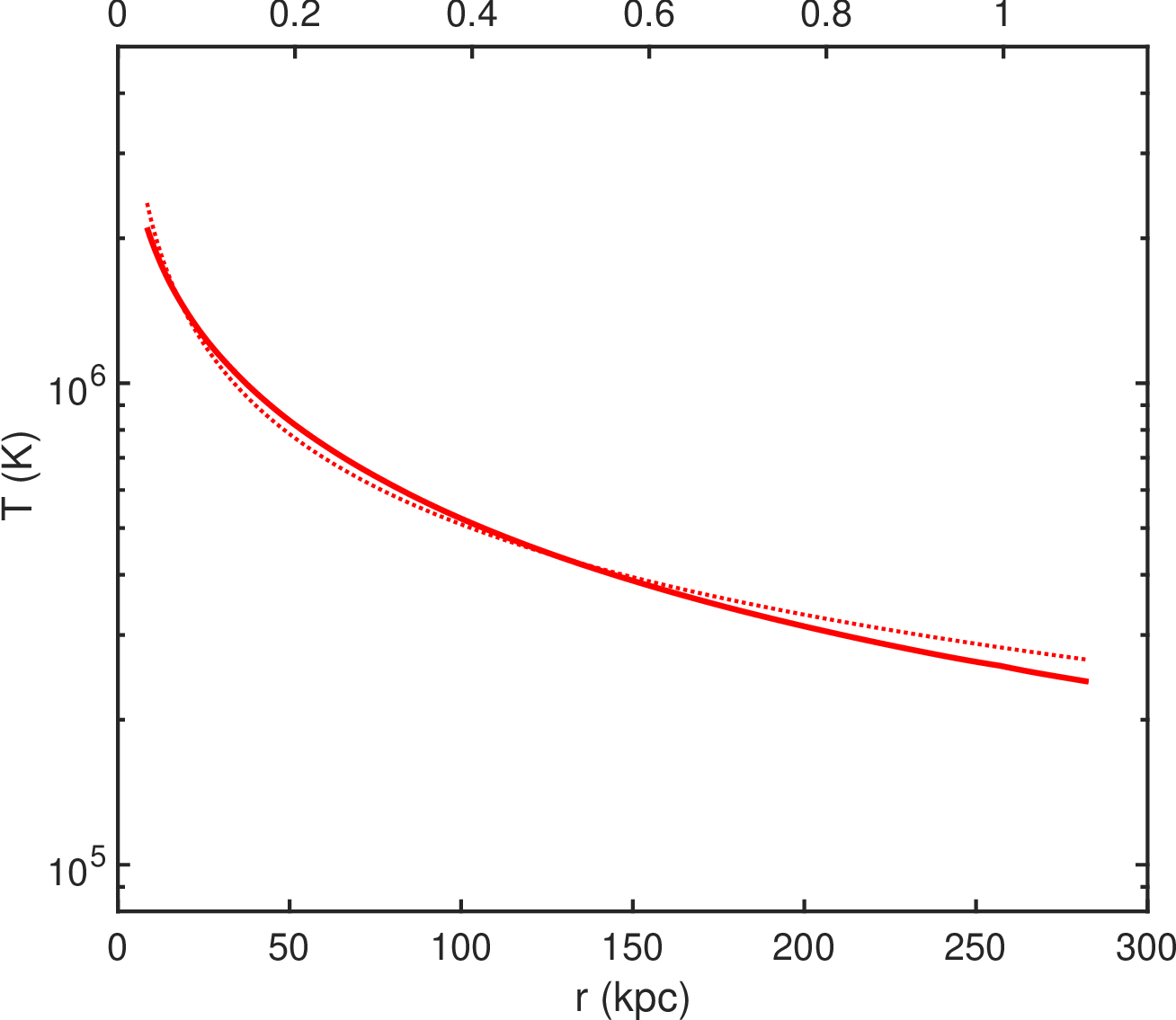} \\
\end{tabular}
\caption{The gas density (left) and thermal temperature (right) profiles in our fiducial model (see \S\ref{subsec_dist}). The solid curves show the profiles resulting from the numerical solution of the HSE equation (Eq. \ref{eq:hse2}), with the boundary conditions at $\rcgm=283$~kpc~set by the input parameters (see Table~\ref{tab:mod_prop}). The dotted curves are power-law approximations of the numerical profiles, with indices of $a_{\rm n}=0.93$ and $a_{\rm T}=0.62$, for the density and temperature, respectively. In this work we show the spatial coordinate in our model both in kpc (bottom axis here), and normalized to the virial radius of the MW, $258$~kpc (shown on top).}
  \label{fig:prof_nt}
\end{figure*}

\begin{figure*}
\begin{tabular}{c c}
 \includegraphics[width=0.45\textwidth]{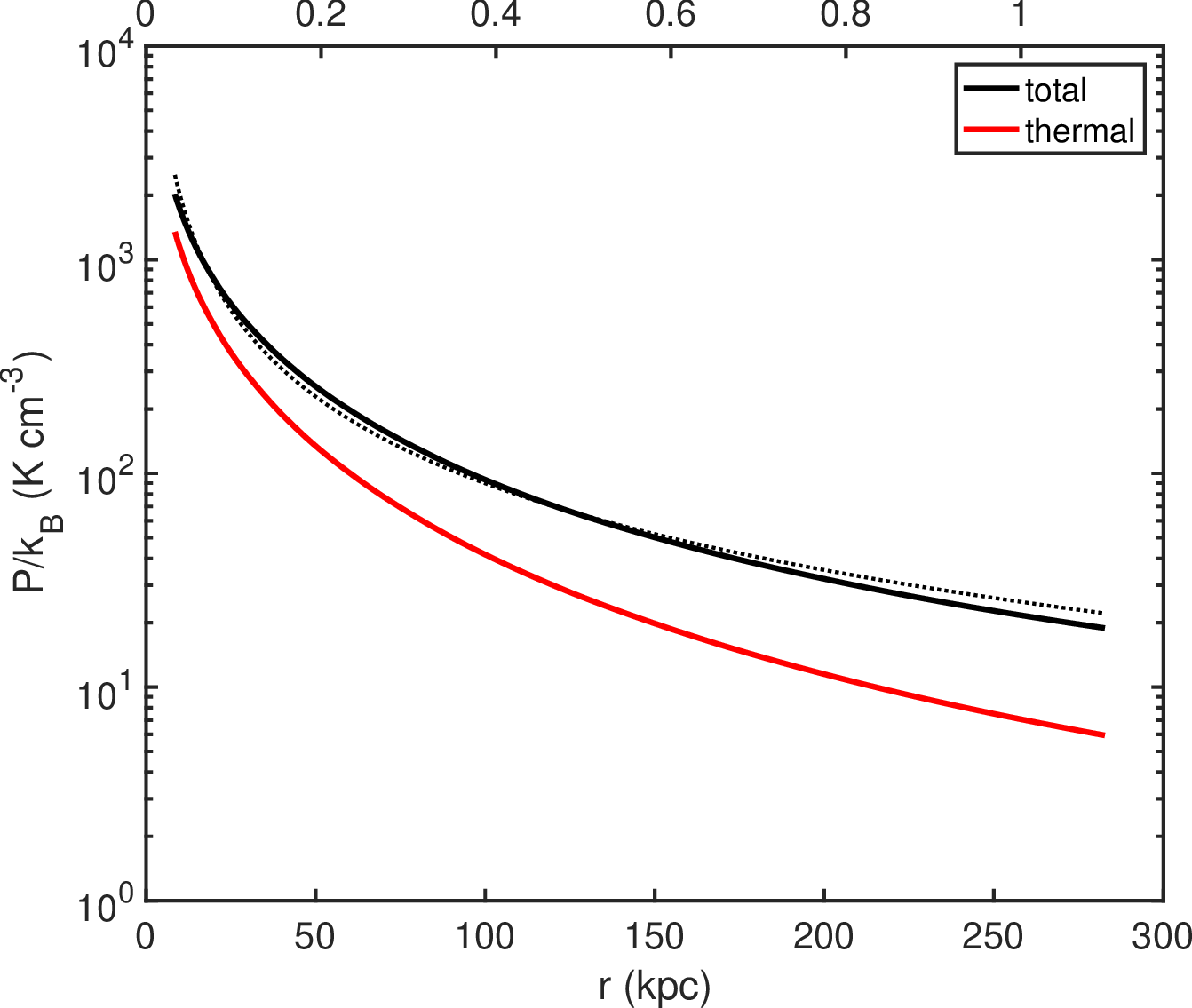} &
 \includegraphics[width=0.44\textwidth]{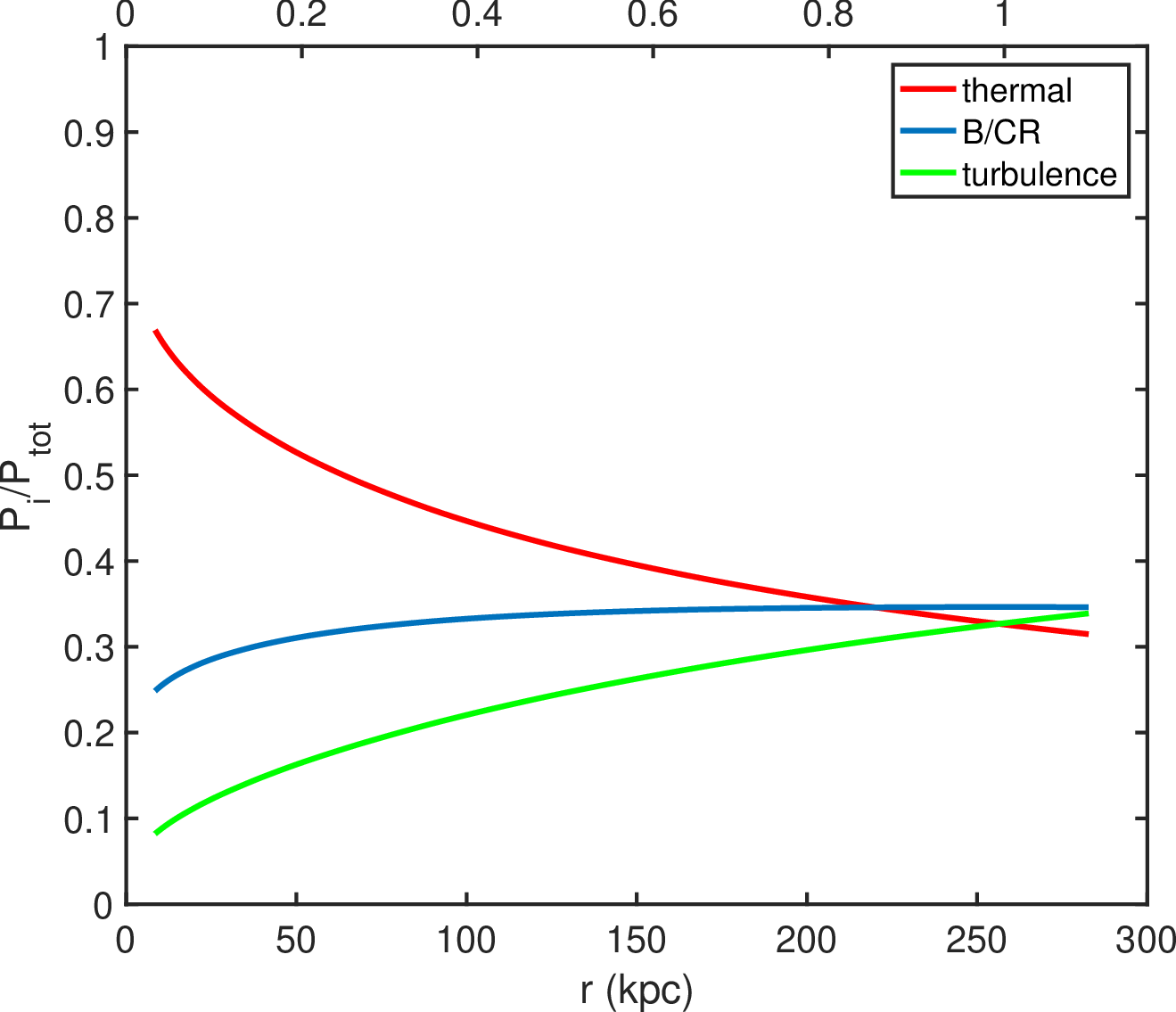}\\
\end{tabular}
\caption{Pressure profiles in the fiducial corona model (see \S\ref{subsec_dist}).
    {\bf Left}: Total (black) and thermal (red) pressure. The total pressure at the outer boundary is set by the gas temperature, density and amount of non-thermal support ($\alp_{\rm OML}$). The thermal pressure in the inner part is $1350$~K~\cmv, near the lower limit of the range estimated in the MW. The total pressure profile can be approximated by a power-law with an index of $a_{\rm P} \approx 1.35$, shown by the dotted curve.
	{\bf Right}: The fractional/relative pressure of the different components in the corona - thermal support (red curve), non-thermal pressure from cosmic rays and magnetic fields (blue) and turbulent support (green). The pressure fraction in each component varies with radius, due to the different equations of state. The thermal support, with an adiabatic index of $\gamma=5/3$, has the steepest profile. The turbulent pressure is parameterized in our model by a constant velocity dispersion (with $\vturb=60~\kms$), and its relative fraction increases with radius.}
  \label{fig:prof_p}
\end{figure*}

\begin{figure}
 \includegraphics[width=0.45\textwidth]{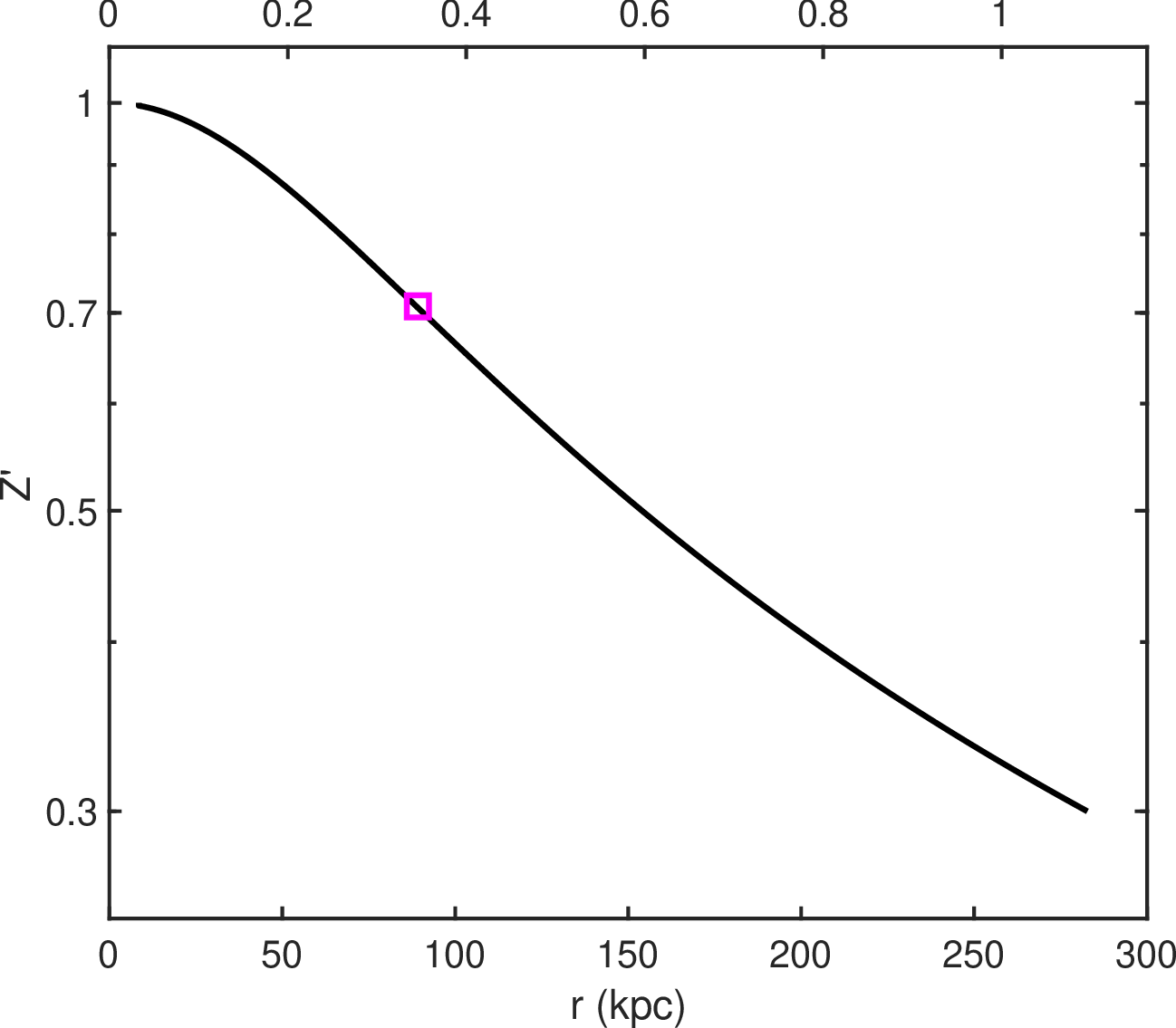}
\caption{The metallicity profile in the warm/hot gas~(see \S\ref{subsec_metallicity} and \S\ref{subsec_dist}). The magenta marker shows the metallicity length scale, $r_Z=90$~kpc, set by the boundary conditions (Equation~\ref{eq:zlength}).}
  \label{fig:metallicity}
\end{figure}

Figure \ref{fig:prof_nt} presents the total hydrogen density and the thermal temperature profiles in the corona (left and right panels, respectively)\footnote{~We present some properties of our model as functions of the physical radius or impact parameter, and others - as functions of the physical scale normalized to the Galactic virial radius. The latter is done mainly when we compare distributions in the model to the measurements in other galaxies, of different sizes and masses. In each case, we present the complementary scale on the top x-axis of the plot. As in \FSMII, we fit and compare our model to the star-forming galaxies in the COS-Halos sample. We note that this sub-sample has a median virial radius of $260$~kpc, very similar to the value we adopt for the MW virial radius ($258 $~kpc).}. For these models we adopt $\rcgm=1.1 \rvir = 283$~kpc. In our fiducial model, the density and temperature at this boundary are $\ncgm = 1.1 \times 10^{-5}$~\cmv~ and $\tcgm = 2.4 \times 10^{5}$~K. Both increase inwards and at \rsun~equal to $2.9 \times 10^{-4}$~\cmv~and $2.1 \times 10^{6}$~K, respectively. The mean hydrogen density within \rcgm~ is $1.8 \times 10^{-5}$~\cmv.

The density, temperature and pressure profiles are well approximated by power-law functions of the radius, $\tilde{p} \times \left( r/\rcgm \right)^{-a}$, where $\tilde{p}$ is the value of the fit at \rcgm. Fits between $\rsun$ and $\rcgm$~give indices of $a_{\rm n}=0.93$ and $a_{\rm T}=0.62$, for the density and the temperature, respectively. These approximations are accurate to within $20\%$ for the density and $10\%$ for the temperature, and they are shown in Figure \ref{fig:prof_nt} as dotted curves. The full approximations, including the normalization factors, are given in Table~\ref{tab:mod_prop}.

Figure \ref{fig:prof_p} shows the pressure versus radius. The left panel shows the total and thermal pressures. The total pressure (black curve) at \rcgm~is $P/\kb = 20$~K~\cmv. This value is consistent with pressure estimates from the accretion rate onto a MW-like galaxy in cosmological simulations. The thermal pressure (red curve) at \rsun~is $1350$~K~\cmv. This is near the lower limit of the range estimated in \FSMII~from observations to be between $\sim 1000$ and $3000$~K~\cmv~(see Section 2 there). The power-law index of the (total) pressure profile is $a_{\rm P} = 1.35$, and this approximation is accurate to within $10\%$.

The right panel of Figure~\ref{fig:prof_p} shows the fractional contribution of each pressure component as a function of radius - thermal support (red), non-thermal pressure from magnetic fields and cosmic rays (blue) and turbulent pressure (green). The ratio of the non-thermal to thermal pressure is parameterized by $P_{\rm nth}/P_{\rm th}=\alp-1$, and to include the contribution of turbulent support, we define $\alp_{\rm OML}=P_{\rm tot}/P_{\rm th}$. With $\alpcgm = 2.1$ and $\alp_{\rm OML}(\rcgm) = 3.2$, the three components have similar contributions to the total pressure at \rcgm. However, due to the higher adiabatic index of the thermal component, the thermal pressure increases more rapidly at smaller radii, and dominates the total pressure at $r < 50$~kpc, with $\alp(\rsun) \sim 1.5$.

We can estimate the strength of the magnetic field implied by the non-thermal pressure. If the cosmic rays and magnetic fields have equal contributions to the energy density, the magnetic field strength in the CGM is given by $B = \sqrt{4 \pi P_{\rm nth}} \propto r^{-2a_n/3}$. In our fiducial model this results in $B \propto r^{-0.62}$, and the field increases from $B \approx 110$~nG at \rcgm~to $940$~nG at \rsun. These values are consistent with the upper limit inferred by \cite{P19sci} for the magnetic field in the CGM of a massive galaxy  at $z \approx 0.36$, with $B \lesssim 500$~nG at an impact parameter of $\approx 30$~kpc.

\begin{figure}
 \includegraphics[width=0.46\textwidth]{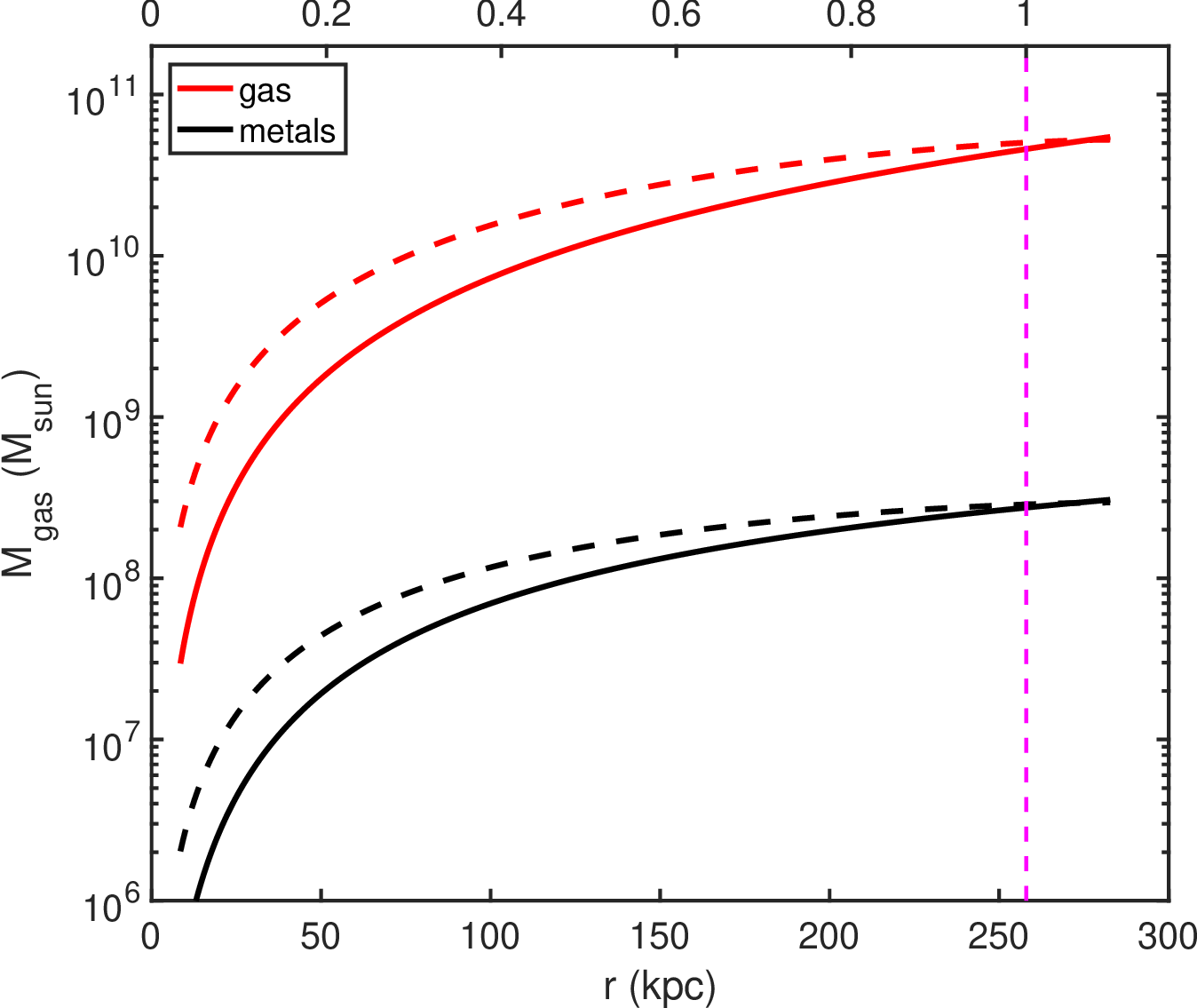}
\caption{The cumulative gas (red) and metal (black) masses in the fiducial model. The solid curves show the spherical mass, enclosed in a radius $r$ (see Table~\ref{tab:mod_prop}), and the dashed curves show the projected mass as a function of the impact parameter. The gas mass within \rvir~(marked by the vertical dashed magenta line) constitutes $\approx 30\%$ of the MW baryonic budget.}
  \label{fig:mass}
\end{figure}

Given the density profile, we calculate the CGM mass and its contribution to the baryonic budget of the Galaxy. The cumulative gas mass distribution is shown in Figure~\ref{fig:mass} for spherically enclosed and projected masses (red solid and dashed curves, respectively). The coronal gas mass inside \rvir~is $4.6 \times 10^{10}$~\msun. Adopting a cosmological baryon fraction of $f_{\rm bar}=0.157$ \citep{Planck16}, this constitutes $\sim 30\%$ of the Galactic baryonic budget. Together with a mass of $\approx 6 \times 10^{10}$~\msun~for the Galactic disk (\citealp{McMillan11,Lic15}), we get a total baryonic mass of $\sim 1.1 \times 10^{11}$~\msun, or $\sim 70\%$ of the Galactic baryons expected within \rvir. The total CGM mass, inside \rcgm, is $5.5 \times 10^{11}$~\msun.

The gas phase metallicity in our model decreases from $Z'=1.0$ at \rsun, to $0.3$ at \rcgm, with a metallicity scale length of $r_Z=90$~kpc, and is plotted in Figure~\ref{fig:metallicity}. The total mass of metals within \rcgm~is $3.1 \times 10^{8}$~\msun. The cumulative metal mass profiles are shown in Figure~\ref{fig:mass} for the spherically enclosed and projected masses (black solid and dashed curves, respectively). These can be useful for comparison with mass estimates from measurements of metal ion column densities (see \S\ref{sec_comparison}).

For example, \citet[hereafter P14]{Peeples14} analyze the COS-Halos OVI measurements to estimate the metal gas mass in the warm CGM. For their preferred model, with an assumed density profile slope of $a_{\rm n}=2$, they infer the projected metal mass inside $150$~kpc and obtain $M_{\rm metals}(h<150~{\rm kpc}) \sim 0.46 \times 10^8$~\msun , with a range of $0.28-1.1 \times 10^8$~\msun. However, P14 find that $a_{\rm n}$ has a significant effect on the total gas mass. For a profile with a slope of $a_{\rm n}=1$, a higher mass profile is allowed by the measurements, with $M_{\rm metals}(h<150~{\rm kpc}) \sim 4 \times 10^{8}$~\msun. In our fiducial model, the projected metal mass in the gas phase within $150$~kpc is $1.9 \times 10^{8}$~\msun, within the range allowed by the different gas density distributions in P14.

\citet{Peek15} find that galactic coronae contain significant amounts of dust, and estimate a dust mass of $M_{\rm dust}\sim 6 \times 10^{7}$~\msun~in the CGM of $0.1-1.0~L^*$ galaxies. This is lower than the gas phase metal mass in our model, but not negligible. Our model does not constrain the dust content of the CGM.

\subsection{Ionization}
\label{subsec_ionization}

The warm/hot gas at each radius in our model is at a constant temperature and density given by the profiles presented in Figure~\ref{fig:prof_nt}. In computing the ionization fractions we include electron-impact collisional ionization and photoionization by the metagalactic radiation field (MGRF). We assume ionization equilibrium. We do not include photoionization by stellar radiation from the Galaxy, since stellar radiation is expected to decrease rapidly as $d^{-2}$ with the distance $d$ from the Galaxy, and is not energetic enough to affect the high oxygen ions we address here (OVI-OVIII). Other Galactic sources may have a contribution to ionizing radiation, although probably on lower ions and at small distances (see \citealp{Cantalupo10} and \citealp{Sander18}). The MW and the COS-Halos galaxies do not have active galactic nuclei and we do not include AGN radition fields (although see \citealp{Oppenheimer18} for consideration of ``fossil" OVI AGN photoionization). 

In collisional ionization equilibrium (CIE), the ion fractions are functions of the gas temperature only \citep{GS07}. When photoionization is included, the ion fractions may also depend on the gas density and radiation field properties, such as intensity and spectral shape \citep{Gnat17}. For a field with a known spectral distribution, the effect of the radiation on the atomic ionization state can be estimated using the ionization parameter, given by $U \equiv \Phi/c n_{\rm H}$. Here $\Phi = 4\pi \int_{\nu_0}^{\infty}{\frac{J_{\nu}}{h\nu}d\nu}$ is the ionizing photon flux, $J_{\nu}$ is the radiation field energy flux density (and $c$ is the speed of light.) In our calculations we consider the \citet[hereafter HM12]{HM12} radiation field, which is a function of redshift only. For the HM12 $z=0$ MGRF, $\Phi \approx 10^4~{\rm cm^{-2}~s^{-1}}$. Scaling the ionization parameter to this value and to the gas density at the outer boundary of our corona model (see Table~\ref{tab:mod_prop}), we get
\begin{equation}\label{eq:upar}
U = 3.3 \times 10^{-2} \left( \frac{\Phi}{10^4~{\rm cm^{-2}~s^{-1}}} \right) \left( \frac{n_{\rm H}}{10^{-5}~\cmv} \right)^{-1} ~~~.
\end{equation}
At $z=0.2$, the median redshift of the COS-Halos galaxies, the ionizing photon flux of the HM12 field is $\Phi \approx 2.3 \times 10^4~{\rm cm^{-2}~s^{-1}}$, and we continue our calculation for $z=0.2$.

The MGRF intensity in the EUV and soft X-rays is uncertain to some extent, with different studies arguing for a stronger \citep{Stern18, FG20} or a weaker radiation field \citep{JBH17}. In this work we adopt the HM12 spectrum and note that the FG20 and HM12 field intensities are within $\sim 30\%$ of each other between 0.1 and 2 keV, the energy range relevant for the ions we address here. The differences may be larger at lower energies and are more relevant for lower ions (see \citealp{Werk16}, \citealp{Prochaska17} and \citealp{Sander18}).

The grey contours in Figure~\ref{fig:ovi_frac} show the OVI ion fraction, $f_{\rm OVI}$, in the temperature-density parameter space, calculated in the presence of the $z=0.2$ HM12 MGRF, using Cloudy 17.00 \citep{Ferland17}. At hydrogen densities above $n_{\rm H} \sim 10^{-3}$~\cmv, the ion fraction is set by collisional ionization. It is then a function of the gas temperature only and peaks at $T_{\rm peak, OVI} \sim 3 \times 10^5$~K, with $f_{\rm OVI} \approx 0.25$. The OVI ion fraction at temperatures far from this peak, at $T<10^5$~K ($T>10^6$~K), is low and oxygen exists in lower (higher) ionization states. At lower densities, below $n_{\rm H} \sim 10^{-5}$~\cmv, the ion fractions clearly deviate from their CIE values due to photoionization.

In general, radiation increases the overall gas ionization, but the change in the fraction of a specific ion, \fion, depends on the gas temperature compared to $T_{\rm peak,ion}$. For $T < T_{\rm peak,ion}$, energetic photons ionize the lower ionization states and increase \fion, compared to CIE. In gas at higher temperatures, radiation ionizes the atom to a higher state and reduces \fion.

We define $U_{\rm crit}$, as the threshold ionization parameter above which an ion fraction deviates by more than $10\%$ compared to the CIE value. While the threshold can vary with temperature, for our qualitative analysis here for the OVI we adopt a single value, of $U_{\rm crit,OVI} \sim 7 \times 10^{-3}$. At $z=0.2$, this corresponds to a density of $n_{\rm H,photo} \sim 10^{-4}$~\cmv, below which photoionization is important. In Figure~\ref{fig:ovi_frac} this critical density is indicated by the vertical green dashed line.

The red line shows the $T \propto n^{2/3}$ temperature-density relation in our model (with $\gamma = 5/3$ for thermal pressure). The black squares mark specific radii, between \rcgm~and the solar circle. For $r \gtrsim 30$~kpc, our model has densities close to the critical photoionization density of $10^{-4}$~\cmv. To compare the ion fraction at a given radius in the model to the fraction in CIE, one can move horizontally (at $T=const.$) from the red curve to a density 1-2 dex above the photoionization threshold and estimate how the ion fraction changes along this line. Since most of the gas in our model is above $T_{\rm peak, OVI}=3 \times 10^5$~K, photoionization {\it reduces} the OVI fraction compared to CIE. This in contrast with models at lower temperatures, where photoionization is invoked as an OVI production mechanism (e.g. \citealp{Stern18}).

The OVII and OVIII ions have photoionization densities similar to the OVI (see also \citealp{Ntormousi10}) and in our model they are also affected by the MGRF. The OVI-OVIII ion fractions as functions of radius in our model are plotted in the left panel of Figure~\ref{fig:metalions}. We also display the NV fraction, and discuss these curves in more detail in \S\ref{subsec_metalions}. 

The total ion densities, $n_{\rm ion} = \fion A_{\rm i} Z' \nh$, are also a function of the gas density and metallicity profiles and of the elemental abundances $A_{\rm i}$. The volume densities of OVI-OVIII and NV are shown in the right panel of Figure~\ref{fig:metalions} as a function of radius. In \S\ref{sec_comparison} we discuss the behavior of the column densities of these ions and compare them to observations.

We note that the measured OVII and OVIII column densities are associated with the MW, at $z=0$. Comparing them to the results of our model, for which we adopt the MGRF at $z=0.2$, may seem inconsistent. However, these column densities, observed from inside the Galaxy, form mostly in the inner, denser part of the corona ($r<30$~kpc), where their ion fractions are set by the gas temperature only (see Figure~\ref{fig:metalions}). Thus, using the $z=0$ MGRF has a very small effect on the OVII and OVIII columns and our comparison is valid (see also \S\ref{subsec_obsmw_abs}).

\begin{figure}
 \includegraphics[width=0.45\textwidth]{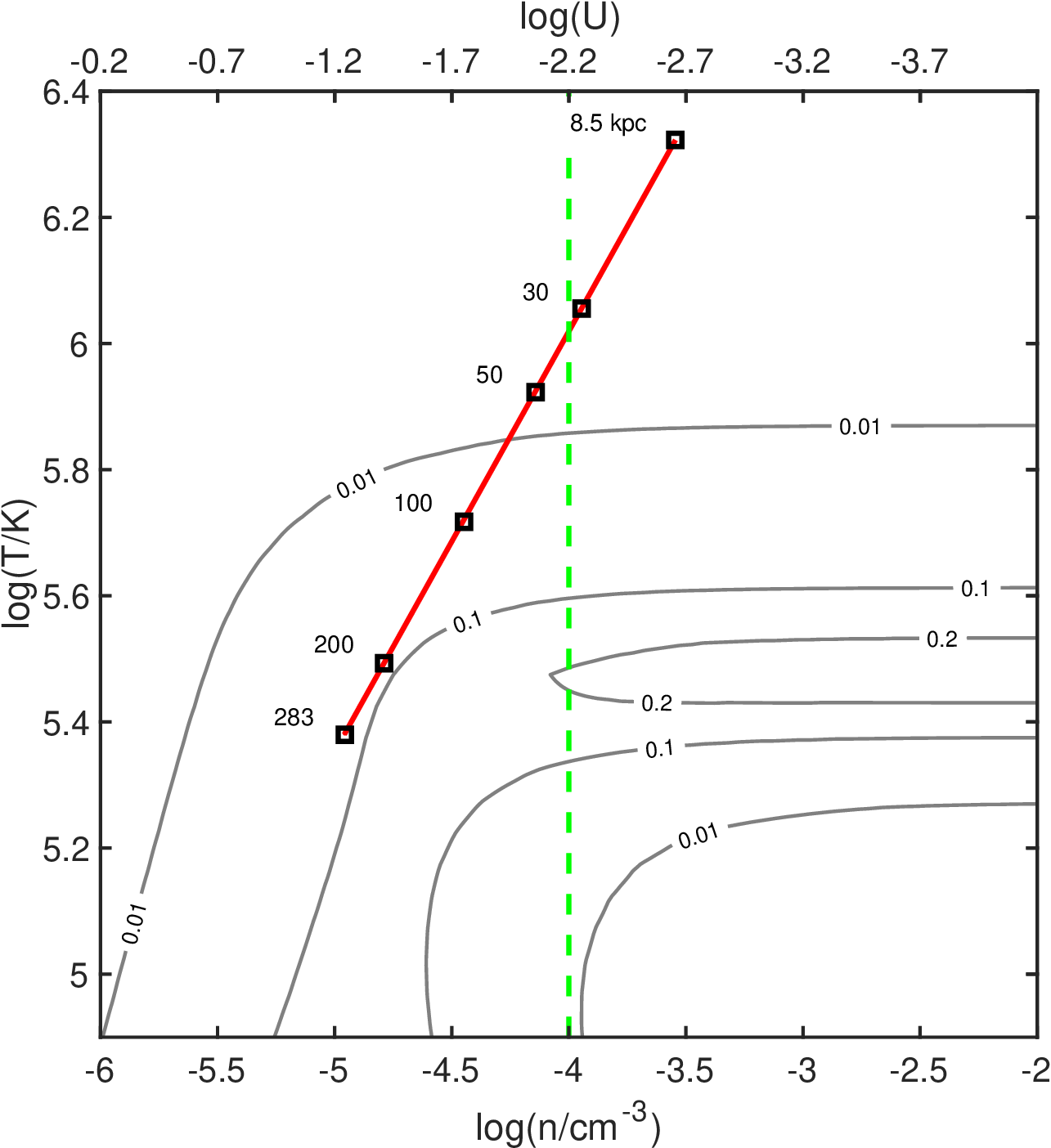}
\caption{OVI ion fraction (grey contours) and the $T \propto n^{2/3}$ (thermal) temperature-density relation for the gas in our fiducial model (red curve). The OVI fraction is calculated in the presence of the $z=0.2$ HM12 metagalactic radiation field using Cloudy, and the gas photoionization parameter $U$ is shown on the top x-axis. The green vertical dashed line marks the density threshold below which photoionization starts to affect the OVI ion fraction (see \S\ref{subsec_ionization}). The black squares along the red curve mark different radii in the corona in kpc. The gas density in the outer region of our model ($r \gtrsim 30$~kpc) is below the threshold density, and the temperature is above the OVI CIE peak ($\sim 3 \times 10^5$~K) out to $r \sim 200$~kpc. In this region of the parameter space, radiation reduces the OVI ion fraction, compared to its value in CIE.}
  \label{fig:ovi_frac}
\end{figure}

We conclude that for the gas properties of our fiducial model, photoionization by the metagalactic radiation field has a non-negligible effect on the ion fractions of the high oxygen ions (OVI-OVIII). We calculate the metal ion fractions as a function of the gas density and temperature using Cloudy 17.00 \citep{Ferland17} and use them to calculate the ion volume and column densities, which we discuss in \S\ref{subsec_metalions} and in \S\ref{sec_comparison}. The MGRF also affects the gas radiative properties, through the metal ion fractions, and we calculate the gas net cooling rate and emission spectrum as a function of the density, temperature and metallicity. In \S\ref{subsec_spectrum}, we use these quantities to calculate the emission properties of the corona.

\subsection{Metal Ions Distributions}
\label{subsec_metalions}

Observations of the CGM reveal the gas distribution through absorption and emission by metal ions, and in this section we describe the spatial distributions shown in Figure~\ref{fig:metalions}. We plot the OVI (solid blue), OVII (green), OVIII (red) and NV (cyan) ions. The ion fractions are shown in the left panel, with the CIE fractions (thin dashed curves) for comparison, and the ion volume densities on the right. For the nitrogen and oxygen ion densities we use the \cite{Asplund09} abundances, with $A_{\rm N} = 6.8 \times 10^{-5}$ and $A_{\rm O} = 4.9 \times 10^{-4}$ for solar metallicity. 

\begin{figure*}
 \includegraphics[width=0.95\textwidth]{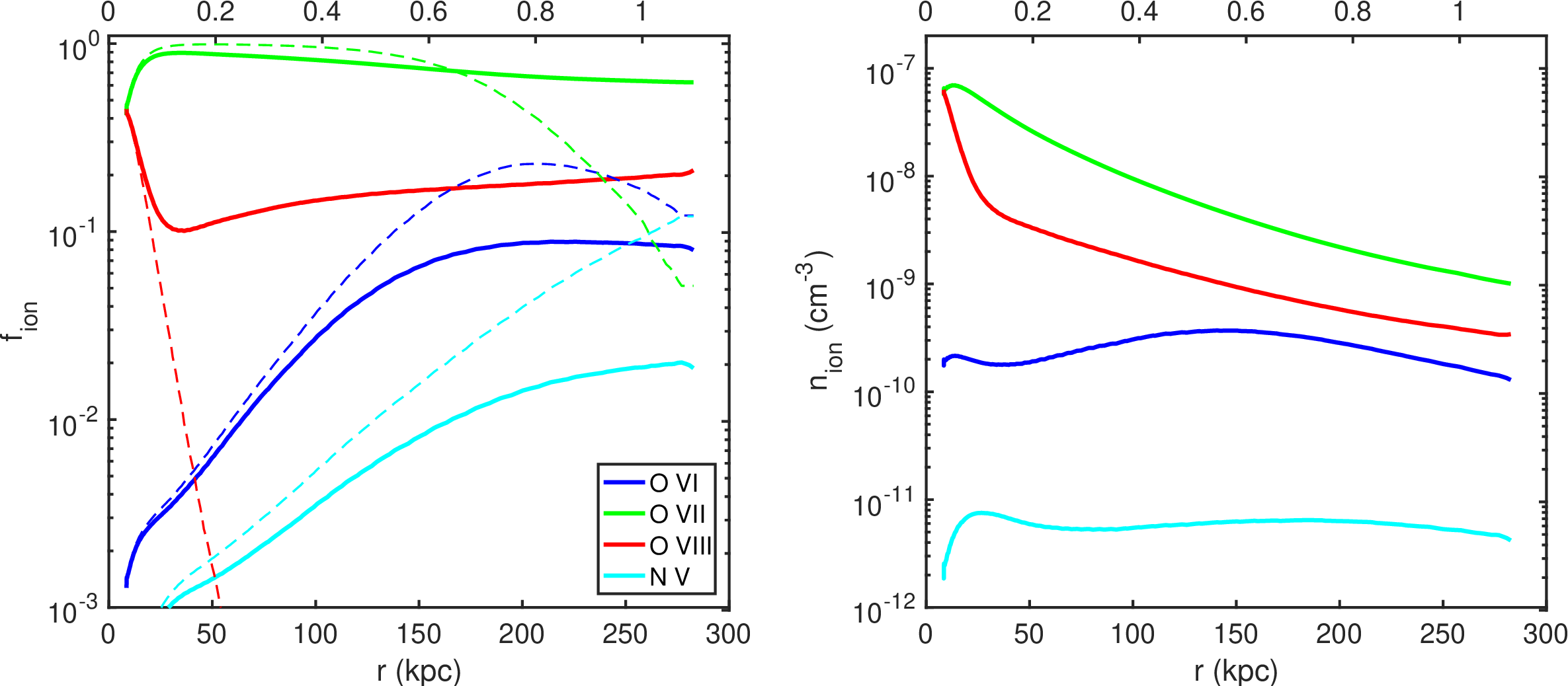}
\caption{The distribution of selected metal ions in our CGM model - OVI (blue), OVII (green), OVIII (red) and NV (cyan). {\bf Left:} The ion fractions, set by collisional ionization and photoionization by the MGRF (solid curves, see \S\ref{subsec_ionization}). The thin dashed curves show the fractions with collisional ionization only, for comparison. The NV and OVI ions are abundant in the outer parts of the corona, where the gas temperature is low. OVIII is created by collisional ionization in the inner part of the CGM ($r \lesssim 30$~kpc), and by photoionization at larger radii. The OVII ion is dominant ($f \sim 1$) at all radii in our model. {\bf Right:} The ion volume densities, given by the product of gas density, metallicity, elemental abundance and ion fraction. The OVII and OVIII trace the gas density profile at $r>30$~kpc, while the NV and OVI densities in the corona are almost constant with radius. The data used to create the right panel of this figure are available.}
  \label{fig:metalions}
\end{figure*}

The NV and OVI ion fractions are most abundant at large radii ($r \gtrsim 150$~kpc), where the gas temperature, with $T \sim 3 \times 10^5$~K, is closest to their CIE peak temperatures ($\approx 2$ and $3 \times 10^5$~K, respectively). The gas density of the CGM at these radii is $\sim 3 \times 10^{-5}$~\cmv, so photoionization is significant and reduces the fractions of both ions. The OVI peak ion fraction, with $f_{\rm OVI} \sim 0.1$, is close to its maximum at CIE ($f\approx 0.25$). The gas temperature is above $T_{\rm peak,NV}$, and this, together with photoionization, lead to lower ion fractions compared to the OVI, with $f_{\rm NV} \lesssim 0.02$.

The effect of photoionization on the OVII varies with radius. At intermediate radii, $30-150$~kpc, the gas temperatures are such that OVII is abundant in CIE, with $f_{\rm OVII} \sim 1$, and photoionization reduces the OVII fraction, but the effect is small ($10-20\%$). At larger radii, where the temperature is below $\sim 5 \times 10^5$~K, photoionization increases the OVII fraction, compared to its CIE values. The OVIII is affected more significantly. In our fiducial model, the gas temperature is high enough to form OVIII collisionally in the central part ($r \lesssim 25$~kpc). Photoionization by the MGRF creates OVIII at larger radii, and its ion fraction increases with the ionization parameter, from $f_{\rm OVIII} \sim 0.1$ at $30$~kpc to $\sim 0.2$ at \rcgm. Overall, OVII is the dominant oxygen ion in our model at all radii, and almost equal to the OVIII fraction at the solar circle.

The resulting densities for our four ions of interest are plotted in the right panel of Figure~\ref{fig:metalions}. The OVII and OVIII ion fractions do not vary strongly with radius. This leads to decreasing ion volume densities, as a result of the density and metallicity distributions in the model. The NV and OVI fractions, on the other hand, increase with radius, resulting in more extended distributions, with almost flat ion density profiles. The OVI volume density is in the range $n_{\rm OVI} \sim 2-4 \times 10^{-10}$~\cmv~for radii between $10$ and $250$~kpc. The nitrogen abundance is $\sim 7$ times lower than that of oxygen, and together with the NV lower ion fraction, this gives volume densities of $n_{\rm NV} \sim 4-8 \times 10^{-12}$~\cmv, $20-60$ times lower than those of OVI. We use the volume densities to calculate the ion column densities through the CGM, and in \S\ref{sec_comparison} we compare these to the measured values and limits.

\subsection{Emission Spectrum and Cooling Rate}
\label{subsec_spectrum}

The left panel of Figure~\ref{fig:emission} shows the predicted emission (cooling) spectrum of the CGM in our fiducial model. The spectrum is given by
\begin{equation}\label{eq:jnu}
J(\nu) = 4 \pi \int_{\rsun}^{\rcgm}{j_\nu(n,T,Z') r^2 dr} ~~~,
\end{equation}
where $j_\nu$ is the emissivity (${\rm erg~\cmv~Hz^{-1}~sr^{-1}}$) of each gas parcel. We used Cloudy 17.00  \citep{Ferland17} to calculate the (optically thin) emissivities as functions of $n$, $T$, and $Z'$, for gas subjected to photoionization by the $z=0.2$ HM12 metagalactic radiation field. (In FSM17 we assumed pure CIE for the emissivities). The resulting spectrum in Fig.~\ref{fig:emission} (denoted by $L_\nu$) is displayed in units of ${\rm \ergs~keV^{-1}~sr^{-1}}$, and consists of collisionally excited metal-ion emission lines, recombination radiation, and bremsstrahlung. The red and black lines show the full and smoothed spectrum, respectively. As shown in Figure~\ref{fig:mass}, the gas in our model is dominated by large radii, where the gas temperature is low, and most of the emission is in the UV. The vertical dashed lines show the $0.4-2.0$~keV band. 

The total luminosity
\footnote{The luminosity includes ionization energies released via recombinations.}
of the warm/hot gas is given by
\begin{equation}\label{eq:lrad}
\Lrad = \int{J(\nu) d\nu \equiv  \int {\cal L} r^2 dr} ~~~ ,
\end{equation}
where ${\cal L}$ is the radiative cooling rate per unit volume (${\rm \ergs~\cmv}$) for each parcel. For our fiducial model $\Lrad=9.4\times 10^{40}$~\ergs. The emission in the $0.4-2.0$~keV band is $\sim 10^{39}$~\ergs, only $\sim 2\%$ of the total cooling luminosity. We also compute the local {\it net} cooling rates given by ${\cal L} - {\cal H} = n_e n_H \Lambda$, where ${\cal H}$ is the heating rate per unit volume\footnote{In our definition of ${\cal H}$ photoionization energy is included in addition to the kinetic energies of the photoelectrons.} due to the MGRF, and $\Lambda~{\rm (\ergs~cm^{3})}$ is the net cooling efficiency. In CIE (${\cal H}=0$), $\Lambda$ is a function of the gas temperature and metallicity only \citep{GS07}. In the presence of radiation, $\Lambda$ is also a function of the ionization parameter and radiation spectral shape \citep{Gnat17}, and we use Cloudy to calculate it. We calculate the volume-integrated net cooling rate
\begin{equation}\label{eq:lcool}
\Lcool = \int{n_e n_H \Lambda r^2 dr} ~~~ 
\end{equation}
and find that in our fiducial model $\Lcool = 7.6 \times 10^{40}$~\ergs. This implies that for our model, $20\%$ of the emitted luminosity is reprocessed MGRF energy.

\begin{figure*}
\begin{tabular}{c c}
 \includegraphics[width=0.45\textwidth]{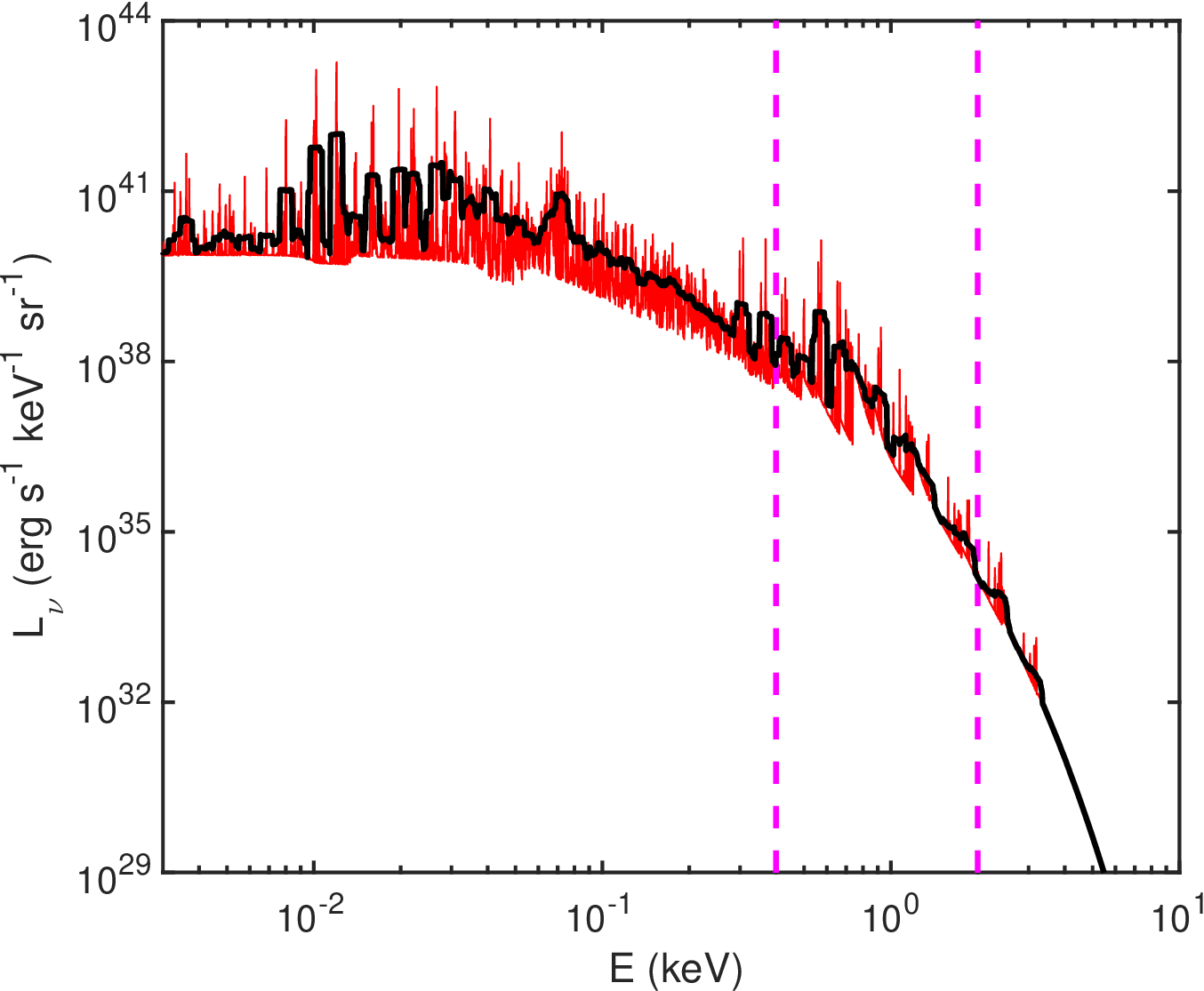} &
 \includegraphics[width=0.45\textwidth]{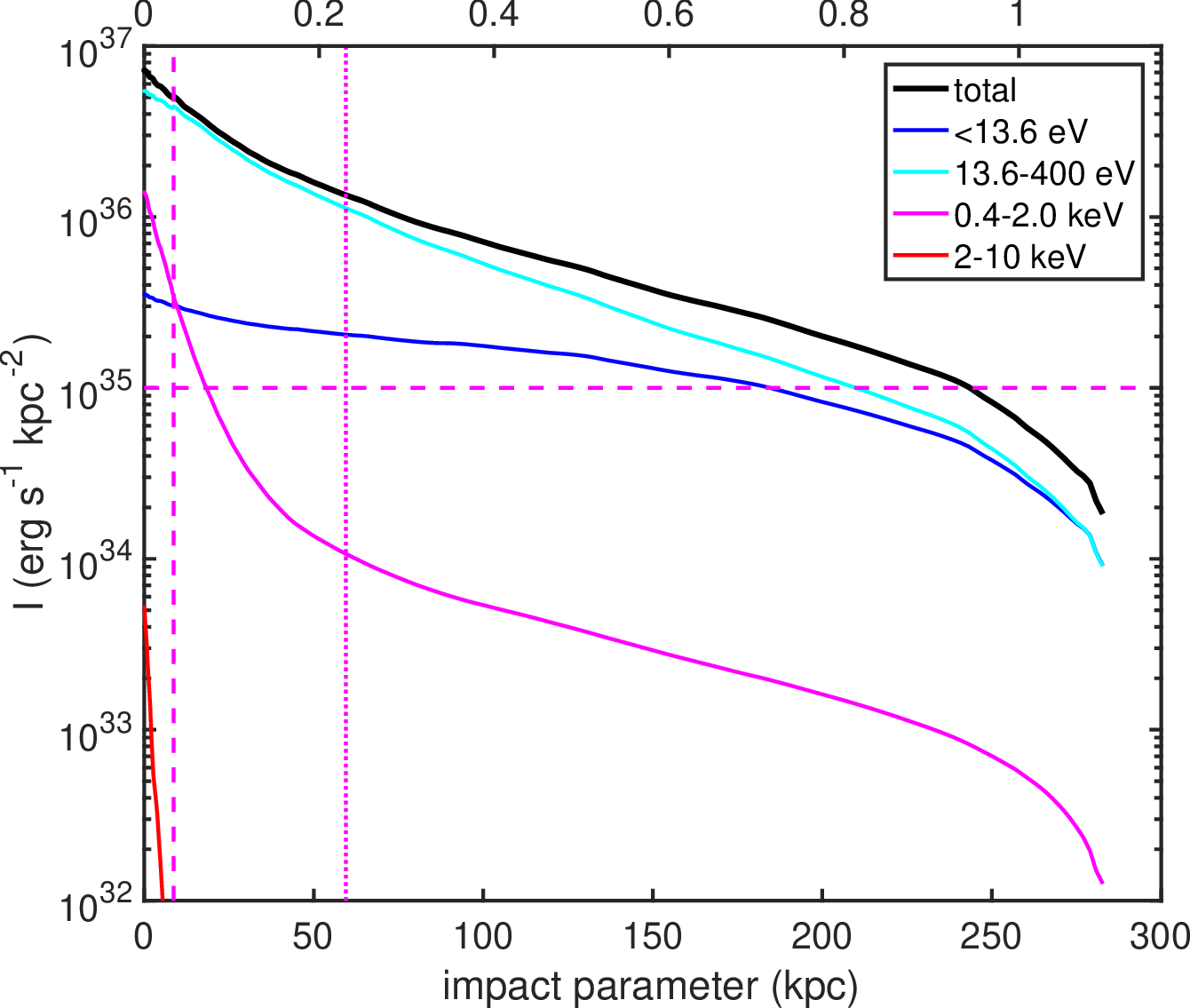} \\
\end{tabular}
\caption{Emission properties of the coronal gas in our fiducial model (see \S\ref{subsec_spectrum}). {\bf Left:} Our predicted total emission spectrum of the coronal gas (red, and smoothed in black). Most of the emission is in the UV, and the luminosity in the 0.4-2 keV band (marked by magenta dashed lines) is $10^{39}$~\ergs, $\sim 2\%$ of the total luminosity. {\bf Right:} The projected emission from the CGM. The emission profile of the total ionizing radiation (black curve) is extended. The soft X-ray emission in the $0.4-2.0$~keV band (solid magenta) is more centrally concentrated, with a half-flux radius of $\rhalf \sim 59$~kpc (vertical dotted line). The horizontal dashed magenta line shows the detection threshold with current instrumentation, $\sim 10^{35}~{\rm erg~s^{-1}~kpc^{-2}}$, as estimated by \cite{LiBreg18}. The vertical dashed line shows the half-flux radius of the emission above this threshold, $\rhalf \sim 9$~kpc.}
  \label{fig:emission}
\end{figure*}

We integrate the spectrum  in different energy bands along lines of sight through the corona to obtain the projected luminosity as a function of the impact parameter, and the result is shown in the right panel of Figure~\ref{fig:emission}. The total projected emission profile (solid black curve) is extended, with a half-flux radius of $\rhalf \sim 100$~kpc. The $0.4-2.0$~keV emission (solid magenta) comes from the hotter gas at smaller radii, and is more centrally concentrated, with $\rhalf \sim 59$~kpc (marked by the vertical dotted line). We note that the instrumental sensitivity and background emission in the X-ray is at the level of the predicted emission. \citet[hereafter L18]{LiBreg18} perform a stacking analysis of the X-ray emission from massive galaxies in the Local Universe, and estimate a background level of $I \sim 10^{35} {\rm erg~s^{-1}~kpc^{-2}}$. This threshold is shown in our plot by a horizontal magenta line. We calculate the half flux radius of the emission above this threshold, and find a value of $\rhalf \sim 9$~kpc, marked by the vertical dashed line in the plot. This demonstrates the challenge in detecting the CGM of MW-like galaxies in emission, given the current instrumental sensitivity and background emission. The other solid curves in the plot show the projected emission in different energy bands - $E<13.6$~eV (blue), $13.6-400$~eV (cyan) and $2-10$~keV (red). As mentioned above, the total emission is dominated by the UV.

\section{Timescales and Energetics}
\label{sec_timescales}

We now address the timescales and energy budget of the coronal gas. First, we present a {\it model-independent} upper limit for the cooling time of OVI-bearing warm/hot CGM, with the full, detailed derivation in the Appendix. We compare the cooling time, \tcool, to the halo dynamical time, \tdyn, and show that for an observed column density of $3 \times 10^{14}$~\cmc~in a MW-sized halo, $\tcool/\tdyn \lesssim 4 $ (with a range of $3-5$ due to variations in the gas distribution). In our fiducial model, we show that at large radii, $\tcool/\tdyn \sim 2-3$. We then calculate the radiative losses and mass cooling rates in the corona, and address the overall energy content of the corona. We find that there is enough energy from SMBH and SNe feedback, and from IGM accretion, to form the CGM and sustain it over $\sim 10$~Gyr.

For ease of comparison to previous work, we adopt the expressions used by \citet{Voit17} for the (isochoric) cooling and dynamical times, given by
\begin{equation}\label{eq:timescales}
\tcool = \frac{3}{2} \frac{n \kb T}{n_{\rm e}n_{\rm H} \Lambda} ~~ , ~~ 
\tdyn = \sqrt{\frac{2 r^3}{GM(<r)}} ~~~,
\end{equation}
where $r$ is the distance from the center of the galaxy/corona, and $M(<r)$ is the total mass enclosed within $r$.\footnote{~In \citet{FSM17} we used the isobaric cooling time, longer by a factor of $5/3$ and $\tdyn = \sqrt{r^3/GM}$, shorter by $\sqrt{2}$. Thus, the ratio $\tcool/\tdyn$ from \FSMII~should be scaled down by $\approx 2.35$ to compare with the values adopted here.}

\subsection{Model-Independent Limit on \tcool/\tdyn}
\label{subsec_tcool_limit}

We now show that the detection of OVI in warm/hot gas implies an upper limit on the gas cooling time. We present a brief version of the analysis here and defer the full derivation to the Appendix. This result is not limited to our model and is relevant for a range of gas distributions.

To obtain an empirical upper limit, we relate the gas cooling time to the OVI-column density. The latter is given by
\begin{equation}\label{eq:novi1}
N_{\rm OVI}(h) = 2 A_{\rm O} \int_0^z{n_{\rm H}(r)Z'(r) f_{\rm OVI}(r) dz'} ~~~,
\end{equation}
where $h$ is the impact parameter and $z'=\sqrt{r^2-h^2}$~\footnote{Here we assume that the warm/hot gas is volume filling, for simplicity. In the Appendix we show that a profile for a non-unity volume-filling factor can be included in the overall functional description of the gas distribution and does not change the final result.}. Assuming the gas properties vary as power-law functions of the radius, we can rewrite this as
\begin{equation}\label{eq:novi2}
N_{\rm OVI}(h) = 2 A_{\rm O} f_{\rm OVI}(h) n_{\rm H}(h) Z'(h) R I_{a} ~~~,
\end{equation}
where $R$ is the outer radius of the gas distribution (i.e. \rcgm~in our model) and $I_{a}$ is a dimensionless integral of order unity for a range of power-law slopes for $f_{\rm OVI}$, $\nh$, and $Z'$.

The key step is to isolate the product, $\nh Z'$, in Equation~(\ref{eq:novi2}) and insert into the cooling time. This gives
\begin{equation}\label{eq:tcool1}
\tcool(h) = 5.8 A_{\rm O} 
\left[ \frac{\kb T(h) f_{\rm OVI}(h)}{\Lambda_{\odot}(T,n)} \right] \frac{R I_a}{N_{\rm OVI}(h)} ~~~,
\end{equation}
where we used $\Lambda = \Lambda_{\odot}(T,n) Z'$, ignoring cooling by hydrogen and helium and resulting in a lower limit for the cooling time. Given the shape of the cooling function and the OVI ion fraction in the temperature-density space in the presence of the HM12 MRGF at $z=0.2$, the term in square brackets is bound from above for gas at $T>10^{5}$~K, with $\kb T f_{\rm OVI}/\Lambda_{\odot} \leq 4.6 \times 10^{10}~{\rm s~cm^3 }$. The maximum occurs at $T \sim 3.5 \times 10^{5}$~K and at densities above $n_{\rm H} \geq 10^{-4}$~\cmv, where $f_{\rm OVI}$ is maximal (see \S\ref{subsec_ionization}). At lower densities radiation suppresses the cooling function but the OVI fraction is reduced even more, so that overall the term is smaller \footnote{~Here, and throughout this work, we assumed that the coronal gas is in equilibrium. For non-equilibrium cooling, the peak OVI ion fraction is reduced by a factor of $\sim 2.5$. The gas cooling rates for densities $\sim 10^{-4}$~\cmv~are similar to their equilibrium values or lower by a factor of $<2$. The resulting limit will be similar to its equilibrium value or even lower.}. Finally, $I_a = 0.5 \pm 0.18$~dex for power-law slopes between $0.5$ and $2.5$ and for impact parameters in the range $0.3 < h/R < 0.9$ (see Appendix). Inserting these into Equation~(\ref{eq:tcool1}), we get
\begin{equation}\label{eq:tcool2}
\tcool(r=h) \lesssim 5.6~\left( \frac{R}{260~{\rm kpc}} \right) \left( \frac{N_{\rm OVI}(h)}{3 \times 10^{14} ~\cmc} \right)^{-1}~{\rm Gyr}  ~~~.
\end{equation}
where we scaled the corona size to the median virial radius of the COS-Halos galaxies, and the OVI column density to the typical OVI column density measured at $h/\rvir \approx 0.6$ by \cite{Tumlinson11}~(see Figure~\ref{fig:new_ovi}). The cooling time range resulting from variation in the underlying gas distributions, affecting the value of $I_a$, is $\pm 30\%$.

For the halo dynamical time in Equation~\eqref{eq:timescales}, we fit the mass distribution in the Klypin profile at large radii (where it is dominated by an NFW profile) as $M \propto r^{0.56}$, and scale it to the MW, resulting in
\begin{equation}\label{eq:tdyn1}
\tdyn(r) \approx 2.8~\left( \frac{r}{260~{\rm kpc}} \right)^{1.22}~{\rm Gyr} ~~~.
\end{equation}

We define the ratio $\zeta \equiv \tcool/\tdyn$ and combine Equations~\eqref{eq:tcool2} and \eqref{eq:tdyn1} to give
\begin{equation}\label{eq:zeta}
\zeta(r=h) < 2.0 ~ \left( \frac{h}{260~{\rm kpc}} \right)^{-1.22} \left( \frac{N_{\rm OVI}(h)}{3 \times 10^{14} ~\cmc} \right)^{-1}  ~~~.
\end{equation}
Our approximation and the derived upper limit are valid for $0.3 < h/R < 0.9$, and the OVI column density we use is measured at $0.6$~\rvir. This implies $\zeta \lesssim 3.7$ (2.8-4.8 uncertainty range), much below the value of $\sim 10$, estimated by \cite{Sharma12a} and \citet{Voit17} for galaxy clusters. A ratio of $\zeta \sim 10$ would require OVI columns lower by a factor of $\sim 2-3$ than observed in the CGM of $L^*$ galaxies by COS-Halos.

The upper limit we derive applies to the gas responsible for the dominant part of the observed OVI absorption. For a  separate, hotter component that contributes a small fraction of the total measured OVI column, the upper limit as given by Eq.~(\ref{eq:tcool1}) will be higher. For example, in our FSM17 model, the OVI column forms mainly in gas that has cooled out of much hotter OVII and OVIII absorbing gas. For the cooled component, $\tcool < 1$~Gyr, and $\tcool/\tdyn<1$,  consistent with our upper limit. However, for the hot component $\tcool/\tdyn \sim 6$ at large radii, slightly exceeding this limit. In the isentropic model in this paper, the bulk of the CGM gas mass is traced by OVI for which the upper limit obtains.

If gas with $\zeta<10$ is thermally unstable and develops multiphase structure, the upper limit we derive on the OVI-bearing gas cooling time implies that cool gas should be present when OVI is detected. The COS-Halos observations seem to be consistent with this prediction, with detections of HI and low metal ions \citep{Werk13,Werk14,Prochaska17}. In Paper III (\FSMIII) we extend our model to include a cool ($T \sim 10^{4}$~K), purely photoionized gas component and compare it to existing observations.

\subsection{Model Timescales}
\label{subsec_tcool_model}

Figure \ref{fig:timescales} shows the timescales in our fiducial model as a function of radius, calculated numerically using the definitions in Equation~\eqref{eq:timescales}. The black curve is the dynamical time of the Galactic halo, given the \cite{Klypin02} potential. The magenta curve is the gas cooling time. At large radii, $r>100$~kpc, the two timescales are well approximated by power-law functions of the radius. The dynamical time is given by Equation~\eqref{eq:tdyn1}, and the cooling time is 
\begin{equation}\label{eq:tcoolapp}
\tcool(r) \approx 6.5~ \left( \frac{r}{\rcgm} \right)^{0.91}~~{\rm Gyr}~~~,
\end{equation}
This approximation is accurate to within $10\%$ between $100$~kpc and \rcgm, and it is shown by the dotted magenta curve in Figure~\ref{fig:timescales}. The fit to the dynamical time is accurate to within $1\%$ in the same range and in the plot, the approximation is indistinguishable from the numerical calculation. Since the two timescales vary similarly with radius, their ratio is almost constant, increasing from $\zeta = 2.4$ at \rcgm~to $3.1$ at $100$~kpc. These values are consistent with the limit derived in \S\ref{subsec_tcool_limit}, and significantly below the value of $\sim 10$, estimated by \citet{Voit17} in clusters of galaxies and adopted by \citet{V19} for the CGM (see also \citealp{Stern19}).

\begin{figure}
 \includegraphics[width=0.45\textwidth]{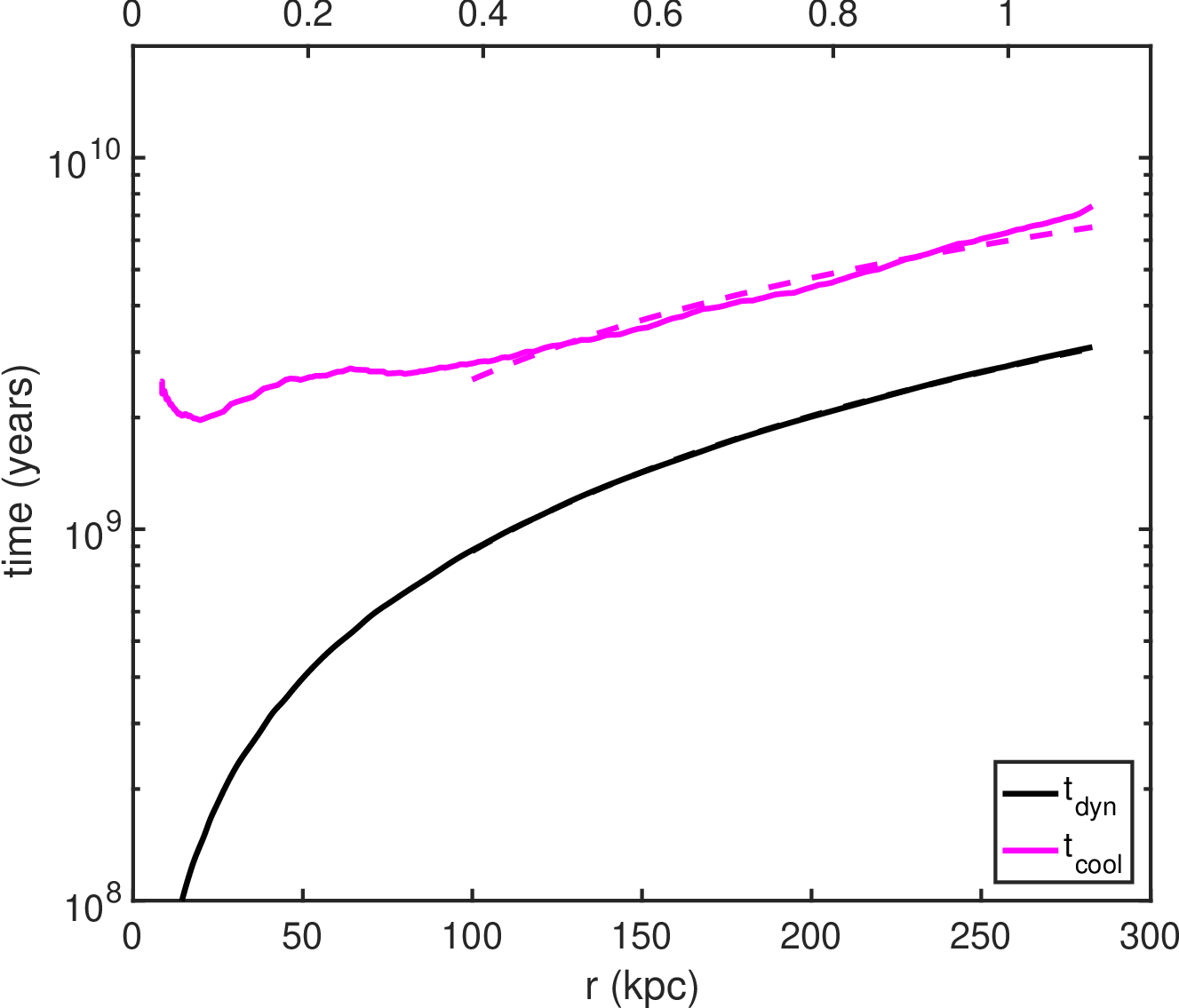}
\caption{The dynamical time (black), and the warm/hot gas cooling time (magenta) in our fiducial model. The decrease in gas density and metallicity leads the cooling time to increase with radius, to $\sim 7$~Gyr at \rcgm. The halo dynamical time has a similar slope at large radii, and the ratio $\zeta = \tcool/\tdyn$ is almost constant there, with $\zeta \sim 2.5$ at $r>100$~kpc (see \S\ref{subsec_tcool_model} for details). The dashed curves show the power-law approximations to the numerical results, and the dynamical time fit is hidden by the solid curve.}
  \label{fig:timescales}
\end{figure}

The mean, global cooling timescale for the corona is the total thermal energy, $E_{\rm th}=8.6 \times 10^{57}$~erg, divided by the net cooling rate, $L_{\rm cool}=7.6 \times 10^{40}$~\ergs, giving $\left< \tcool \right> \equiv E_{\rm th}/L_{\rm cool} = 3.6$~Gyr. As discussed in \S\ref{sec_model}, our model assumes a steady state, so that (most of) the radiative losses are offset by heating (see \S\ref{subsec_energetics} for energy budget estimates) and the CGM is stable on a $\sim 10$~Gyr timescale. We now briefly discuss the gas cooling properties other than \tcool, which may be useful to study the energy budget of the CGM.

\begin{figure*}
 \includegraphics[width=0.95\textwidth]{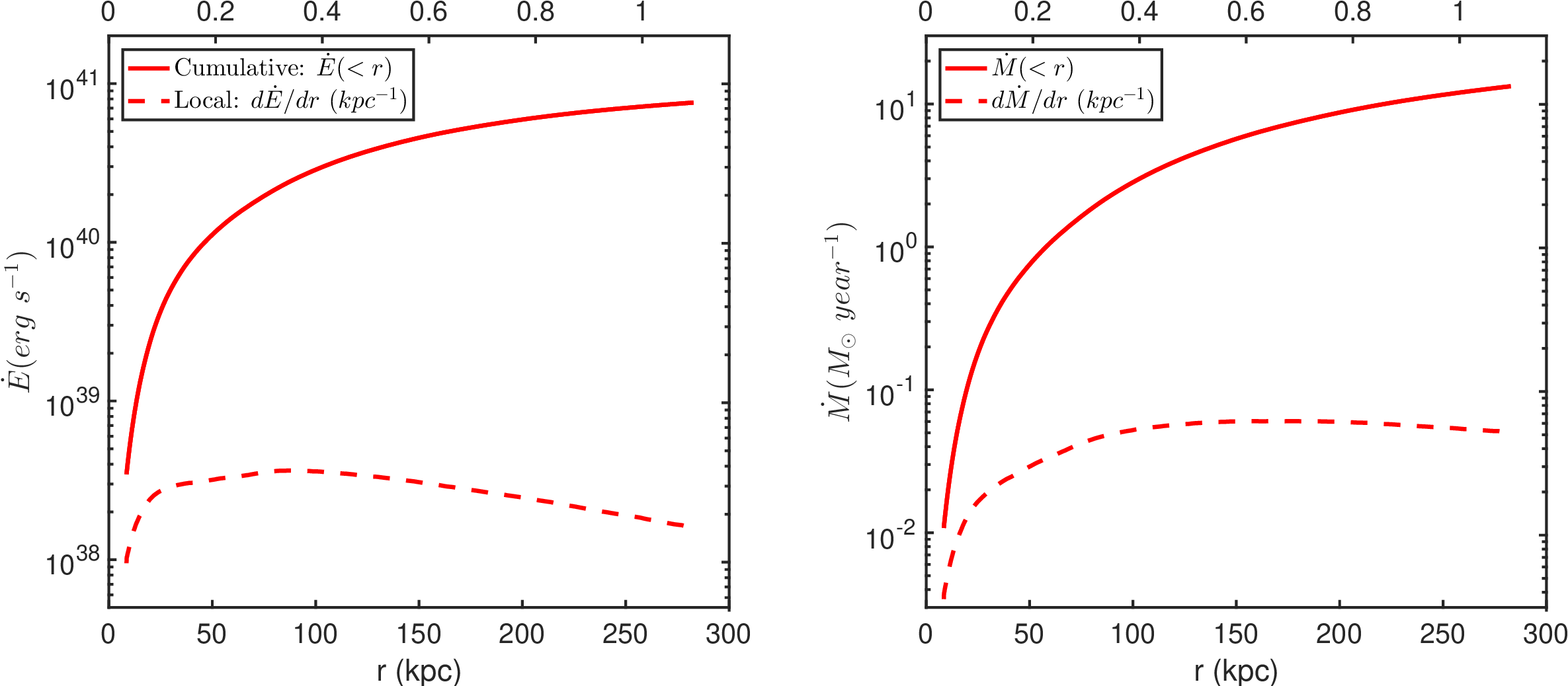}
\caption{Radiative losses (left) and mass cooling rates (right) in the CGM. The dashed curves show the local rates in a shell, per unit length (in kpc), and the solid lines are the cumulative quantities for the volume enclosed by $r$. The mass cooling rates are calculated as the gas mass divided by the cooling time. In our model we assume the radiative losses are mostly offset by heating, so that the actual gas accretion rate onto the galaxy is significantly lower (see \S\ref{subsec_tcool_model}).}
  \label{fig:rates}
\end{figure*}

Figure~\ref{fig:rates} shows the distribution of local cooling rates and mass cooling rates from the CGM. The left panel shows the gas radiative cooling rate, $\mathcal{L} = n_{\rm e} n_{\rm H} \Lambda$, as a function of radius. The dashed curve shows the local rate per unit length, $4\pi r^2 \mathcal{L}$, between $1.0$ and $3.6 \times 10^{38}~\ergs~{\rm kpc^{-1}}$. This is the distribution of energy injection rate needed to keep the CGM in a steady state and is a constraint for the mechanisms that can provide this energy. The solid curve is the integrated, cumulative value inside $r$, with a total of $7.6 \times 10^{40}$~\ergs~inside \rcgm.

Given the gas mass distribution (Figure~\ref{fig:mass}) and its cooling time (Figure~\ref{fig:timescales}), we can calculate the gas mass cooling rate as $\dot{M}(r) = M_{gas}(r)/\tcool(r)$. Without energy input into the CGM, this gives an upper limit on the gas mass cooling out of the corona and accreting onto the galaxy (see also \citealp{Joung12}, \citealp{Arm16} and \citealp{Stern20}). The right panel in Figure~\ref{fig:rates} shows the mass cooling rate, similar to the (energy) cooling rate in the left panel, with the local value per unit length as the dashed curve, and the integrated value in solid. The local rate is almost flat with radius, with a mass cooling rate of $d\dot{M}/dr \approx 0.05~{\rm \msuny~kpc^{-1}}$ between $70$~kpc and \rcgm. The global mass cooling rate, given by $\dot{M}_{\rm tot} = \mgas/\Lcool$ is $ \approx 13.3$~\msuny. Integrating this over $10$~Gyr gives $\approx 1.3 \times 10^{11}$~\msun, a factor of $\sim 2$ higher than the $z=0$ mass of the MW disk.

\subsection{CGM Energetics}
\label{subsec_energetics}

We now address the energy budget of the CGM in our model, and the energy sources available for creating the warm/hot corona and maintaining it in a steady state.

First, we calculate the total energy in the CGM by integrating the different pressure components in our fiducial model (see Fig.~\ref{fig:prof_p}) over the volume of the corona
\begin{multline}\label{eq:etot}
E_{CGM} = \int{\left( E_{\rm th} + E_{\rm nth} + E_{\rm turb} \right) dV} = \\
\int{\left( \frac{3}{2}P_{\rm th} + 3P_{\rm nth} + \frac{3}{2}\sigma_{\rm turb}^2 \mbar n \right) dV} ~~~.
\end{multline}
This gives a total of $2.9 \times 10^{58}$~erg, with $29\%$ in thermal energy, $51\%$ in CR/B and $20\%$ in turbulent energy. Given the gas radiative cooling rate (see Eq.~\ref{eq:lcool}), we can estimate how much energy was lost from the CGM. Taking $t_{\rm life} \sim 10$~Gyr for the galaxy lifetime, we get $E_{\rm cool} = L_{\rm cool} \times t_{\rm life} \approx 2.4 \times 10^{58}$~erg. Thus, a total of $E_{\rm tot} \equiv E_{\rm CGM} + E_{\rm cool} \approx 5.3 \times 10^{58}$~erg is required to form the CGM and balance its radiative losses over the lifetime of the galaxy.

Possible energy sources for the CGM include feedback from supernovae (SNe) and the central super-massive black hole (SMBH), and accretion onto the halo from the IGM. All three mechanisms can (i) generate shocks that heat the gas and accelerate cosmic rays, and (ii) drive turbulence that amplifies magnetic fields and heats the gas when it dissipates. SNe- and SMBH-driven outflows can advect magnetic fields and cosmic rays from the galactic disk to the CGM. We use MW data to estimate the energy budget available to fuel the CGM.

We estimate the total energy injected into the CGM by SNe over the lifetime of the galaxy as
\begin{equation}\label{eq:esne}
E_{\rm SNe} \approx M_{*, \rm  MW} \times R_{\rm  SN} \times E_{\rm SN,s} \times f_{\rm SN} ~~~, 
\end{equation}
where the terms on the right-hand side are the MW present-day stellar mass, the mean number of SNe per unit of stellar mass, the total energy for a single SN event, and the fraction of that energy that goes to the CGM. We take $M_{*,\rm  MW} = 5 \times 10^{10}$~\msun, $R_{\rm  SN}$ of $1$ SN event per $100$~\msun, and $E_{\rm SN,s} = 10^{51}$~erg. The coupling factor, $f_{\rm SN}$, is uncertain, and for $f_{\rm CGM} = 0.1$, we get $E_{\rm SNe} \approx 5.0 \times 10^{58}$~erg. Next, we estimate the energy from SMBH feedback. The mass of the MW SMBH, Sgr~$A^*$, is $M_{\rm MW,SMBH} = 4.1 \times 10^{6}$~\msun~\citep{Grav19}. Assuming $f_{\rm SMBH} = 0.03$ of the rest mass was converted to kinetic energy by the accretion disk~\citep{SG17}, we get $E_{\rm SMBH} \approx 2.2 \times 10^{59}$~erg.

We estimate the contribution of gas accretion from the IGM onto the galactic halo to the energy budget. We do this by calculating the gravitational energy in the CGM, $E_{\rm acc} = f_{\rm acc} \times \left(GM_{\rm halo}/\rcgm \right)\times M_{\rm CGM}$, where $f_{\rm acc}$ is a numerical coefficient of order unity accounting for the growth of the Galaxy with cosmic time. For our fiducial model this results in $E_{\rm acc}/f_{\rm acc} \approx 1.7 \times 10^{58}$~erg. While it is subdominant compared to the energy outputs of SNe and the SMBH, $E_{\rm acc}$, accretion provides an energy source at the outer boundary of the halo, different from SNe and SMBH feedback, which inject energy at the base of the corona.

The energy from SNe and SMBH is injected from within the galaxy. The observations show that the CGM at large radii ($>100$~kpc) is metal-enriched \citep{Tumlinson11, Prochaska17}, providing evidence for extended galactic outflows. Numerical simulations also suggest that winds driven by SNe and SMBH feedback can transport energy and metals to large radii in the halo \citep{MN07, Bower17, Fielding17a, LiT20}. However, we do not consider specific transport processes in this paper.

For the gas densities in our model, the equilibrium temperatures for heating by the MGRF are $\sim 2 \times 10^4$~K, significantly lower than the gas temperatures in the model. As noted in \S\ref{subsec_spectrum}, only $20\%$ of the total computed luminosity (shown in Figure~\ref{fig:emission}) is reprocessed MGRF energy.

To summarize, our estimate shows that there is enough energy available in a MW-like galaxy to form the CGM and power the CGM. We have shown that SNe and energy injection by Sgr $A^*$ can offset the radiative losses over $\sim 10$~Gyr. The energy from SNe events is sufficient to power the CGM for a coupling constant of $f_{\rm SN}=0.1$. The energy from SMBH feedback is a factor of $\sim 4$ higher than SNe, and if a significant fraction of this energy was injected into the CGM, it could have ejected gas beyond the virial radius.

One important feature of the COS-Halos OVI measurements \citep{Tumlinson11} is the bi-modality in the presence of OVI absorption, with detections around SF galaxies, and only upper limits in passive galaxies, with ${\rm sSFR} \lesssim 4 \times 10^{-12}~{\rm yr}^{-1}$. Our model framework does not directly relate the OVI column to the SFR in the galaxy, but it is consistent with several possible explanations for such a relation. One option is that since the OVI-bearing gas, at $T \sim 3 \times 10^5$~K, has a high cooling efficiency, it requires energy input to prevent it from cooling, as discussed above. In our model we assume that the radiative losses of the CGM are balanced and we have shown that SNe, as driven by star formation, can indeed offset these losses. Another scenario, explored by \citet{Opp16} using the EAGLE simulations, is that both the SFR and the OVI are independently correlated to the halo (and stellar) mass. In this case, the halo mass with the peak star formation rate happens to be the same as the halo mass at which the CGM temperature corresponds to the OVI CIE peak. Similarly, in our model, the gas temperature at the outer boundary is related to the halo mass and size (see Eq.~\ref{eq:tvir1}). In either scenario, (a) halos hosting galaxies with low star formation, or (b) higher mass halos may have hotter coronae with lower cooling efficiencies and longer cooling times. These coronae may be detectable through UV and X-ray absorption of higher metal ions.

\section{Comparison to Observations}
\label{sec_comparison}

In \FSMII~we presented a summary of observational data, mainly UV/X-ray emission and absorption, probing warm/hot gas around the MW and other $L^*$ galaxies in the nearby universe (see Section~2 and Table~1 there). We start by addressing these data, first of the MW (\S\ref{subsec_obsmw}), and then of other galaxies (\ref{subsec_obsext}). For the latter, we consider additional observations, including measurements of OVI absorption around and beyond the virial radius \citep{Johnson15} and NV absorption \citep{Werk13,Werk16}.

Table~\ref{tab:mod_res} summarizes the values of the quantitites we discuss, comparing our fiducial isentropic model to observations and our \FSMII~isothermal model.

\begin{table*}
\centering
	\caption{Fiducial model - Comparison to Observations}
	\label{tab:mod_res}
	\begin{tabular}{| l || c | c | c | c |}
	\midrule
	                   & \FSMII~(Isothermal) & Isentropic Model (this paper)   & Observations	& References     \\
	\midrule
	$\sigma_{\rm oxygen}$ (~\kms) 		&	$72$ & $60$	   & $67.2$ ($54.5 - 79.7$)	& (a) (b)      \\
	$\Psun$~(K~\cmv)	                	  		&	$2200$	& $1350$   & $1000-3000$            & (c) (d)     \\
	$DM$ (LMC)~(${\rm \cmv~pc})$ 		&	$17.4$  & $8.8$	   & $\lesssim 23$			& (e) (f)    \\
	$\nh(50-100 \rm~{kpc})~(\cmv)$		& $0.83-1.3 \times 10^{-4}$ & $0.35-0.72 \times 10^{-4}$  & $\sim 10^{-4}$  & (g) (h)  \\
	$\left< \nh\right>_{250~{\rm kpc}}~(\cmv)$ & $4.6 \times 10^{-4}$ & $2.0 \times 10^{-5}$& $\sim 2.5\times10^{-5}$ & (i)   \\
	\midrule
	\multicolumn{5}{| c |}{Milky Way Absorption (\S\ref{subsec_obsmw_abs})} 		    \\
	\midrule
	$N_{\rm OVII}~(\cmc)$ & $1.6 \times 10^{16}$ & $1.2 \times 10^{16}$~($\lscale=33.2$~kpc) & $1.4~(1.0-2.0) \times 10^{16}$ & (j) (k) (o) \\
	$N_{\rm OVIII}~(\cmc)$ & $3.8 \times 10^{15}$ & $3.4 \times 10^{15}$~($\lscale=11.6$~kpc) & $3.6~(2.2-5.7) \times 10^{15}$ & (l) (o) \\
	$\rm OVII/OVIII$~ratio      		& $4.5$ & $3.6$     & 	$4.0~(2.8-5.6) $    & (b) (k) (l) \\
	\midrule
	\multicolumn{5}{| c |}{Milky Way Emission (\S\ref{subsec_obsmw_em})} 		    \\
	\midrule
    $\s04$~(${\rm \ergs~\cmc~deg^{-2}}$) & $0.82 \times 10^{-12}$ & $0.33 \times 10^{-12}$ & $2.1~(1.9-2.4) \times 10^{-12}$ & (m) \\
	$I_{22~\rm \AA}$~(L.U.)\footnote{~L.U.$ = \liun$.\label{ftn:liun}}
	                            & $1.2$ & $0.57$~($\lscale=5.2$~kpc)   & $2.8~(2.3-3.4) $       & (n) \\
	$I_{19~\rm \AA}~(L.U.)^{\rm \ref{ftn:liun}}$ & $0.33$  & $0.17$~($\lscale=3.2$~kpc)   & $0.69~(0.58-0.83)$    & (n) \\
	$\lia/\lib$~ratio			& $3.6$    & $3.3$			   &	$4.3~(3.4-5.5) $ 				    & (b) (n) \\
	
    \midrule
    References:
    & \multicolumn{4}{| l |}{(a) \cite{Tumlinson2011b}, (b) \cite{FSM17} (c) \cite{Wolfire03} (d) \cite{Dedes10}} \\
    & \multicolumn{4}{| l |}{(e) \cite{Anderson10} (f) \cite{PZ19} (g) \cite{Grcevich09} (h) \cite{Salem15}} \\
    & \multicolumn{4}{| l |}{(i) \cite{Blitz00} (j) \cite{Bregman07} (k) \cite{Fang15} (l) \cite{Gupta12}} \\
    & \multicolumn{4}{| l |}{(m) \cite{Henley10} (n) \cite{Henley10lines} (o) \cite{Das19a}} \\

	\bottomrule
		\end{tabular}
		\vspace{0.5 cm}
\end{table*}

\subsection{Milky Way}
\label{subsec_obsmw}

\cite{Blitz00} estimate the CGM density needed to explain the observed dearth of gas in MW dwarf satellite galaxies and find a mean value of $\left< \nh \right> \approx 2.5 \times 10^{-5}$~\cmv~inside $250$~kpc. \cite{Grcevich09} perform a similar analysis for satellites at distances of $50-100$~kpc and find densities around $10^{-4}$~\cmv. \cite{Salem15} use simulations to reproduce the distribution of ram-pressure stripped gas around the LMC (at a distance of $50$~kpc), and estimate a coronal gas density of $1.10^{+0.44}_{-0.45} \times 10^{-4}$~\cmv. In our fiducial model, the mean density of warm/hot gas inside $250$~kpc is $\left< \nh \right>_{250~{\rm kpc}} \approx 2.0 \times 10^{-5}$~\cmv. The gas densities at $50-100$~kpc in our model are in the range $\sim 0.4-0.7 \times 10^{-4}$~\cmv, a factor of $\sim 2$ lower than the values estimated by \cite{Grcevich09} and \cite{Salem15}. They are also lower a factor of $\sim 2$ than the densities in our \FSMII~isothermal model. We note that estimates from ram-pressure stripped systems may be biased towards the denser regions of the corona.

\cite{Manchester06} present dispersion measure (DM) measurements to pulsars in the Large Magellanic Cloud (LMC). \cite{Anderson10} discuss these and estimate an upper limit of ${\rm DM} \leq 23$~\cmv~pc for the CGM component. \cite{PZ19} use the same observations and estimate ${\rm DM} = 23 \pm 10$~\cmv~pc. In our model, the computed dispersion measure in the corona to the LMC is ${\rm DM}=8.8$~\cmv~pc, consistent with the \cite{Anderson10} upper limit.

\subsubsection{OVII and OVIII Absorption}
\label{subsec_obsmw_abs}

The OVII and OVIII column densities for our model are $N_{\rm OVII} = 1.2 \times 10^{16}$~\cmc~and $N_{\rm OVIII} = 3.4 \times 10^{15}$~\cmc. These are consistent with the observed values, of $1.4~(1.0-2.0) \times 10^{16}$ and $3.6~(2.2-5.7) \times 10^{15}$~\cmc, respectively ($1$-$\sigma$ error ranges).  The ratio of the column densities in our model is $N_{\rm OVII}/N_{\rm OVIII} = 3.6$, close to the value we estimated in \FSMII~from observations, of $4.0~(2.8-5.6)$.

To quantify where most of the column is formed, we define the scale length, \lscale, as the distance along the line of sight from the solar circle to the point where the column density is half of its total value at \rcgm. For our fiducial model, the OVII and OVIII scale lengths are $\approx 33$ and $12$~kpc, respectively. The length scales are smaller than in our \FSMII~isothermal model ($\sim 50$~kpc for both ions), for two reasons. First, the metallicity in our new model decreases outwards, compared to the constant metallicity we assumed in \FSMII. Second, the temperature gradient leads to a different distribution for each ion (see Figure \ref{fig:metalions}). OVII is abundant for a wide range of temperatures and therefore extends to larger radii, resulting in $\lscale \sim 30$~kpc. OVIII, on the other hand, forms mostly in the inner hot part of the corona and has a more compact distribution. Since radiation affects the ion fractions mostly at large radii ($r>\lscale$, see \S\ref{subsec_metalions}), adopting the $z=0.2$, rather than the $z=0$ radiation field does not make a significant difference here. We verify this by by re-calculating the fiducial model using the $z=0$ MGRF and find that the OVII/OVIII columns for an observer inside the galaxy change by less than $3\%$.

\citet[hereafter D19a]{Das19a} detected $z \approx 0$ absorption from the highly ionized species NVII, NeIX, and NeX) along a single sight line towards the blazar ${\rm 1ES1553}$. They attributed these absorptions to an additional hot $\sim 10^7$~K CIE component with the CGM of the Galaxy. However, they commented that it is uncertain how ubiquitous such hot gas might be. Our model does not include gas at such high temperatures.

\subsubsection{X-Ray Emission - OVII/OVIII lines and 0.4-2.0 keV band}
\label{subsec_obsmw_em}

For an observer at \rsun, the X-ray emission along a line of sight in our model is centrally concentrated, with length-scales of $\sim 3-5$~kpc. These are smaller than the emission length-scales in \FSMII, mainly since the higher temperature gas in the inner part of the corona is more emissive in the X-rays compared to the cooler gas at larger radii. The decreasing metallicity profile also contributes to the decrease in the gas emissivity, since metal ions constitute a significant fraction of the total emission at $\sim 1$~keV.

The $22~{\rm \AA}$ and $19~{\rm \AA}$ feature emission intensities in our fiducial model are $I_\lia=0.58$ and $I_\lib=0.18$~L.U.~(line units - \liun). These account for $\sim 20-25\%$ of the observed values, with $I_\lia = 2.8~(2.3-3.4)$ and $I_\lib = 0.69~(0.58-0.83)$~L.U. ($1$-$\sigma$ errors). The line intensities ratio, $I_\lia/I_\lib = 3.3$, is also below the observed value of $4.3~(3.4-5.5)$. The X-ray emission intensity in the 0.4-2.0 keV band in our model is $3.3 \times 10^{-13}~\fluxun$.  This constitutes $16\%$ of the emission intensity measured by \citet{Henley10}, with $2.1~(1.9-2.4) \times 10^{-12}~\fluxun$~(see Table 1 in \citealp{FSM17}).

Since the emission is centrally concentrated, its intensity depends strongly on the density (or pressure, for similar temperatures) near the solar circle. To test the conditions needed to reproduce the observed emission, we construct a higher pressure model, with $P(\rsun)/\kb \sim 3000$~K~\cmv, that still reproduces the OVI-OVIII observations (by keeping the gas density-metallicity product constant). In this model, the emission intensities are $I_\lia \approx 1.6$ and $I_\lib \approx 0.5$~L.U, and the band emission is $S_{0.4-2.0} = 1.0 \times 10^{-12}~\fluxun$. This is higher by a factor of $\sim 3$ compared to our fiducial model, and closer to, but still below, the MW values. However, reproducing the measured values requires $P(\rsun)/\kb \sim 4500$~K~\cmv, significantly higher than suggested by observations of high velocity clouds above the MW disk, and a factor of $>3$ higher than in our fiducial model, with $P/\kb = 1350$~K~\cmv~\footnote{~Our isothermal model in \FSMII~was normalized to a thermal pressure of~$P/\kb = 2200$~K~\cmv~at \rsun.}. Furthermore, such a model exceeds the DM upper limit estimated to the LMC, with ${\rm DM} \approx 30~{\rm cm^{-6}~pc}$. As we discussed in \FSMII, an alternative explanation is that most of the X-ray emission originates in the hot ISM in the Galactic disk, not included in our model.

\subsection{External Galaxies}
\label{subsec_obsext}

\subsubsection{OVI and NV absorption}
\label{subsec_obsext_abs}

 \begin{figure*}
 \includegraphics[width=0.95\textwidth]{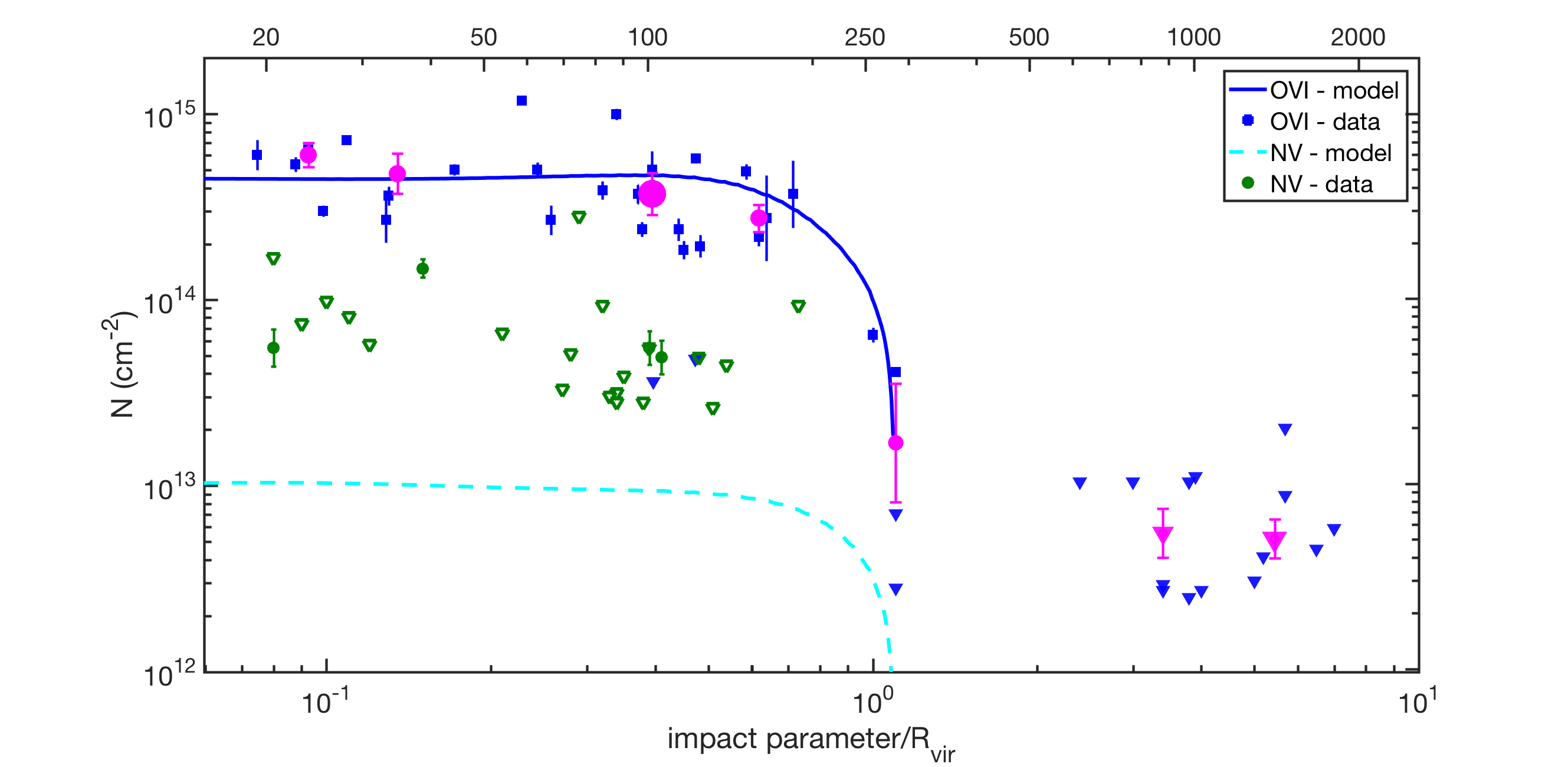}
\caption{OVI and NV column density profiles - observations and model. The data are shown as a function of the impact parameter normalized by the virial radius on the bottom axis, and the physical impact parameter for our model on top. The blue markers are the OVI column density measurements (squares) and upper limits (filled triangles), taken from the COS-Halos \citep{Tumlinson11} and eCGM surveys \citep{Johnson15}. The magenta circles show the OVI binned data, with the marker size and error bars indicating the number of objects and the scatter in the bin. The green markers are the NV data, taken from COS-Halos \citep{Werk13}, with filled circles and empty triangles for the measurements and upper limits, respectively. The solid blue and dashed cyan lines show the OVI and NV column density profiles in our fiducial model (see \S\ref{subsec_obsext}). OVI absorption is detected out to $h \sim \rvir$, with only upper limits at larger impact parameters. The value of $\rcgm=1.1\rvir$ in our models is chosen to reproduce this distribution. Our model predicts that the NV column densities in the CGM of MW-like galaxies are a factor of $\sim 3-10$ below the current upper limits.}
   \label{fig:new_ovi}
 \end{figure*}

For the OVI absorption data, we combine two sets of observations. The first are measurements from the COS-Halos survey described in \FSMII \footnote{We note that the COS-Halos star-forming galaxies have a relatively narrow halo mass range - $90\%$ of the galaxies with OVI detections have masses between $3 \times 10^{11}$ and $2 \times 10^{12}$~\msun~(see Figure~3 in \citealp{MW18}). Thus, modeling the COS-Halos SF galaxies using the MW halo is a reasonable approximation. The virial temperature scales as $M_{\rm vir}^{2/3}$, giving a factor of $(20/3)^{2/3} \approx 3.5$, or $\pm 0.27$~dex. This is similar to the scatter in the OVI column density measurements in the CGM of these galaxies (see Figure~\ref{fig:new_ovi} here).}, probing impact parameters of $h \lesssim 0.6$~\rvir. The second data set are measurements from the eCGM survey, presented by \cite{Johnson15} and extending out to $5-10$ virial radii of the observed galaxies. Beyond \rvir, \cite{Johnson15} report mostly upper limits for the OVI column densities, typically below $10^{13}$~\cmc~(see Figure 3 there). Since we aim to model MW-like galaxies, we select from this sample isolated, star-forming (SF) galaxies with stellar masses above $\sim 3 \times 10^{9}$~\msun, similar to the SF galaxies in the COS-Halos sample. This results in 18 measurements, and the combined data set (COS-Halos and eCGM) is shown in Figure \ref{fig:new_ovi} by the blue markers, with measured columns as squares and upper limits as filled triangles.

The OVI data can be well approximated by a simple step function. Within approximately the virial radius ($h \lesssim \rvir$), the profile is consistent with a constant column density of $\sim 4 \times 10^{14}$~\cmc. At $h \sim \rvir$, the column density drops sharply, with only non-detections at larger impact parameters. For $3<h<7$~\rvir, the median upper limit is $\sim 7 \times 10^{12}$~\cmc, a factor of $\sim 50$ lower than the typical column density measured by COS-Halos. For a clearer comparison to our model, we bin the individual measurements in radius in logarithmic intervals, taking the median column density in each bin and estimating the error as the scatter. The binned data are shown by the magenta markers, with the marker size proportional to the number of objects in the bin. The OVI column density profile for our model is the blue solid curve in Figure~\ref{fig:new_ovi}, consistent with the binned data points at $h/\rvir>0.1$. The CGM distribution in our model ends at $\rcgm = 1.1 \rvir$, chosen to be consistent with the few OVI detections at $h\sim 1.1 \rvir$, and the non-detections at larger impact parameters. 

In the COS-Halos sample, \cite{Werk14} search for NV absorption and report upper limits for most sightlines. For the star-forming galaxies in the sample, 20 out of 24 sightlines have upper limits for the NV column densities. The COS-Halos NV data is shown in Figure~\ref{fig:new_ovi} by green markers, with the non-detections as empty triangles and measured columns as filled circles. The median upper limit on the NV column density is $5 \times 10^{13}$~\cmc, with a scatter of $<0.3$~dex. The cyan dashed curve shows the NV column density profile in our fiducial model. The profile is almost constant with impact parameter, with $N_{\rm NV} \approx 10^{13}$~\cmc, consistent with the measured upper limits. Thus, our model predicts that the NV column densities in the CGM of MW-like galaxies at low redshift are a factor of $\sim 3-10$ below the current upper limits.

The four sightlines with detected NV absorption have column densities between $\sim 0.5$ and $\sim 1.5 \times 10^{14}$~\cmc. Three of the galaxies associated with these sightlines have stellar masses below $1.5 \times 10^{10}$~\msun. This is a factor of $3-4$ lower than the MW stellar mass and below the median stellar mass of the COS-Halos star-forming subsample, $M_* \approx 2 \times 10^{10}$~\msun. In CIE, the NV ion fraction peaks at $\sim 2 \times 10^5$~K. These galaxies may have lower halo masses and virial temperatures closer to this value than our fiducial model, leading to an increase in the NV column. The sightline associated with the fourth galaxy has an impact parameter of $h/\rvir \sim 0.15$ and the detected absorption may also be contaminated by gas associated with the galactic disk.

\subsubsection{X-ray Emission}
\label{subsec_obsext_em}

The emission properties of our model are shown in Figure~\ref{fig:emission}, with the computed emission spectrum presented in the left panel, and the projected intensity profile for an external observer - on the right.

Observationally, only a handful of galaxies have been detected in X-ray emission so far, all in the Local Universe and more massive than the MW. Some detections of X-ray emission around massive spirals have been attributed to high star formation in the disk \citep{Strickland04a,Tullman06b}. More recently, \citet{Das19b} combine X-ray imaging and spectral analys and report the detection of extended emission around NGC3221, a massive spiral galaxy with $SFR \sim 10$~\msuny, out to impact parameters of $\sim 150$~kpc. Focusing on galaxies with SFR similar to the MW, \cite{Pedersen06} use Chandra to measure X-ray emission around NGC~5746, and report a $0.3-2.0$~keV luminosity of $L_X \sim 4.4 \times 10^{39}$~\ergs. \cite{Rasmussen09} re-analyze these observations with updated calibration data and add observations of NGC~5170. They do not detect significant emission in either galaxy, and place a $3-\sigma$ upper limit of $4.0 \times 10^{39}$~\ergs~on the X-ray luminosity. The projected integrated, bolometric luminosity in our model inside $40$~kpc is $1.5 \times 10^{40}$~\ergs. The emission in the $0.4-2.0$~keV band is $4.4 \times 10^{38}$~\ergs, consistent with the limit by \cite{Rasmussen09}. 

L18 measure the X-ray emission intensity profiles of several massive galaxies ($M^* > 1.5 \times 10^{11}$~\msun), observed as part of the CGM-MASS survey. They use stacking analysis and detect emission in the $0.5-1.25$~keV band at the level of $\sim 10^{35}-10^{36}~{\rm \ergs~kpc^{-2}}$ out to $\sim 150$~kpc from the galaxies, or $h \sim 0.3-0.4$~\rvir. They find that the projected intensity profile decreases as a power-law function of the impact parameter, scaling as $I \propto h^{-a}$, with $a=1.4-1.5$, in the range of $h/\rvir\sim 0.03-0.6$. The emission in this band in our model has similar intensities in the inner part but a slightly steeper profile, with a power-law slop of $a=1.7$.

We note two important differences between the MW CGM and that of more massive galaxies. First, the halo virial temperature scales with the halo mass and radius as $\tvir \propto \mvir \rvir^{-1} \propto \mvir^{2/3}$ (see Eq. \ref{eq:tvir1}). Higher gas temperatures can produce the overall stronger emission reported by \cite{Pedersen06}. Second, for median cosmological halos, the halo concentration, $C\equiv \rvir/\rs$ (where~\rs~is the halo scale radius, see \citealp{SMW02}) decreases with halo mass \citep{DM14}. For a given halo mass, lower concentrations result in more extended dark matter distributions and flatter gravitational potentials. The combination of lower concentrations and higher gas temperatures may lead to flatter gas density distributions and emission profiles, compared to the MW, consistent with the results by L18. Additional parameters in our model may vary with galaxy mass, such as the ratio of thermal to non-thermal support, the turbulent velocity scale in the CGM, etc. An exploration of the variation in halo mass and its effect on the properties of the CGM is beyond the scope of this paper.

\section{Predictions for Future Observations}
\label{sec_predict}

In this section we present observational predictions of our model. We calculate column densities of different metal ions that are present in the warm/hot gas, and can be observed in UV and X-ray absorption. We use the calculated spectrum of the corona to predict the emission intensity profiles in different energy bands. We predict the dispersion measure for observations of pulsars and/or fast-radio-bursts (FRBs), and we calculate the radially dependent Compton $y$-parameter, for comparison to Sunyaev-Zeldovich distortions inferred from CMB measurements. We show these quantities for observations of the MW and other galaxies. The results presented here are available online, for comparison to other models and observational data.

\subsection{Milky Way}
\label{subsec_predmw}

\begin{figure*}
\includegraphics[width=0.97\textwidth]{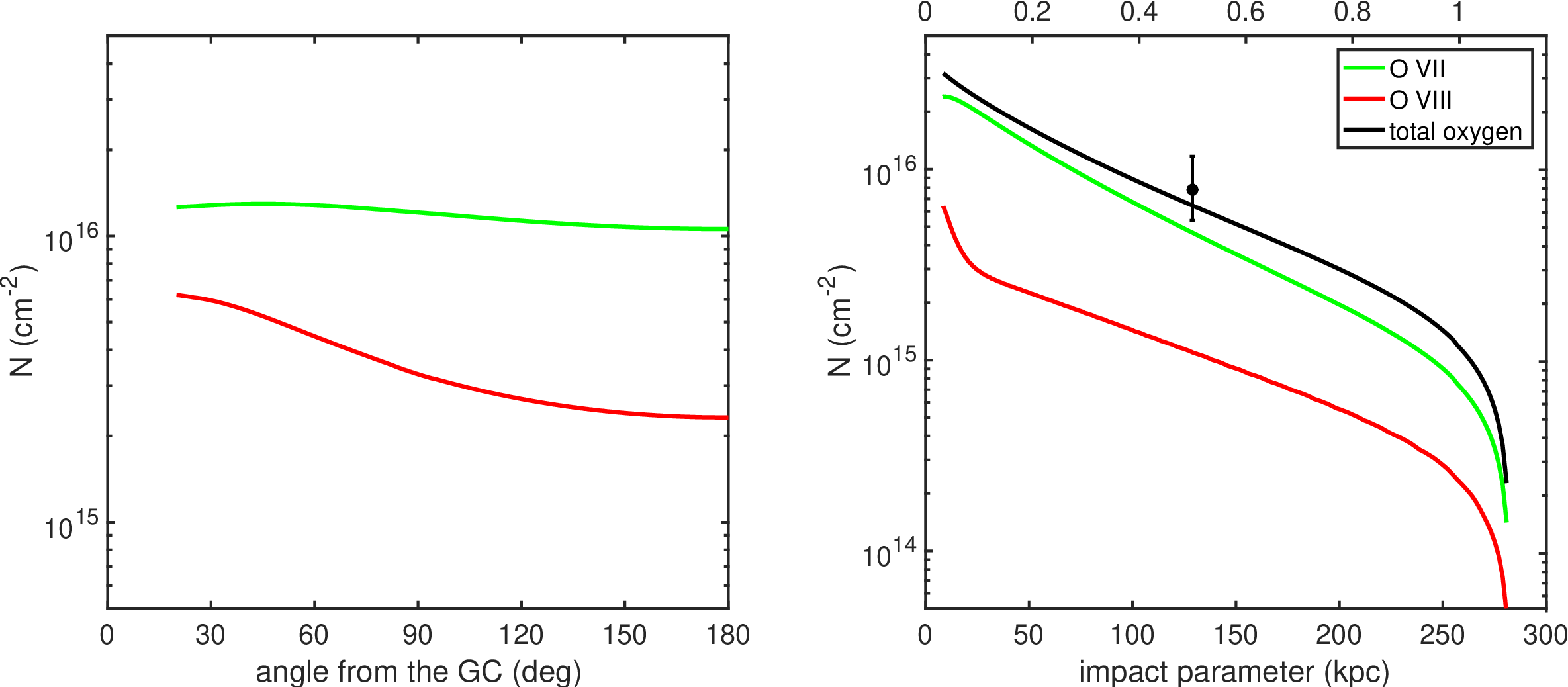}
\caption{The OVII (green) and OVIII (red) column densities in our fiducial model. {\bf Left:} The columns as a function of the angle from the Galactic center for an observer inside the galaxy, at $\rsun=8.5$~kpc. The volume OVII distribution is extended, with a length-scale of $\sim 30$~kpc, and the column density does not vary strongly with $\tgc$. The OVIII is formed mainly in the inner, high-temperature part of the CGM, and the column density away from the GC is lower (see \S\ref{subsec_predmw}). {\bf Right:} The columns for an external observer, looking through the CGM at an impact parameter $h$. The OVII and OVIII columns at $\sim 10$ kpc are a factor of two higher than the observed values for the MW. The marker shows the total oxygen column measured by \cite{Nicastro18} in the WHIM, with a nearby galaxy at a projected distance of $h=129$~kpc. The total oxygen column in our model (black curve) at this impact parameter is consistent with the measurements, suggesting that a significant fraction of the detected absorption may originate in the CGM rather than the IGM (see \S\ref{subsec_predext_abs}). The data used to create the right panel of this figure are available.}
  \label{fig:predxr}
\end{figure*}

The OVII and OVIII absorption at $z\sim0$ has been measured in the X-ray-brightest QSOs, with $\sim 30-40$ OVII detections and a handful of sightlines with OVIII \citep{Bregman07,Gupta12,Miller13}. \cite{Fang15} searched for a correlation of the OVII column density with Galactic latitude or longitude and found that existing data are consistent with a constant column density profile. However, current absorption observations in the X-ray often have significant uncertainties, due to limited sensitivity and spectral resolution. Future X-ray observatories will provide measurements for a larger number of sightlines with higher accuracy \citep{Athena1,Lynx1,Arcus1}, and we calculate the OVII/OVIII column distributions in our model to be tested by these observations. 

We plot the predicted OVII and OVIII column densities in the left panel of Figure~\ref{fig:predxr}. Since our model is spherically symmetric, for an observer inside the Galaxy the column densities, as well as other quantities, are a function only of the angle, \tgc, from the Galactic Center (GC). As described in \S\ref{subsec_metalions} and \S\ref{subsec_obsmw_abs}, the OVII ion is abundant at all radii in the CGM, and its half-column length-scale, $\lscale \sim 30$~kpc, is relatively large compared to \rsun. Thus, for an observer at $r \sim 10$~kpc, the OVII column density (green curve) is almost constant with \tgc, consistent with current observations \citep{Fang15}. The OVIII ion, on the other hand, is formed mostly in the central part of the CGM and its length scale is smaller ($\sim 10$~kpc). Thus, the column density at small \tgc, with $6 \times 10^{15}$~\cmc, is higher by a factor of $\sim 2-3$ than at large angles from the center (red curve).

\begin{figure*}
\includegraphics[width=0.97\textwidth]{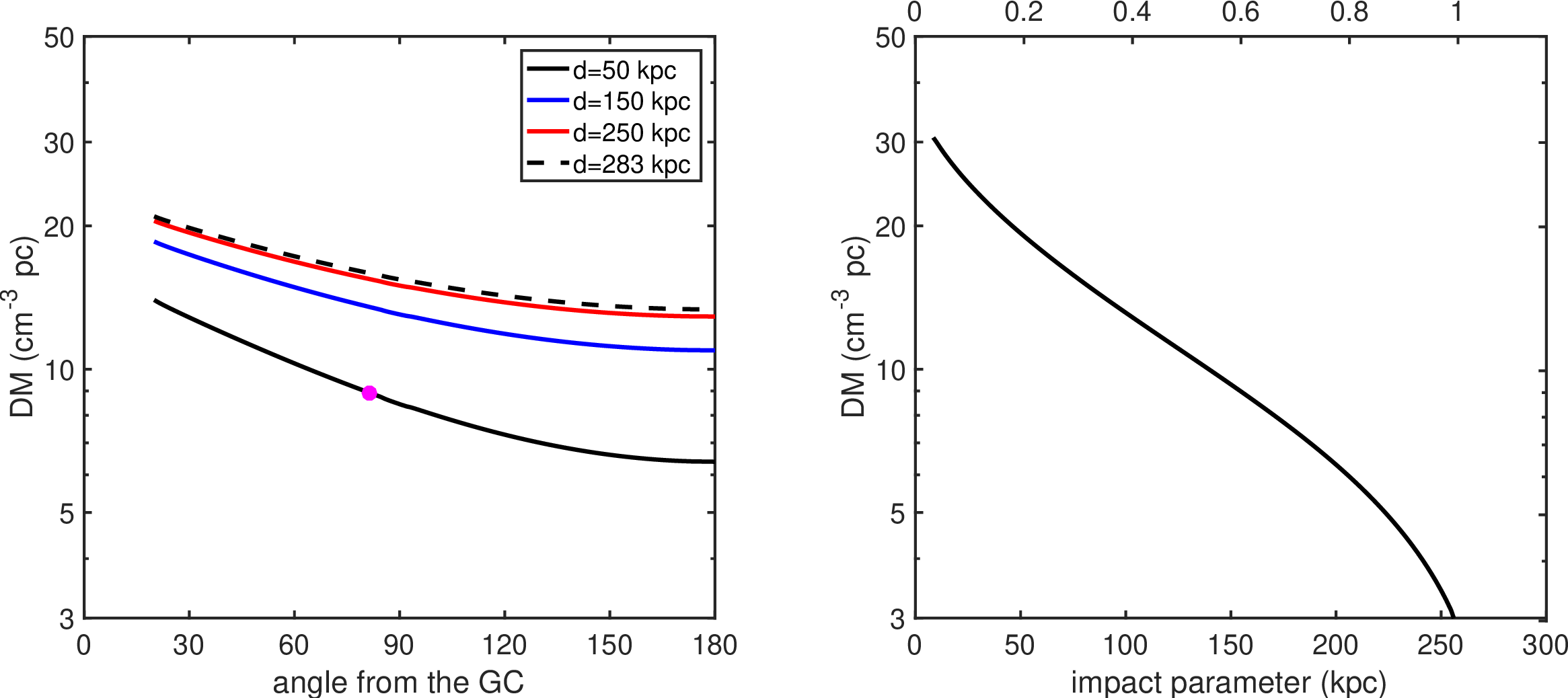}
\caption{The dispersion measure in our model. {\bf Left:} The DM as a function of the angle from the Galactic center for an observer inside the galaxy, at $\rsun=8.5$~kpc (see \S\ref{subsec_predmw}). The solid curves show the DM for sources inside the halo, at distances of $50$, $150$ and $250$ kpc from the GC. The magenta circle on the $d=50$~kpc curve marks the angle to the LMC. {\bf Right:} The DM for an external observer, looking through the CGM at an impact parameter $h$ (see \S\ref{subsec_predext_em}). The data used to create the right panel of this figure are available.}
  \label{fig:preddm}
\end{figure*}

Dispersion measure can provide a strong constraint on the total gas column, since it is independent of the gas metallicity. Today, DM has been measured for pulsars in the LMC/SMC, at a distance of $\sim 50$~kpc \citep{Crawford01,Manchester06,Ridley13}. Upcoming facilities (LOFAR and SKA, for example, \citealp{LOFAR10,SKA15}), with higher sensitivities, may be able to find pulsars in other, more distant satellites of the MW, and measure their dispersion measures. In the left panel of Figure \ref{fig:preddm} we show the dispersion measure in our model as a function of the angle from the GC, for distances of $d=50$, $150$ and $250$~kpc from the GC (solid black, blue and red curve, respectively). The magenta circle marks the LMC, at $\tgc=81^{\circ}$, with $8.8$~\cmv~pc. Future DM measurements for extragalactic sources (FRBs, for example) may provide constraints on the total DM of the MW CGM. In our model, the contribution from $r>250$~kpc is small, and integration out to \rcgm~gives values of DM $\sim 13-21$~\cmc~pc (dashed black curve), close to the values at $250$~kpc. 

\cite{PZ19} estimate the DM of the MW CGM at $\sim 50-80$~\cmv~pc, integrating to the virial radius. However, this results from models with a large CGM mass\footnote{~The density profiles in \cite{PZ19} are scaled to give a total CGM mass of $0.75\Omega_b/\Omega_m M_{halo} \sim 1.8 \times 10^{11}$ for $M_{\rm halo}=1.5 \times 10^{12}$~\msun, estimated for the MW.}, and hence gas density. Their CGM mass is a factor of $\sim 3$ higher than in our model, and scaling down their values for the DM by the same factor gives $17-27$~\cmv~pc. This is similar to the range in our model when integrated to \rvir, as shown by the solid red curve in Figure~\ref{fig:preddm}. This demonstrates the usefulness of (accurate) DM measurements to constrain the gas density and total mass in the CGM.

\subsection{External Galaxies}
\label{subsec_predext}

\subsubsection{UV and X-Ray Absorption}
\label{subsec_predext_abs}

UV and X-ray absorption from hot gas has been detected for the MW galaxy. However, in the X-ray, absorption observations at $z \sim 0$ lack kinematics, due to the limited spectral resolution of current instrumentation. In the UV, the detected absorption lines are spectrally resolved and their kinematics are measured. Nevertheless, the exact location of the absorbing gas is still unclear due to the complex dynamics of the disk-CGM interface \citep{Zheng15,Zheng19,Martin19}.

Measurements of OVII and OVIII absorption in other galaxies (similar to the COS-Halos OVI observations) will better determine the extent of the hot CGM and be more sensitive to low surface density gas compared to emission observations. In the UV, ions such as NV, OVI, NeVIII and MgX, probe different gas temperatures and can be helpful in constraining the CGM properties. We use our model to predict the column densities for such future observations \citep{Athena1,LUVOIR1}. We present the column density profiles for an external observer both as a function of the physical impact parameter, and normalized to \rvir.

The right panel of Figure~\ref{fig:predxr} shows the column density profiles of OVII and OVIII versus impact parameter $h$ in our fiducial model (green and red curves, respectively). As discussed above, the OVII ion fraction is high and almost constant across the wide range of temperatures in our model, and the resulting OVII column density profile is extended. It is well fit by an exponential profile, $log({N_{\rm OVII}}) \propto \left(-h/L_N \right)$, with a scale of $L_N \sim 0.63$~\rvir, set by the metallicity gradient and the gas density profile. The OVIII column (red), on the other hand, has a two-part profile. In the inner regions ($h<25$~kpc), the OVIII ion fraction is controlled by collisional ionization and decreases rapidly with temperature. This gives a column density profile that has a strong dependence on the impact parameter, with $N \sim 7 \times 10^{15}$~\cmc~at $10$~kpc. In the outer part, the OVIII fraction is set by photoionization and increases with radius between $30$~kpc and \rcgm~(see Figure~\ref{fig:metalions}). The resulting column density profile in the outer part is well fit by an exponential function with $L_N \approx \rvir$, flatter than the OVII.

\citet[hereafter N18]{Nicastro18} report the discovery of OVII absorption in the warm/hot intergalactic material (WHIM). They present measurements of total oxygen column densities for two absorbers, with $7.8^{+3.9}_{-2.4} \times 10^{15}$ and $4.4^{+2.4}_{-2.0} \times 10^{15}$~\cmc, at $z=0.434$ and $z=0.355$, respectively. N18 search for possible associations of these absorbers to galaxies, and for the first system they find a spiral galaxy at a similar redshift and a projected distance of $129$~kpc (although see \citealp{Johnson19}, suggesting that this absorption is associated with the blazar environment). Assuming this galaxy is indeed associated with the absorber and it is similar to the MW, we can compare the measured column to our model. The total oxygen column density from N18 is shown by the black marker in the right panel of Figure \ref{fig:predxr} (including the $1$-$\sigma$ errors reported by the authors). The black curve shows the total oxygen column density in our fiducial model. At an impact parameter of $ h=129$~kpc, our model predicts $N_{O}=3.0 \times 10^{15}$~\cmc, dominated by the OVII ion, with $N_{\rm OVII}=2.3 \times 10^{15}$~\cmc. This suggests that a non-negligible fraction of the observed absorption could originate in the warm/hot CGM of the galaxy adjacent to the line of sight, rather than the IGM. Information regarding the stellar or total mass of this galaxy will allow scaling the impact parameter to the virial radius and performing a better comparison to our model. Furthermore, separating the CGM contribution from the total column will allow a better estimate of the IGM properties. The closest galaxy to the second absorber found by N18 is at a projected distance of $633$~kpc, and a similar association to the CGM is less likely.

\begin{figure}
\includegraphics[width=0.47\textwidth, height=7 cm]{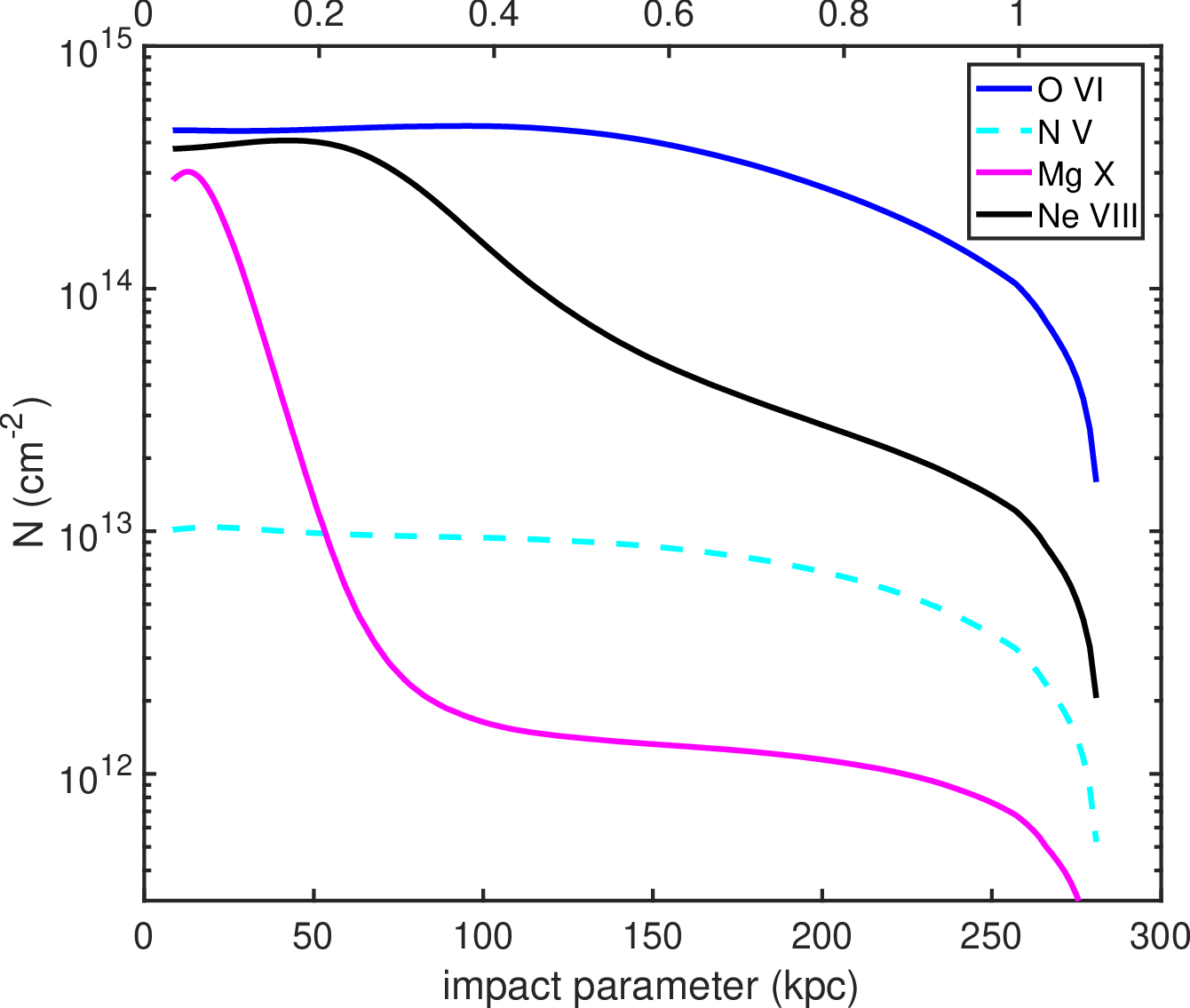}
\caption{Predicted column density profiles for selected metal ions, observable in the UV, for external galaxies. The OVI and NV column density profiles (solid blue and dashed cyan) are identical to those shown in Figure \ref{fig:new_ovi}, and measurements are available from the COS-Halos and eCGM surveys. The NeVIII and MgX columns (solid black and magenta, respectively) have a two-part structure, with collisional and photo-ionization dominating in the inner and outer parts, respectively (see \S\ref{subsec_predext_abs}).}
  \label{fig:preduv}
\end{figure}

\begin{figure}
 \includegraphics[width=0.47\textwidth, height=7 cm]{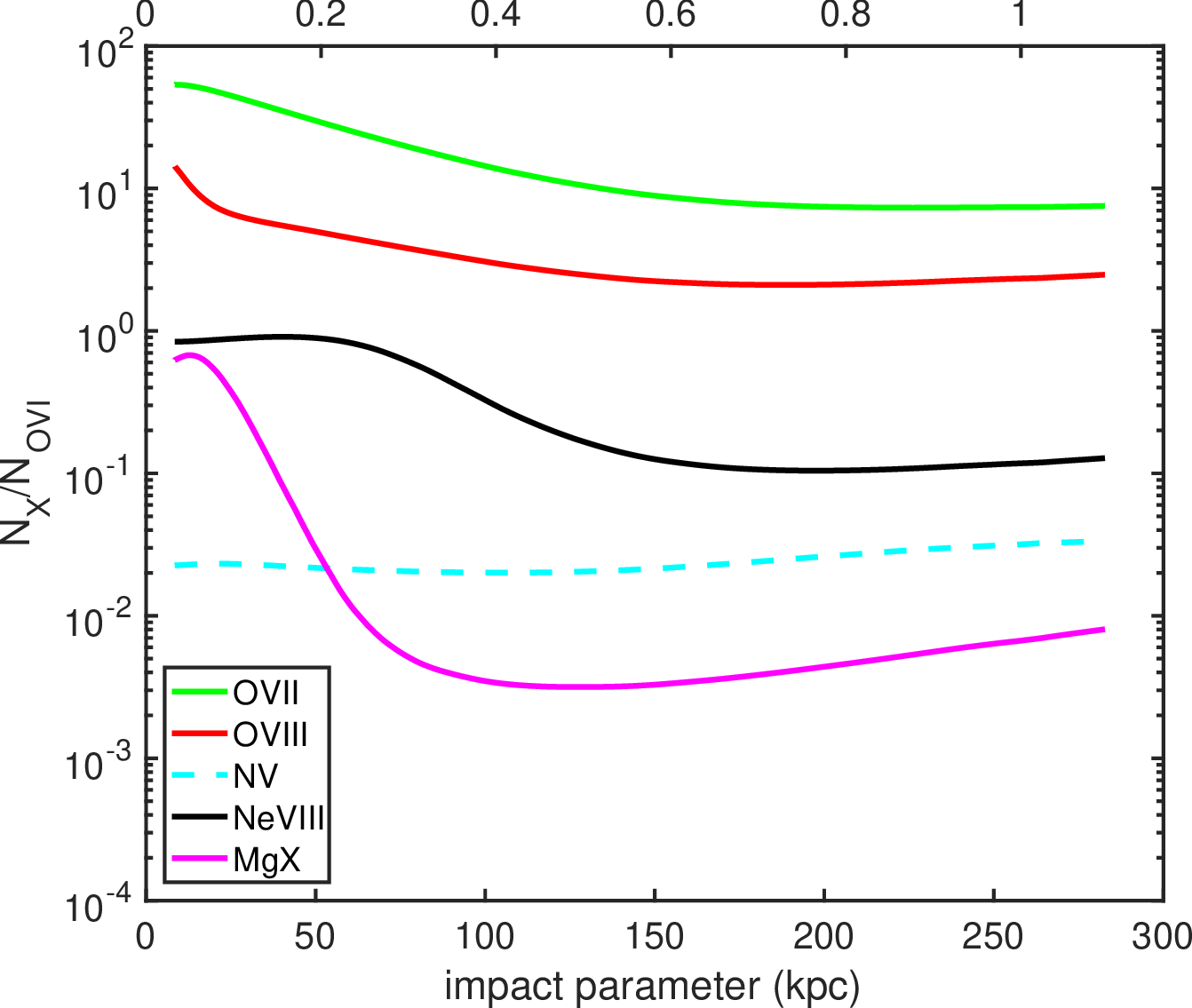}
\caption{Predicted ratios of the ion column densities to the $N_{\rm OVI}$ column density profile (see \S\ref{subsec_predext_abs}). These are independent of the metallicity normalization in a specific galaxy, but do depend on the shape of the metallicity profile.}
  \label{fig:predrat}
\end{figure}

Figure \ref{fig:preduv} shows the column densities of several other metal ions, observable in the UV. We select NV, OVI, NeVIII, and MgX - ions present in gas at temperatures between $\sim 2\times 10^5$ and $1.2 \times 10^6$~K. First, the NV and OVI profiles (dashed cyan and solid blue curves, respectively) are identical to those presented in Figure~\ref{fig:new_ovi}. As discussed in \S\ref{subsec_obsext_abs}, the OVI and NV ions are abundant mainly in the outer parts of the corona (where \tgas~is low), resulting in flat column density profiles. The COS-Halos NV absorption measurements give upper limits for a large fraction of the observed sightlines. We predict that the actual column densities are $\sim 0.5-1.0 \times 10^{13}$~\cmc, a factor of 3-10 below the existing upper limits.

The NeVIII and MgX ions (solid black and magenta curves) probe hotter gas, at $T \sim 10^6$~K, and their ion fractions peak at smaller radii. Thus, their column density profiles have a two part structure, with high columns at small impact parameters, and lower values at larger (projected) distances. Current instrumentation limits observations of NeVIII to $0.5<z<1.0$ \citep{Meiring13,Hussain15,Burchett19}. To compare current observations with this work, we can assume that the halo and CGM properties of these higher redshift galaxies are not very different from the MW/COS-Halos galaxies. Our model then predicts that the column density in the central part of the profile, controlled by collisional ionization of NeVIII, will not change significantly with redshift. In the outer part of the corona, NeVIII is created by the MGRF, and for a field intensity higher by a factor of 3-5, the column density at large impact parameters may be higher by a similar factor. Current detections of MgX absorption are rare and are at higher redshifts than our model, $z>1.0$ \citep{QB16}.

\subsubsection{X-Ray Emission and Dispersion Measure}
\label{subsec_predext_em}

The left panel in Figure~\ref{fig:emission} shows the predicted emission spectrum of the warm/hot gas in our fiducial model, and the magenta solid curve in the right panel shows the projected emission intensity profile in the $0.4-2.0$~keV band. The horizontal magenta line shows the background level estimated by L18, of $10^{35}~{\rm \ergs~kpc^{-2}}$, in their stacking analysis. The emission intensity in our model is above this level out to $\sim 20$~kpc. For an external observer, this would not extend much beyond the size of the MW disk, and can be challenging to define clearly as CGM emission. The power-law slope of the emission profile is $a\approx 1.7$. As discussed in \S\ref{subsec_obsmw_em}, the emission intensity profile may be slightly flatter for higher mass galaxies, due to the lower gas temperature and more compact mass distributions of the dark matter halos.

In the right panel of Figure \ref{fig:preddm} we show the dispersion measure as a function of the impact parameter. The DM through the CGM is $\gtrsim 20$~\cmv~pc at impact parameters below $\sim 50$~kpc, and decreases to $\lesssim 5$~\cmv~pc at $h>200$~kpc. For a sightline through the halo of an $L^*$ galaxy, \cite{PZ19} estimate a DM between $10-150$~\cmv~pc, for impact parameters between $\sim 15$~kpc and \rvir. Scaling down the DM with the CGM mass, by a factor of $\sim 3$, brings their prediction into agreement with our fiducial model (see \ref{subsec_predmw}). Future FRB campaigns may allow to probe the CGM of galaxies in the Local Universe and beyond through DM measurements \citep{Bandura14,McQuinn14}.

\subsubsection{Sunyaev-Zeldovich (SZ) Effect}
\label{subsec_predext_sz}

We calculate the spatially resolved SZ signal through the corona at an impact parameter $h$ as 
\begin{equation}\label{eq:ypar_res}
y(h) = \frac{\sigma_T}{m_e c^2} \int{P_{\rm e, th}(r) dz} = \frac{\sigma_T \kb}{m_e c^2} \int{n_{\rm e}(r) T_{\rm th}(r) dz}~~~,
\end{equation}
where $P_{\rm e, th}$ is the electron (thermal) pressure, $n_{\rm e}$ is the electron density and $dz$ is the element along a line of sight. The resulting $y$-parameter is shown in Figure~\ref{fig:ypar}, and the profile decreases from $\sim 10^{-8}$ at small impact parameters, to $\sim 2-3 \times 10^{-10}$ at \rvir. Current CMB observations do not have this sensitivity, and this prediction can be compared to future, spatially-resolved CMB measurements of galactic halos (see \citealp{Singh15}).

\begin{figure}
 \includegraphics[width=0.47\textwidth, height=7 cm]{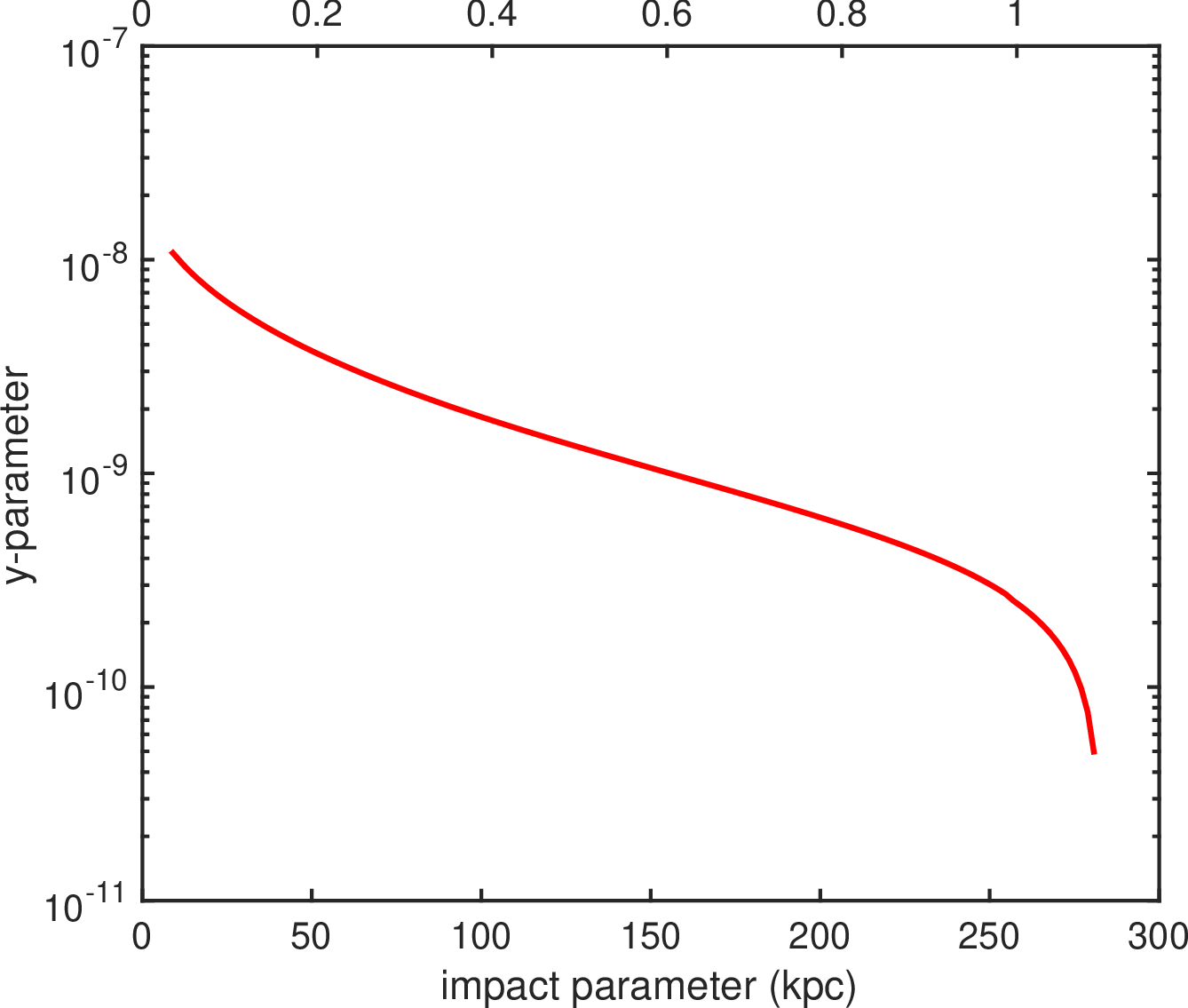}
\caption{The predicted Compton y-parameter profile of the coronal gas in our fiducial model (see \S\ref{subsec_predext_sz}).}
   \label{fig:ypar}
 \end{figure}

\citet[hereafter P13]{Planck13_SZ} search for the SZ signal from gas in galaxies by stacking CMB measurements of locally brightest galaxies (LBGs). They report the Comptonization parameter normalized to a distance of $500$~Mpc, giving the intrinsic integrated SZ signal and defined as
\begin{equation}\label{eq:ypar_tot}
\ypar \equiv \frac{\sigma_T}{m_e c^2} \frac{E^{-2/3}(z)}{(500~{\rm Mpc})^2} \int^{R_{500}}{P_{\rm e, th} dV} ~~~,
\end{equation}
where $E^2(z) = \Omega_m(1+z)^3+\Omega_{\Lambda}$. The signal is calculated out to $R_{500}$ of the halo and \ypar~is usually reported in square arcminutes. We calculate it at $z=0$, and $E(z)=1$. 

P13 detect a signal from systems with stellar masses above $\sim 10^{11}$~\msun. For the lowest mass bin with a $3 \sigma$ detection, $M_{*} = 2 \times 10^{11}$~\msun~($M_{500} \sim 2 \times 10^{13}$~\msun), they report $\ypar \approx 5 \times 10^{-6} ~{\rm arcmin}^2$. To estimate the signal from a MW-mass galaxy, we can use the $\ypar-M_{500}$ relation, usually fit by $\ypar \propto M_{500}^{a_M}$.  In their analysis, P13 adopt the slope predicted by the self-similar solution for the gas distribution in a halo, of $a_M = 5/3$. 
Using this value to calculate the SZ signal for a MW-mass galaxy, with $M_{500} \sim 7 \times 10^{11}$~\msun, gives a $\ypar \sim 2 \times 10^{-8}~{\rm arcmin}^2$. However, P13 note that a single power-law is not a formally acceptable fit to the measured $\ypar-M_{500}$ relation. This may be a result of the gas distributions in galaxies differing from those in clusters (see also \citealp{Bregman18}). Thus, the actual SZ signal for MW-mass galaxies may be different from the extrapolated value. For the MW, $R_{500}=135$~kpc, and in our fiducial model, $\ypar = 0.5 \times 10^{-8}~{\rm arcmin}^2$.

The angular resolution of the Planck maps used in the P13 stacking analysis is $10'$ (FWHM, see \S 5.1 there), and the MW $R_{500}$ will not be resolved at distances above $\sim 50$~Mpc. For spatially unresolved CMB observations, we integrate the SZ signal in our model out to \rcgm~and get $\Tilde{Y} = 1.2 \times 10^{-8} ~{\rm arcmin}^2$. This is similar to the estimate by \cite{Singh15}, of $\ypar \sim 10^{-8}$, for the warm, OVI-bearing CGM.

\section{Comparison to \FSMII}
\label{sec_fsm17}
    
We now address the similarities and differences between our isentropic corona described in this work and our isothermal model prsented in \FSMII.
    
Starting with the similarities, both models require significant non-thermal support to reproduce the observed OVI column density profile. For isothermal, the ratio of total to thermal pressure is independent of radius, with a value of $\alpha \approx 2$. For isentropic, the ratio varies with radius between $\alp(\rcgm) \approx 3$ and $\alpha(\rsun) \sim 1.5$.
The result of the non-thermal support is that the gas density profiles have shallow slopes, with similar power-law indices, of $0.93$ and $\approx 0.90$, in the isentropic and isothermal models, respectively. The extent of the CGM in both models is similar, with $280$ and $250$~kpc, as suggested by the OVI absorption studies of $\sim L^*$ galaxies in the low-redshift Universe.
    
The two models differ in several important aspects.
First are the gas temperature distributions. In \FSMII~we included a local (isobaric) lognormal distribution of temperature (and density), but the local mean gas temperature does not vary with radius. In our isentropic model, locally, the gas has a single temperature, but it decreases from $\sim 2 \times 10^6$~K at \rsun~to $\tvir \sim 2 \times 10^5$~K at \rcgm. The temperature variation results from adopting the adiabatic equation of state, with $T \propto \rho^{\gamma-1}$ and constant entropy. 
    
Second, the mean gas density in our isentropic corona is a factor of $\sim 3$ lower than in \FSMII, and the total gas mass inside \rvir~is also lower by a similar factor. With a baryonic overdensity density of $\sim 20$, this is closer to the values predicted by structure formation theory. As a result of the lower density and temperature, the total pressure at the outer boundary is $\sim 20$~\cmv~K, $\sim 10$ times lower than that in our isothermal model, and similar to the IGM pressure in cosmological simulations. The pressure in the inner part of our isentropic model is also lower, with $P/\kb = 1350$~K~\cmv, compared to $2200$~K~\cmv~in the isothermal model.
    
For the higher gas densities in \FSMII, pure CIE is a good approximation, and photoionization by the MGRF has a negligible effect on the gas ionization state. In our isentropic model, the gas density at large distances from the galaxy is low enough for photoionization to reduce the ion fractions of the NV and OVI ions at large radii. For other ions, OVII and OVIII, radiation may increase the fractions locally in some parts of the corona, but does not have a significant effect on the total column densities for an observer inside the Galaxy.
    
In \FSMII, the gas metallicity is constant, with $Z' = 0.5$~solar, while in our current model we adopt a varying metallicity profile. This is motivated by enrichement of the CGM by the Galaxy through outflows and metal mixing, and in our fiducial model, the metallicity varies from $Z'(\rsun)=1.0$ to $Z'(\rcgm)=0.3$. The combination of temperature and metallicity gradients leads to shorter lengthscales for the OVII and OVIII (see \S~\ref{subsec_obsmw_abs}). Furthermore, in \FSMII~the OVII and OVIII ions had the same half-column lengthscale for an observer inside the galaxy. The temperature gradient in the isentropic model leads to a different spatial distribution of these ions - the OVII is more extended, while the OVIII is more compact (see Figure~\ref{fig:predxr} and Table~\ref{tab:mod_res}).
    
Finally, both in \FSMII~and in this work, the origin of the OVI is in warm, collisionally ionized gas. However, the properties of this gas in the two models are different. In our isothermal corona, the warm gas is a separate phase that condenses out of the hot, $2 \times 10^6$~K, gas. Since we assume the two phases are in pressure equilibrium, the warm gas density is higher than in the hot phase. This, together with its higher cooling rate leads to short cooling times, of $\tcool \sim 2 \times 10^8$~years (without heating). In our isentropic model, the OVI is formed in virialized gas at lower densities. The combination of the lower gas density and metallicity leads to a longer cooling time, with $\tcool \geq 3 \times 10^9$~years at $r>100$~kpc, and this gas can be long-lived even without constant energy injection. Furthermore, the total luminosity of the isentropic corona is lower by $\sim 20$ and the radiative losses per gas unit mass are $10$ times lower. Similar to the isothermal model, we assume a stable heating/cooling equilibrium in our current model.

\section{Discussion}
\label{sec_disc}

Many recent works studied the CGM in detailed simulations of galaxy formation and evolution. \citet{Opp16} addressed the origin of the OVI-SFR correlation in the EAGLE simulation suite, and \citet{Hafen19} studied the properties of the CGM in the FIRE simulations. \citet{Nelson18} explored the distribution of highly ionized oxygen (OVI-OVIII) in the `Illustris' cosmological simulation, and \citet{RF19} focused on the evolution of OVI with redshift in zoom-in simulations. \citet{LiT20} examine the impact of SNe-driven outflows on the structure of the CGM. \citet{Hummels19} and \citet{Peeples19} use zoom-in simulations to check how numerical resolution affects the CGM properties. In this section we compare our model to recent analytical models of the CGM. 

\citet[hereafter MB13]{Miller13} fit the observed OVII and OVIII column densities for a constant temperature. They assume a power-law radial density distribution, and find a best fit power-law index of $a_{\rm n} \sim 1.7$. They adopt $\rcgm=\rvir=200$~kpc and get a total CGM mass of $1.2^{+1.7}_{-0.2}~\times 10^{10}$~\msun. There are two caveats to this estimate. First, the value they adopt for the virial radius is smaller than what is usually taken for the MW, with $\rvir \sim 250 $~kpc for a halo of $\sim 1.5 \times 10^{12}$~\msun~(see Table 8 in \citealp{BHG16}). Second, the mean hydrogen density inside $200$~kpc is $1.1 \times 10^{-5}$~\cmv, lower than the estimate by BR00 for $r<250$~kpc, with $\sim 2.5 \times 10^{-5}$~\cmv. MB13 address this discrepancy by adding an `ambient' component, with a constant density of $n_{\rm e} = 10^{-5}$~\cmv, and say that its mass is within their mass uncertainty. We now re-estimate the MB13 gas mass to compare it to our model. A constant density component with $r=258$~kpc and $n_{\rm e}= 10^{-5}$~\cmv~has a total mass of $2.0 \times 10^{10}$~\msun. Scaling the intrinsic gas mass by a factor of $(258/200)^{1.3} \approx 1.4$ (since $n \propto r^{-1.7}$) and summing the two components, gives a total mass of $\sim 3.6 \times 10^{10}$~\msun. This is closer to the warm/hot gas mass inside \rvir~in our fiducial model, with $4.6 \times 10^{10}$~\msun. This calculation shows the sensitivity of the result to the value of \rcgm~and the importance of density constraints at large distances from the MW for estimates of the total CGM mass. Furthermore, the apparent difference between the initial steep profile of the MB13 fit, inferred from the X-ray absorption measurements, and the density estimate of BR00, may be evidence for lower CGM temperatures at larger distances from the Galaxy.

\citet[hereafter QB18a]{QB18a} and \citet[hereafter QB18b]{QB18b} construct a CGM model for halos with masses between $3 \times 10^{10}$ and $2 \times 10^{13}$~\msun. In their model, the temperature is constant as a function of radius. They assume a virial temperature of $\tvir \sim 7 \times 10^5$~K for a $10^{12}$~\msun~halo, a factor of $\sim 3$ higher than the temperature of the shocked gas at \rcgm~in our fiducial model. They consider two main model versions - (i) an isothermal model with a single temperature at each radius (with and without radiation) and (ii) a model with a local temperature/density distribution function that is proportional to the gas cooling time.

For their fiducial galaxies, QB18a assume a power law density profile with a slope of $1.5$ and a constant metallicity of $Z'=0.3$~solar, and calculate the column densities through the CGM for several high metal ions - OVI-OVIII, NeVIII and MgX. For MW-mass halos, they get $N_{\rm OVI} \sim 5 \times 10^{13}$~\cmc~at an impact parameter of $h/\rvir=0.3$~(see their Figure 5), and including photoionization reduces the OVI column, similar to the effect in our model. Adopting a local temperature distribution increases the OVI column to $\sim 1.5 \times 10^{14}$~\cmc~, still below the values observed by COS-Halos. To fit the OVI-OVIII columns observed in the MW, QB18 construct a different model, with a higher metalllicity, of $Z'=0.5-1.0$, and $\tvir \sim 2 \times 10^6$~K. Thus, the gas temperature inferred by QB18a for the MW is higher by a factor of $\sim 2-3$ than that of their fiducial galaxies at similar halo masses. The temperature of the hot phase and the gas metallicity in our \FSMII~model are similar to the QB18a MW fit. However, in our analysis, other MW-like galaxies in the low-redshift Universe have similar values for these properties.

The total gas mass in the QB18a fiducial models is low compared to the stellar mass of these galaxies, with $M_{\rm CGM} \sim 1-2 \times 10^{10}$~\msun~for a $10^{12}$~\msun~halo (see their Figure 18). Including the stellar mass gives an almost constant baryon fraction for halos with $\mhalo > 5 \times 10^{11}$~\msun, with $f_{\rm b} \approx 0.05-0.06$, or $30-40\%$ of the cosmic budget. Given the density profile, extending the CGM distribution to twice the virial radius increases the coronal gas mass only by a factor of 2-3. For these CGM masses, the mean coronal gas density inside the virial radius is small, with $\left<\nh \right> \sim 10^{-5}$~\cmv, and the actual density at large radii is lower by a factor of $\sim 3$. This is similar to the problem discussed by MB13 for their model. A similar solution, adding an ambient, constant density component, will increase the total gas mass and result in a gas density profile with an effective shallower slope. A key difference between our models (isothermal and isentropic) and MB13/QB18 is the slope of the density profile. Our models have flatter profiles that result from including non-thermal pressure support.

\citet[hereafter S18]{Stern18} construct a two zone model for the CGM, with the two regions separated by the virial shock, located at $r_{\rm shock} \approx 0.6$~\rvir. The inner CGM consists of hot gas, at $T \sim 5 \times 10^{5}$~K, and the outer part is cool, photoionized gas, at $T=3 \times 10^{4}$~K. In this model, the OVI is formed in the cool, photoionized gas outside $r_{\rm shock}$.

\citet[hereafter V19]{V19} presents CGM models with gas in hydrostatic equilibrium and entropy that increases as a function of radius. We now compare the properties of his fiducial model (named pNFW/Zgrad) and our isentropic model. The dark matter halo of the pNFW model has a isothermal core and an outer NFW part. The gas distribution at small radii has a constant cooling to dynamical time ratio, $\tcool/\tdyn = 10$ (or higher), motivated by precipitation limited models (see also \citealp{V18}). The gas density behaves as $n \propto r^{-1.2}$ at small radii, similar to our density slope. At large radii, the gas density profile steepens, with $n \propto r^{-2.3}$. The gas density range between $30$ and $\sim 250$~kpc is similar to ours (see their Figure 1), and the total CGM mass inside $r_{200} \approx 261$~kpc is $5 \times 10^{10}$~\msun. The gas metallicity profile in pNFW/Zgrad is also similar to ours, decreasing from $Z/Z_{\odot}=1.0$ in the vicinity of the disk to $0.3$ in the outer halo.

The gas temperatures in the pNFW model are $T \sim 8.5 \times 10^5$~K at $r_{200}$, and $\sim 3 \times 10^6$~K at $\sim 10$~kpc, higher than in our fiducial model. As a result, the OVII/OVIII column ratio in this model is $\sim 1$  for an observer inside the galaxy, higher than estimated for the MW CGM, $\sim 4$ (see Table~\ref{tab:mod_res}. Furthermore, the OVI column densities for an external observer are $<10^{14}$~\cmc, lower than measured in the COS-Halos survey. To solve this, V19 invokes local temperature fluctuations (as we introduced in \citealp{FSM17}), and shows that a wide distribution ($0.3-0.4$~dex) can increase the OVI fractions and columns by a factor of up to $\sim 5$.

V19 assumes CIE in his calculations of the ion fractions for the MW, and predicts that in lower mass halos OVI will be created by photoionization. We check this by scaling down the temperature at $r_{200}$ in his MW model to $M_{\rm halo} = 3 \times 10^{11}$~\msun, which gives $T(r_{200}) =  8.5 \times 10^{5}~{\rm K} (3 \times 10^{11}/2 \times 10^{12})^{2/3} \approx 2.5 \times 10^{5}$~K, just below the OVI peak. As we show in \S~\ref{subsec_ionization} (Figures~\ref{fig:ovi_frac} and~\ref{fig:metalions}), at this temperature, and densities of $10^{-5}-10^{-4}$~\cmv~(or pressures of $P/\kb \sim 2-20$~K~\cmv), photoionization already reduces the OVI fraction to $f_{\rm OVI} \approx 0.1$. At higher temperatures, for halo masses between $3 \times 10^{11}$ and $2 \times 10^{12}$~\msun, the OVI will be even lower. Photoionization will increase the OVI fraction for lower temperatures at the outer boundary, at $T \lesssim 2 \times 10^{5}$~K and $n_{\rm H} < 10^{-4}$~\cmv.

In the V19 framework, the gas density in the inner region of the CGM is regulated by its cooling time, so that $\tcool/\tdyn$ is above some threshold value, chosen to be $10$ or higher. We note that in our isentropic model, the same result is obtained naturally - the gas temperature in the inner part is high ($\sim 2 \times 10^{6}$~K) and the dynamical time is low, leading to $\tcool/\tdyn > 10$ at $r<30$~kpc (see Figure~\ref{fig:timescales} here). However, the ratio is different in the two models at large radii, $\sim 200$~kpc, with $\tcool/\tdyn \sim 10$ in V19 and $\sim 3$ in our isentropic model, consistent with the upper limit (see \S\ref{subsec_tcool_limit}). The gas densities and metallicities are similar (for pNFW/Zgrad), and the reason for this difference is the gas temperature. In our model, we set the temperature at $\rcgm$~to be roughly the virial gas temperature at that radius, which happens to be close to the peak in the gas cooling efficiency, at $\sim 3 \times 10^5$~K. V19 chooses $T(r_{200}) \approx 8.5 \times 10^5$~K, where the cooling efficiency of the gas is lower by a factor of $2-3$ and the cooling time is long.

The gas properties in our model are similar to those of the MW-mass ($10^{12}$~\msun) halo in the idealized simulations by \citet[hereafter F17]{Fielding17a}. The gas densities between $0.1$~\rvir~and \rvir~are in the range $10^{-5} - 4 \times 10^{-3}$~\cmv, and the CGM temperature is in the range of $3 \times 10^5-2 \times 10^6$~K (see their Figure 7), similar to the densities and temperatures in our model. F17 find that for a $10^{12}$~\msun~halo, the feedback strength does not affect the CGM properties outside the central part of the halo (at $r/\rvir<0.1$). \cite{Loch20} analyze these simulations and find that when turbulent support is included, the CGM at large radii is close to hydrostatic equilibrium. The density profile in the simulated CGM is steeper than ours, with $a_{\rm n} \sim 1.5$. However, the simulations do not include feedback from the central black hole, magnetic fields and cosmic rays. We have shown that non-thermal pressure support is important for reproducing the observed OVI column density profile. This is especially true at large radii in our isentropic model, where the value of \alp~increases with distance from the Galaxy. Recent simulations also show that cosmic ray pressure is significant in MW-mass halos at $z<1$ \citep{Ji19,KQ20}.

These comparisons uncover an interesting point. Models of MW-sized halos that adopt gas temperatures of $\sim 10^6$~K produce OVI column densities of $\sim 10^{14}$~\cmc, a factor of $3-5$ lower than measured by COS-Halos in $z \sim 0.2$ galaxies. One solution for this is to invoke temperature fluctuations, as we did in FSM17, and \citet{V19} finds that wide distributions are required to reproduce the observed OVI. In our current model, the gas temperature at the outer CGM boundary is lower than the values adopted by QB18a and V19. This gives high OVI columns without local temperature distributions. We emphasize that the warm/hot CGM undoubtedly has some temperature fluctuations, and such fluctuations were important in the FSM17 model\footnote{~In FSM17 we invoked isobaric temperature/density fluctuations in the gas, which resulted in shorter cooling times for a fraction of the hot gas mass ($\sim 20\%$), and provided a physical mechanism for the formation of the warm, OVI-bearing phase. The small fluctuations had a minor effect on the actual ion fractions and column densities.}. In the present model, we have chosen to omit them since they add an additional parameter and they do not affect our model significantly unless they are larger than those in FSM17, which were about $0.15$ dex. To summarize, the two main suggested mechanisms for creating high $N_{\rm OVI}$ in warm/hot gas are (i) a global variation in gas temperature, with $T \sim 3 \times 10^5$~K at the outer boundary, and (ii) a wide local temperature distribution in $T \sim 10^6$~K gas. \citet{Stern18} suggest a different scenario, in which the OVI is created in low-density, cool, photoionized gas outside the virial shock.

\section{Summary}
\label{sec_summary}

In this paper we present a new phenomenological isentropic model for the circumgalactic medium of $L^*$, Milky-Way-like galaxies. Our model reproduces a wide range of absorption measurements, in the UV and X-ray, of the MW and the $0.1 < z < 0.4$ galaxies observed in the COS-Halos/eCGM surveys. We assume that the CGM is in hydrostatic equilibrium and adopt an adiabatic equation of state for the virialized gas, which results in a temperature variation as a function of radius (see Figure~\ref{fig:prof_nt}). We also introduce a decreasing metallicity profile, motivated by gas enrichment of the CGM by the galaxy (\S\ref{sec_model}). 

In \S\ref{sec_fiducial} we described our fiducial corona, defined by a specific set of parameters chosen to reproduce the highly ionized oxygen observations in absorption (see Table~\ref{tab:mod_prop}). The gas density and pressure at the outer boundary of the corona, $\rcgm \approx 283$~kpc, are low, with $\nh \sim 10^{-5}$~\cmv~and $P_{\rm tot}/\kb \sim 20$~K~\cmv, consistent with a picture of (quasi-)static corona. The total gas mass inside the virial radius (\rcgm) is $4.6 \times 10^{10}$ ($5.5 \times 10^{10}$)~\msun. Together with the Galactic disk, this constitutes $\sim 70\%$ of the galactic baryonic budget of the Milky Way. 

Our model is tuned to reproduce the OVI-OVIII absorption observations, and these do not directly constrain the total gas mass. In our model, a given temperature distribution sets the density profile shape and the gas mass is then proportional to the density at the outer boundary, or the pressure at the solar radius. The ion fractions are also set by the temperature and for a fixed value of \rcgm, the column densities constrain the product of the gas metallicity and density (or pressure). However, each of these properties individually can vary, and we scale the CGM mass in our model, with $P/\kb=1350$~K~\cmv~at the solar circle, \rsun, to the observationally estimated range of $\sim 1000 - 3000$~K~\cmv~(see \ref{subsec_bounds}). This results in $0.34 - 1.0 \times 10^{11}$\msun~for the gas mass inside \rvir, or, including the disk mass, between $60\%$ and $100\%$ of the Galactic baryonic budget for a $10^{12}$~\msun~halo. Cool, $\sim 10^4$~K, gas may be an additional significant component.

For the gas densities and pressures in our fiducial model, photoionization by the metagalactic radiation field affects the metal ion fractions (\S\ref{subsec_ionization}). This is in contrast to the \FSMII~model, where due to the higher gas densities and temperatures, pure CIE was a valid assumption. In our calculations we include the effect of the MGRF on the ion fractions and cooling functions, and adopt the HM12 field at $z=0.2$, the median reshift of the COS-Halos galaxies.

We derive a {\it model-independent} upper limit on the cooling time of OVI-bearing warm/hot gas in \S\ref{sec_timescales}, with the detailed calculation presented in the Appendix. We show that for the typical column density measured in the COS-Halos survey, $N_{\rm OVI} \approx 3 \times 10^{14}$~\cmc, the cooling time at large radii in the CGM ($r/\rvir \sim 0.6$) is less than $5.6 \times 10^9$~years. For a MW-mass halo, this results in a ratio of $\lesssim 4$ for the cooling to dynamical times, below the value of $\approx 10$ estimated in previous works for galaxy clusters, and invoked in precipitation models for the CGM. This suggests that cool gas may form by condensation out of the warm/hot phase, in agreement with observations of low metal ions in the CGM, and we address these in our next paper. In our fiducial model, $\tcool/\tdyn \sim 2.5$ at $r>100$~kpc, consistent with the limit we derive. Our equilibrium model assumes that most of the radiative losses are offset by heating of the CGM, requiring an energy input of $\sim 8 \times 10^{40}$~\ergs.The total (thermal, non-thermal and turbulent) energy in our fiducial corona model is $\sim 2.5 \times 10^{58}$~erg, similar to the energy radiated at the present-day luminosity over $\sim 10$~Gyr. We estimate that the total energy available over this epoch in the MW from SMBH feedback, SNe events and IGM accretion is $\sim 2.5 \times 10^{59}$~erg, a factor of $\sim 5$ higher than needed to form and balance the radiative losses of the the CGM.

We compare our model to existing CGM observations in \S\ref{sec_comparison}. It reproduces the OVI column density profile of the COS-Halos/eCGM galaxies (Figure~\ref{fig:new_ovi}), and the OVII-OVIII columns measured in the MW (Table~\ref{tab:mod_res}). The NV column densities in the model are $\sim 10^{13}$~\cmc, a factor of $\sim 5$ below the upper limits reported in COS-Halos. Our computed dispersion measure, DM$=8.8$~pc~\cmv, is consistent with the estimated upper limit $\lesssim 23$~pc~\cmv~to the LMC. The X-ray emission intensities in the model constitute $\sim 20\%$ of the values measured in the MW. Reproducing these requires high pressure at the solar radius, of $\sim 4500$~K~\cmv. As shown in \FSMII, a Galactic disk origin may be a plausible explanation for this emission.

Finally, in \S\ref{sec_predict}, we present predictions of our model for future observations in the UV and X-ray. We calculate the column densities of different metal ions (NV, NeVIII, MgX, etc.), and the emission intensity profiles in different energy bands. We find that in the X-ray, the emission detected today may be very compact due to instrumental sensitivity and backgrounds. We show predicted profiles for the CGM dispersion measure for pulsar and FRBs observations and the Compton $y$-parameter, for measurements of the Sunyaev-Zeldovich effect. We plot our predictions as a function of the angle from the Galactic center, for the MW and the impact parameter through the CGM, for external galaxies. We hope these will be useful for testing our model, improving our understanding of the CGM and studying the physical processes that shape its structure and evolution.

\vspace{0.5 cm}

The manuscript is accompanied by two data files in machine-readable format. The files list the model properties and outputs as a function of the radius and impact parameter, to allow comparison to models and observations. The provided data were used to produce Figures 1-3, 6-9 and 11-15.

\begin{acknowledgements}

We thank Yuval Birnboim, Greg Bryan, Avishai Dekel, Drummond Fielding, Shy Genel, Orly Gnat, Jerry Ostriker, Kartick Sarkar, David Spergel, and Jonathan Stern for fruitful discussions and helpful suggestions during the course of this work. We thank Joss Bland-Hawthorn, Sebastiano Cantalupo, Filippo Fraternali, Aryeh Maller, Smita Mathur, Mike Shull, Benny Trakhtenbrot, and the anonymous referee for their helpful comments on the manuscript.

This research was supported by the Israeli Centers of Excellence (I-CORE) program (center no. 1829/12), the Israeli Science Foundation (ISF grant no. 857/14), and DFG/DIP grant STE 1869/2-1 GE625/17-1. C.F.M. is supported in part by HST grant, HST-GO-12614.004-A. C.F.M. and Y.F. thank the Center for Computational Astrophysics and the Flatiron Institute, Simons Foundation, where some of this research was carried out with A.S., for hospitality and funding.

\end{acknowledgements}



\renewcommand{\theequation}{A-\arabic{equation}}
\renewcommand{\thefigure}{A-\arabic{figure}}
\setcounter{equation}{0}
\setcounter{figure}{0}
\setcounter{section}{0}
\section*{APPENDIX - \\ Cooling to Dynamical Time Ratio of OVI-bearing Gas}  
\label{sec_apptime}

In this Appendix we present a full derivation of our analytical estimate for the cooling time of OVI-bearing gas and derive an upper limit for the ratio of cooling to dynamical time for a MW-mass galaxy. We argue that the OVI columns observed in the COS-Halos survey by \cite{Tumlinson11} imply cooling to dynamical time ratios significantly lower than estimated in galaxy clusters by \cite{Voit17}.

\subsection*{I\MakeLowercase{on} C\MakeLowercase{olumn} D\MakeLowercase{ensity}}

The column density of ion $i$ at an impact parameter $h$ in a spherically symmetric halo is
\begin{equation}\label{eq:Ax1}
N_i(h) = 2 A_{\rm i} \int_0^z{n_{\rm H}(r)Z'(r) f_{V}(r) f_{\rm ion,i}(r) dz'} ~~~,
\end{equation}
where $r^2=h^2+z^2$, $A_{\rm i}$ is the solar abundance of the element corresponding to ion~$i$, $Z'$ is the metallicity relative to solar, $f_{V}$ is the volume filling factor of the gas containing ion $i$ and $f_{\rm ion,i}$ is the ion fraction. We assume a power-law variation of the density, $n_{\rm H} \propto r^{-a_n}$, metallicity, $Z' \propto r^{-a_Z}$, filling factor, $f_{V} \propto r^{-a_V}$, and ion fraction, $f_{\rm ion,i} \propto r^{-a_f}$. We then have
\begin{equation}\label{eq:Ax2}
N_i(h) = 2 A_{\rm i} f_{\rm ion,i}(h) n_{\rm H}(h) Z'(h) f_{\rm V}(h) \int_0^z{\frac{dz'}{(r/h)^{a}} } ~~~,
\end{equation}
where $n_{\rm H}(h) = n_{\rm H}(r=h)$, etc., and $a = a_n + a_Z + a_V + a_f$ ($a>0)$. Let
\begin{equation}\label{eq:Ax3}
y' \equiv \frac{z'}{h} = \left( \frac{r^2}{h^2} - 1 \right)^{1/2}  ~~~,
\end{equation}
and let $R$ be the virial radius of the Galaxy, close to the outer radius of the CGM. We then define
\begin{equation}\label{eq:Ax4}
I_a(y) \equiv \frac{1}{R} \int_0^z{\frac{dz'}{(r/h)^{a}}} = 
\frac{1}{(1+y^2)^{1/2}} \int_0^y{\frac{dy'}{(1+y'^2)^{a/2}}} ~~~,
\end{equation}
and get
\begin{equation}\label{eq:Ax5}
N_i(h) = 2 A_{\rm i} f_{\rm ion,i}(h) n_{\rm H}(h) Z'(h) f_{\rm V}(h) R I_{a} ~~~.
\end{equation}

If we restrict our attention to normalized impact parameters in the range $0.3<h/R<0.9$, which contains most of the COS-Halos measurements (see Figure~\ref{fig:new_ovi}), then for $a$ between $1$ and $2$, $I_a = 0.50 \pm 0.13$~dex (or $0.50 \pm 0.18$~dex for $a=0.5-2.5$).

\subsection*{L\MakeLowercase{imit on the} C\MakeLowercase{ooling} T\MakeLowercase{ime}}

Let the rate of radiative net cooling per unit volume be $n_{\rm e} n_{\rm H} \Lambda$. We assume that the gas is irradiated by the metagalactic radiation field (MGRF). The cooling function, $\Lambda$, is then a function of the gas density, temperature, and metallicity (see \S\ref{sec_timescales} here and \citealp{Gnat17}). The isochoric gas cooling rate is then
\begin{equation}\label{eq:Ax7}
\tcool = \frac{3 n \kb T}{2 n_{\rm e} n_{\rm H} \Lambda(T,n,Z)} 
\end{equation}
where we have adopted $n_{\rm He} = n_{\rm H}/12$, and assumed that the gas is fully ionized. For the metallicity scaling, $\Lambda = Z'\Lambda_{\odot}$, we neglect cooling due to H and He, so this an upper limit on the cooling time. Inserting the expression for $n_{\rm H}(h)Z'(h)$ from Equation \ref{eq:Ax5}, we get
\begin{equation}\label{eq:Ax8}
\tcool(h) = 5.8 A_{\rm i} f_{\rm V}(h) 
\left[ \frac{\kb T(h) f_{\rm ion,i}(h)}{\Lambda_{\odot}(T,n)} \right] \frac{R I_a}{N_i(h)} ~~~.
\end{equation}
In this expression the uncertain metallicity $Z’$ does not appear, and the cooling time is inversely proportional to the observable column density.

We now apply this to OVI. We assume that the warm/hot gas is volume filling, so that $f_{\rm V}=1$ and $a_{\rm V}=0$. This gives an upper limit for the cooling time, consistent with the rest of our analysis here. The filling factor of the warm/hot, OVI-bearing gas in our model is unity. The solar abundance of oxygen is $A_{\rm O} = 4.9 \times 10^{-4}$, and as we estimated above, $I_a \approx 0.50$. For our estimate here we take $R=260$ kpc, the median virial radius of the COS-Halos star-forming galaxies, and close to the MW virial radius in our model (see \S\ref{sec_fiducial}). Given the shape of the cooling function and the OVI ion fraction in the density-temperature space, the expression $\kb T f_{\rm ion,i}/\Lambda_{\odot}$ is bound from above for gas at $T>10^{5}$~K. For the HM12 MRGF at $z=0.2$, $\kb T f_{\rm ion,OVI}/\Lambda_{\odot} \leq 4.6 \times 10^{10}~{\rm s~cm^3 }$, and the maximum occurs at $T \sim 3.5 \times 10^{5}$~K, at densities above $n_{\rm H} \geq 10^{-4}$~\cmv, where the OVI is in CIE and $f_{\rm ion,OVI}$ is maximal (see \S\ref{subsec_ionization}). We insert this value into Eq. \eqref{eq:Ax8} to obtain a {\it model-independent} upper limit for the cooling time at $r=h$
\begin{equation}\label{eq:Ax9}
\tcool(r=h) \lesssim 5.6~{\rm Gyr}~
\left( \frac{R}{260~{\rm kpc}} \right) \left( \frac{N_{\rm OVI}(h)}{3 \times 10^{14} ~\cmc} \right)^{-1}  ~~~.
\end{equation}
This approximation is valid for $0.3<h/R<0.9$, through $N_{\rm OVI}(h)$, and we scaled the column density to the value measured in COS-Halos at $h/R \sim 0.6$ (see Figure~\ref{fig:new_ovi}). 

\subsection*{C\MakeLowercase{omparison with the} D\MakeLowercase{ynamical} T\MakeLowercase{ime}}

The dynamical time used by \cite{Voit17} is $\sqrt{2r/g(r)}$, where $g(r)$ is the acceleration due to gravity. Scaling gives
\begin{equation}\label{eq:Ax11}
\tdyn(r) = 2.8~{\rm Gyr} \left( \frac{r}{260~{\rm kpc}} \right)^{3/2} \left( \frac{M(r)}{10^{12}~\msun} \right)^{-1/2} ~~~.
\end{equation}
We can fit the Klypin MW mass profile at large radii, where it is approximately an NFW profile, with a power law, giving
\begin{equation}\label{eq:Ax12}
M(r) \approx 10^{12}~\msun~\left( \frac{r}{260~{\rm kpc}} \right)^{0.56} ~~~ ({\rm 130~kpc} < r < {\rm 260~kpc}) ~~~.
\end{equation}
Inserting this into Equation~\eqref{eq:Ax11} results in
\begin{equation}\label{eq:Ax13}
\tdyn(r) \approx 2.8~{\rm Gyr}~\left( \frac{r}{260~{\rm kpc}} \right)^{1.22} ~~~.
\end{equation}
We note that unlike our expression for the cooling time upper limit, this approximation for \tdyn~is valid all the way out to $R$.

We can then define $\zeta(r) \equiv \tcool(r)/\tdyn(r)$ and write the upper limit of this ratio for the typical column of OVI as
\begin{equation}\label{eq:Ax14}
\zeta(r=h) < 2.0 ~ \left( \frac{h}{260~{\rm kpc}} \right)^{-1.22} \left( \frac{N_{\rm OVI}(h)}{3 \times 10^{14} ~\cmc} \right)^{-1} ~~~.
\end{equation}
Accounting for the uncertainty factor in the value of $I_a$ gives ratios in the range of $1.5-2.6$ for $1<a<2$, corresponding to a factor of $1.3$ uncertainty; for $0.5<a<2.5$, the range is $1.3-3.0$, or a factor $1.5$ uncertainty. Our approximation and the derived upper limit are valid for $0.3 < h/R < 0.9$, and the column density we used is measured at $h/R=0.6$, corresponding to $h=156$~kpc. Inserting this impact parameter we get an observed upper limit $\zeta < 3.7$ (in the range $2.8-4.8$ for $1<a<2$). This is significantly lower than values of $\zeta \sim 10$, found by \cite{Sharma12a} in simulations and by \cite{Voit17} in observations of galaxy clusters.

To summarize, we find that for warm/hot gas with an OVI column density of $\sim 3 \times 10^{14}$~\cmc~at large impact parameters, observations set an upper bound $\zeta \lesssim 5$. This limit includes the uncertainty in the underlying ion volume density distribution. It is also independent of the exact gas metallicity, as long as the gas cooling in the relevant temperature range is dominated by metals ($Z' \gtrsim 0.1$). A ratio $\zeta \sim 10$ would require OVI columns significantly lower than observed in the CGM of $L^*$ galaxies.


\addcontentsline{toc}{section}{References}

\bibliographystyle{yahapj}


\end{document}